\documentclass{emulateapj-rtx4}
\usepackage{lscape}

\shorttitle{Color Gradients of Elliptical Galaxies}
\shortauthors{Kim \& Im}
\begin{document}
\title{Optical-Near Infrared Color Gradients and Merging History of Elliptical Galaxies}

\author{\textbf{Duho Kim}\altaffilmark{1,2} \& \textbf{Myungshin Im}\altaffilmark{1}}

\altaffiltext{1}{Center for the Exploration of the Origin of the Universe (CEOU), 
Astronomy Program, Department of Physics and Astronomy, Seoul National University, 
1 Gwanak-rho, Gwanak-gu, Seoul 151-742, South Korea}
\altaffiltext{2}{School of Earth and Space Exploration, Arizona State University, P.O. Box 871404, Tempe, AZ 85287-1404, USA}

\begin{abstract}
It has been suggested that merging plays an important role in the formation
and the evolution of elliptical galaxies. While gas dissipation by star formation
is believed to steepen metallicity and color gradients of the merger products, mixing of stars through dissipation-less merging (dry merging)
is believed to flatten them. In order to understand the past merging history of
elliptical galaxies, we studied the optical-near infrared
(NIR) color gradients of 204 elliptical galaxies. 
 These galaxies are selected from the overlap region of the Sloan Digital Sky Survey (SDSS) Stripe 82 and the UKIRT Infrared Deep Sky Survey (UKIDSS) Large Area Survey (LAS). The use of the optical and the NIR data ($g, r$, and $K$) provides large wavelength baselines, and breaks the age-metallicity degeneracy, allowing us to derive age and metallicity gradients. The use of the deep SDSS Stripe 82 images make 
it possible for us to examine how the color/age/metallicity gradients 
are related to merging features.
 We find that the optical-NIR color and the age/metallicity gradients of elliptical galaxies with tidal features are consistent
 with those of relaxed ellipticals suggesting that the two populations underwent
a similar merging history on average and that mixing of stars was more or less
completed before the tidal features disappear. 
Elliptical galaxies with dust features have steeper color gradients than the other two types, even after masking out dust features during the analysis, which can be due to a process involving wet
merging. More importantly, we find that the scatter in the color/age/metallicity gradients of the relaxed
 and merging feature types, decreases as their luminosities (or masses) increase at 
$\mathrm{M} > 10^{11.4} \, \mathrm{M_{\odot}}$ but stays to be large 
at lower luminosities. 
 Mean metallicity gradients appear nearly constant over the explored mass range, 
 but a possible flattening is observed at the massive end.  
 According to our toy model that predicts how the distribution of  metallicity gradients changes as a result of major dry merging, 
the mean metallicity gradient should flatten by 40\%  
 and its scatter become smaller by 80\% per a mass doubling scale 
 if ellipticals evolve only through major dry merger. 
  Our result, although limited by a number statistics at the massive end, is consistent with
 the picture that major dry merging is an important 
 mechanism for the evolution for ellipticals at $\mathrm{M} > 10^{11.4} \, \mathrm{M_{\odot}}$,  but less important at the lower mass range.  
\end{abstract}

\keywords{galaxies: elliptical and lenticular, cD --- galaxies: evolution --- galaxies: interactions --- galaxies: peculiar}

\section{INTRODUCTION}

  Elliptical galaxies are featureless, elliptical stellar systems, predominantly
  made with old stars. 
  They are a prime object in the currently popular hierarchical galaxy formation 
  models under the $\Lambda$CDM universe, since in a cold-dark matter dominated
  universe, galaxies grow through merging with other galaxies, and it is 
  well-known from simulations that merging between galaxies with comparable masses
  (called major merger) produce galaxies with the characteristics of ellipticals
  \citep{too77,bar92,naa06}. Therefore, studying elliptical galaxies is important to test our 
  current theoretical understanding of the galaxy formation and the evolution mechanism.
 
  Many observational works have been performed for elliptical galaxies near and far,
  but our understanding on the evolution of elliptical galaxies is still far from 
  complete. 
  On one hand, there are pieces of observational evidence for a complex history
  in the evolution of elliptical galaxies in support of the hierarchical merging picture.
  For example, very deep imaging data reveal more than 70\% of elliptical galaxies with tidally 
  disrupted features that are indicative of their past merging history \citep[e.g.,][]{tal09,van05,she12}. 
  Small levels of recent star formation are found as blue cores in some ellipticals \citep[e.g.,][]{im01,lee06}
 and the UV excess light \citep{yi05} or as excessive MIR fluxes \citep{ko09,ko12,shi11,hwa12}.
  These results break a simplistic view of the monolithic collapse model \citep{egg62,lar74} that
 ellipticals are old galaxies with little merging nor recent star formation activities.
  While these observational evidence point toward the hierarchical build-up of elliptical
 galaxies, other properties of ellipticals do not bode well with the merger scenario.
  Tight scaling correlations relations are observed for early-type galaxies, such 
 as the fundamental plane relation \citep[e.g.,][]{djo87,jun08}, 
 and the color-magnitude relation \citep{bau59,vis77}.
  Also, the number density of early-type galaxies does not evolve as drastically as 
 some of the simple hierarchical merging models have predicted \citep{im99,im02,sca07}. 
  
   The observational results that appear contradictory to each other can 
 be reconciled by introducing dry merging in the evolution of elliptical galaxies. 
   Dry (or dissipation-less) merging is a process where two gas-poor galaxies with old stellar 
  population merge together to create a more massive, old galaxy.
   Although a dissipative merging between gas-rich galaxies (wet merging) 
  inevitably introduces 
  scatters in various scaling relations and too many blue ellipticals, 
  the dry merging offers a way to avoid such a situation while allowing 
  frequent merging activities in the hierarchical merging models. This is possible
  since dry merging preserves the red colors of the already old pre-mergers, 
  avoiding the introduction of additional scatters in the scaling relations
  arising from new star formation as is the case for
   wet merging. Dry merging is also gaining popularity as a possible mechanism to explain
  the observed rapid growth in sizes of early-type galaxies from high to low redshifts  
  \citep{van08,tru06,dad05}

   One of interesting consequences of dry merging
  can be found in the radial gradient of various properties, such as the metallicity,
  the age, and the colors of ellipticals.
   It has been known that centers of low redshift elliptical galaxies are redder than 
 their outskirts \citep[e.g.][]{fra90,pel90,tam03}, and
 this radial change of color is called the color gradient. 
   The color gradient is in general attributed to a gradient in metallicity 
  although a small amount of the age gradient is known to contribute to the color gradient,
  where low redshift ellipticals show the higher metallicities 
  and the younger stellar ages at the center
  than in the outskirt. According to line indices analysis studies, 
  the metallicity gradients have a mean value of about $-0.2$ to $-0.3$ per a decade change in radius
  while the slope of the age gradient is known to be of order of $-0.1$ or less \citep{oga05,spo10,kun10}.
   The metallicity and the age gradients can be derived from color profiles and such studies 
  tend to implicate the metallicity gradients of order of $-0.4$ \citep{lab10,tor12},
  steeper than the spectroscopically determined metallicity gradients. The origin of the discrepancy 
  between the spectroscopic and the photometric studies 
  is not clear, but could be related to a difference in the explored range of radius.
   In the spectroscopic studies, the gradients are determined within $r_{eff}$ (effective radius),
  while the photometric studies extend the radius coverage to a few $\times$ $r_{eff}$.

   When a monolithic collapse of the proto-galactic gas creates an elliptical galaxy, star formation can occur 
  over several generations at center using the gas that is retained by a strong gravitational potential
  in the central part of the galaxy.
  The gas is enriched by previous episodes of star formation, so stars in the central part 
  are expected to have a high metallicity.  On the other hand, the star formation
  truncates earlier in the outer region of an elliptical galaxy since the gas is lost easily due 
  to a shallow gravitational potential preventing further episodes of star formation.
   The net result is that an elliptical would have
  stars with higher metallicity and a bit younger mean stellar age at the inner region than
  the outer region, creating the metallicity, the age, and the color gradients that are   
  in qualitative agreement with the observed results. A similar situation is expected to
  occur after mergers of two gas-rich galaxies. 
   However, according to the formation models of ellipticals through monolithic collapses of
  proto-galactic gas, the metallicity gradient is predicted to be quite steep, 
  with $−0.5$ to $−1$ in the logarithmic value of metallicity per one dex change in radius 
  \citep{lar74,car84,kaw03}. Simulations that implement merging activities predict 
  that the metallicity gradient is shallower at
  $-0.2$ -- $-0.3$ \citep[e.g.,][]{kob04,hop09}.

    Dry merging modifies the above theoretical picture in the following way.
  When two gas-poor galaxies with a comparable mass merge (``major dry merging''), a re-distribution of stars
   dilute the metallicity gradient, therefore flattening it in the post-merger \citep[e.g.][]{whi80,ko05,dim09}.
    An opposite trend is suggested to occur during minor dry merging, i.e., merging between
   ellipticals with very different masses (mass ratios of 5--10 to 1 or larger), where 
   stars in a minor companion galaxy get deposited at a large distance from the center of
 the main pre-merging galaxy, avoiding stars to be mixed throughout the main galaxy, contrary  
 to the case of major dry merging \citep{ber11}. 
    Observations offer somewhat mixed views about the importance of dry merging, and 
   the results from some of such studies are presented below.   
    \citet{roc10} find that color gradients of the brightest cluster galaxies (BCGs) are 
  flatter that those of non-BCGs. They conclude that a great number of dry merging events flattened
  C-M relation and individual BCG.     
    \citet{ko05} find color-gradients are flatter for elliptical galaxies in the cluster environment 
  than in field, suggesting more frequent merging history in the high density environment. 
   \citet{spo10} derived the metallicity gradients of 51 nearby early-type galaxies
  using their spectroscopic data and find a flattening trend in the metallicity gradients toward 
  a large galaxy mass but with a large scatter.
   Their interpretation of the result is that galaxies with a few $\times$ $10^{10} \, \mathrm{M_{\odot}}$ are
  created through wet merging, but evolved via dry merging thereafter. The increase in the scatter
  of the metallicity gradient is also found in other studies \citep[e.g.,][]{oga05}, and some works  
  interpret that the increase in the scatter is suggested is caused by 
  dry merging of objects with a variety of metallicity gradients
   \citep[e.g.,][]{dim09}. However, such an interpretation requires 
  galaxies with steep gradients to create merger products with a large scatter, and 
  the number of such galaxies with steep gradients should decrease as galaxies experience
  successive dry merging events. This will inevitably reduce the scatter in the mass-metallicity
  gradient relation at high mass end, which is inconsistent with the observed trend.   Also,
  \citet{pip10} show that the large dispersion can be explained under the frame work of 
  the monolithic collapse models where various star formation efficiencies produce 
  the difference in gradients with the most efficient star formation creating the steepest
  gradient.

  In this paper, we will examine the color/age/metallicity gradients and their connection to 
  dry merging, analyzing the optical and the NIR imaging data of
  a sample of 204 nearby elliptical galaxies
  that are divided into three different types -- ones showing no merging feature,
  ones with tidal-feature (as a result of merging), and ones with dust-feature. 
  We will also present a simple model that predicts the evolution of the mass-metallicity
  gradient relation under major dry merging. This model will be used to interpret our result.  
   
  The use of the optical and the NIR data allows us to break the age-metallicity degeneracy
  that usually hampers the interpretation of colors of early-type galaxies with
  optical data alone \citep{wor94,sag00,bel00,li07}. The large wavelength 
  baseline between the optical and the NIR are also advantageous in deriving color gradients
  which are known to be shallow in optical bands \citep{hin01,lab10}. 
  Also, NIR data, especially $K$-band, of an elliptical galaxy is known to be a good indicator
  of stellar mass, since mass-to-light ratios ($M/L$) in NIR are 
  less sensitive to galaxy types
  or ages than in the optical \citep{bel01,pah95,jun08}. Here, we will use 
  the $K$-band light as a mass indicator of elliptical galaxies.

   Our sample will be drawn from a sample of early-type galaxies in 
  \citet{kav10} which classifies early type galaxies based on merging features. 
  They are galaxies with very deep imaging data reaching down to the surface brightness (SB)
  limit of $\mu_{r} \sim26~\mathrm{mag}~\mathrm{arcsec}^{-2}$, enabling us to subdivide and examine
  elliptical galaxies into three different types that reflect different stages of merging.
   The relaxed types are objects without strong merging feature. If they had undergone
  merging in the past, and they should have done so a long time ago ($>$ several Gyrs). 
  Presumably, such objects should have well-established gradient properties with 
  tighter scatters.
   The tidal feature types are objects that underwent merging relatively
  recently ($<$ several Gyrs). They may or may not share the gradient properties
  similar to the relaxed types, but we can expect the gradient properties to be
  less established than the relaxed-types with a somewhat larger scatter.
   Finally, the dust-feature types are objects with dust and gas, and they are 
  suggested to be involved in a merger with a gas-rich companion \citep[e.g.][]{kav12}.
   Therefore, such objects need to be considered separately when examining dry merging.
   

Throughout this paper, we adopt a concordance cosmology of 
$\Omega_{m} = 0.3$, $\Lambda = 0.7$, and $H_{0} = 70$ km sec$^{-1}$ Mpc$^{-1}$ 
\citep[e.g.][]{im97}, and AB system for magnitudes.

\section{DATA}

\subsection{The Sample}

The elliptical galaxies that we study in this paper are selected from a catalog of nearby ($z < 0.05$), luminous ($M_{r}< -20.5$) early-type galaxies in the SDSS Stripe 82 imaging area which is located along the celestial equator in the Southern Galactic Cap ($-50{\degr}<{\alpha}<59{\degr}$,$-1.25{\degr}<{\delta}<1.25{\degr}$; \citealt{kav10}; hereafter K10). The reason why we chose the catalog is that it accompanies images that are $\sim$2 magnitudes deeper than SDSS single scan images, enabling us to identify faint merging features like tidal tails for sub-classification of early-type galaxies. In K10, they classified galaxies into various morphological classes (early-type galaxies and late-type galaxies and 'Sa-like' systems which are bulge-dominated galaxies with faint spiral features) through a  visual inspection of both monochromatic Stripe 82 and multi-color SDSS standard-depth image. In K10, early-type galaxies are divided further into 5 types : early-types without any merging features, early-types with tidal features, dust features, both the dust and the tidal features, and early-types that are interacting. For our study, we regrouped the early-type galaxies into three types -- 238 objects without merging features (relaxed type), 38 objects with tidal features (tidal-feature type) and 22 objects that exhibit dust features including cases showing both tidal and dust features (dust-feature type). We excluded the interacting types
in K10 since the number of such object is small and they in general pose a challenge during the surface brightness fitting process. Overall, the ``relaxed'' early-types are considered as objects that experienced mergers long time ago so that their merging features disappeared. The ``tidal-feature'' early-types are considered as objects that experienced a dry merger or a wet merger relatively recently ($\lesssim$ a few Gyrs ago) and the early-types with dust-lanes are considered as objects that experienced a wet merging recently. 
Figures 1 -- 3 show example images of the three types of early-type galaxies. 

In the next step, we cross-matched the K10 early-type galaxies with the NIR data from UKIDSS LAS. During this process, we also excluded 18 elliptical galaxies that have bright objects in the vicinity 
(to an extent that they cannot be masked out) since a reliable fitting of their surface brightness profiles becomes difficult under the presence of the bright neighbors. These neighbors are foreground stars or galaxies. 
Example images of the early-types with bright neighbors are presented in Figure 4 (the top two panels).

 A visual inspection of the K10 early-type galaxies reveal objects with disks (23\% of the sample). 
The presence of the disk makes our color gradient analysis complicated since the stellar populations in disks 
may significantly differ from those in bulges. Also, objects that underwent major merging events are likely to have  a non-significant disk component. Therefore, we excluded early-type galaxies with a significant disk component (which we suspect mostly are S0's or S0a's). This is done by matching the K10 early-type galaxies with those in a catalog created by \citet{sim11} which provides the bulge to total light ratio ($B/T$) of SDSS galaxies from a 2-dimensional surface brightness fitting using GIM2D \citep{sim99}. In general, elliptical galaxies have $B/T \gtrsim 0.5$ in the optical wavelengths longer than 0.4 micron \citep[e.g.][]{im02,sim86}, therefore we select early-types with $B/T \gtrsim 0.5$, and call them ``elliptical galaxies''. 
Most of the K10 early-type galaxies are found in the catalog of \citet{sim11}, but some were not, 
because they are too bright and extended for the automated GIM2D pipeline to handle them properly. 
For such early-types that are not in the catalog of \citet{sim11}, we performed a 2-dimensional surface brightness fitting using GALFIT \citep{pen10}, and rejected those with $B/T < 0.5$ (5 such cases; see \S\ 3.1). 
See Figure \ref{disky} for disky early-types rejected by this procedure.

After the rejection of the disky early-type galaxies, 206 ``elliptical'' galaxies remain 
in the K10-LAS matched sample. Additionally 2 elliptical galaxies are excluded from the sample since they are too small (the outer boundary for the color gradient range is smaller than $\sim7$ arcsec) for deriving reliable color gradients.  
Images of these small early-types are given in Figure \ref{rej} 
(the bottom two panels).
Finally, we have 162 relaxed, 32 tidal-feature and 10 dust-feature type elliptical galaxies in the final sample
for our analysis as indicated in Table \ref{tbl1}.

\subsection{Optical and NIR Data}

We used the SDSS Stripe 82 imaging data for the study of optical properties. The Stripe 82 data are taken from the SDSS Data Release 7 \citep{aba09}. A co-addition of about 50 SDSS single scans provides $ugriz$ images $\sim$2 mags deeper (SB limits of $\mu_{r} \sim26~\mathrm{mag}~\mathrm{arcsec}^{-2}$) than the main SDSS images of individual scan. The $g$-band and $r$-band images were downloaded from the SDSS Data Archive Server (DAS), and they were used to  derive surface brightness profiles. We find that the optical band seeing is not uniform with the range of $1.0$ -- $1.5\arcsec$. This point will become important when deriving optical-NIR color gradient as we shall describe later. The pixel scale of the SDSS images is 0.3961\arcsec

For the NIR images, we used the Data Release 8 of UKIDSS LAS \citep{law07}. Most but not all of the SDSS Stripe 82 region was covered ($-25\degr\ < \alpha < 60 \degr\ , -1.25\degr\ < \delta < 1.25\degr\ $) in $YJHK$-bands by UKIDSS LAS. Typical seeing of an LAS image is 1\arcsec ~FWHM in $K$-band. The depth of the UKIDSS LAS data is shallower than the SDSS Stripe 82 data (SB limit is $\mu_{K}\simeq 20~\mathrm{mag~arcsec^{-2}}$), so the fitting range for an optical-NIR color gradient is limited by the NIR data. We limited ourselves to the use of only $K$-band data, since $K$-band provides the maximal wavelength baseline in conjunction with the optical bands, 
and the use of $J$ or $H$-bands appear redundant for the stellar population analysis
\citep[e.g.][]{gildepaz02}. 
Although inclusion of more passbands, 
especially those in UV, is likely to improve the analysis, we limit ourselves to 
the use of $g, r$, and $K$-band data for simplicity here. 
The $K$-band images of our sample galaxies are downloaded from the WFCAM science archive\footnote{http://surveys.roe.ac.uk/wsa}. 
The pixel size of an LAS image is 0.4\arcsec, comparable to but a bit larger than the pixel size of the SDSS images.

\section{ANALYSIS}

\subsection{Surface Brightness Profile}

\begin{figure}
\epsscale{1.0}
 \includegraphics[scale=0.70]{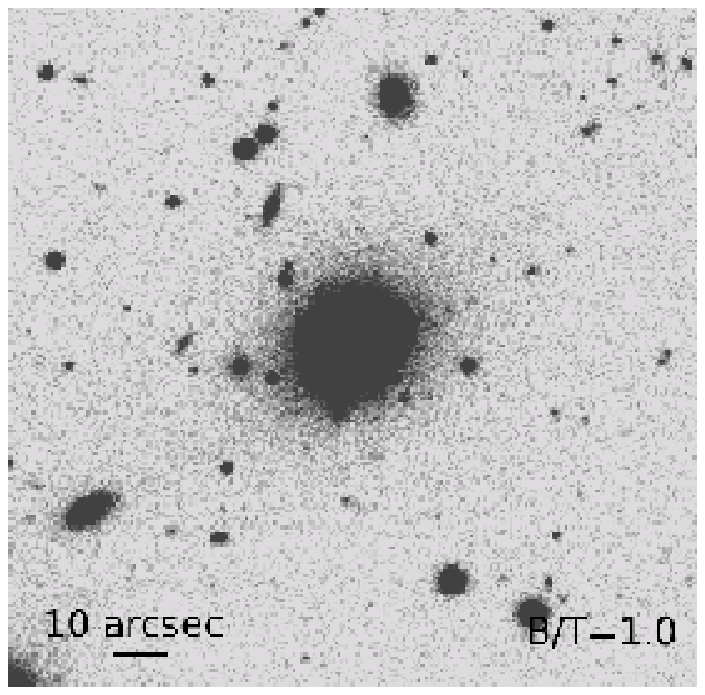}
 \includegraphics[scale=0.70]{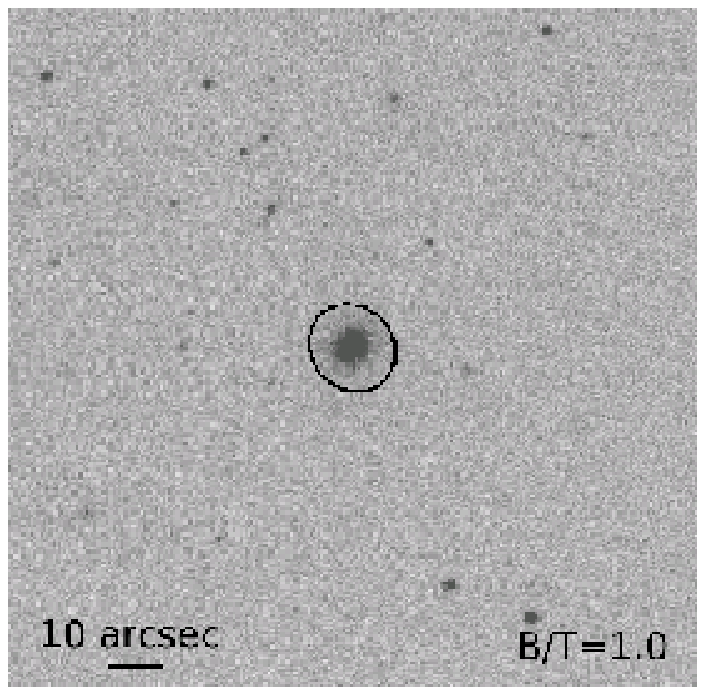}
 \includegraphics[scale=0.42]{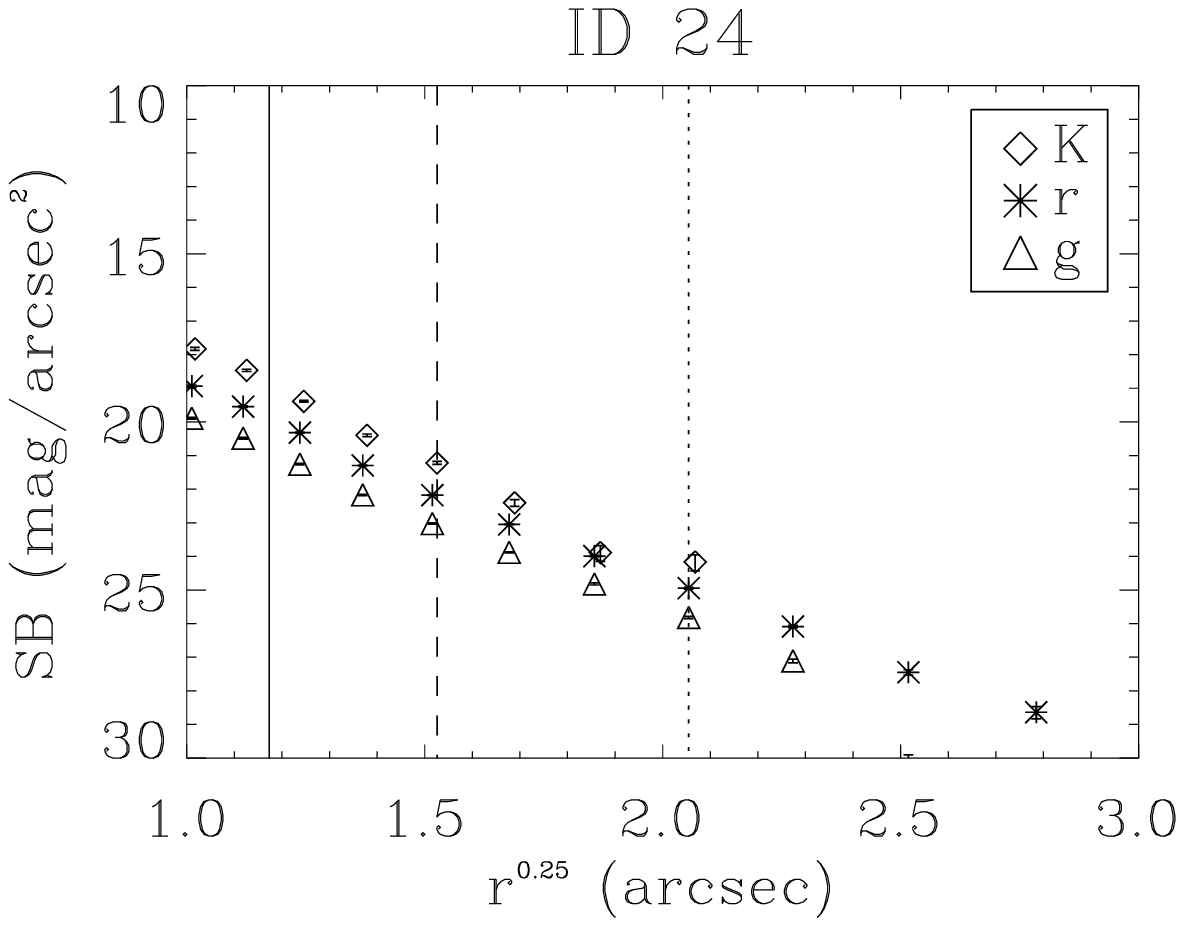}\\
 \includegraphics[scale=0.70]{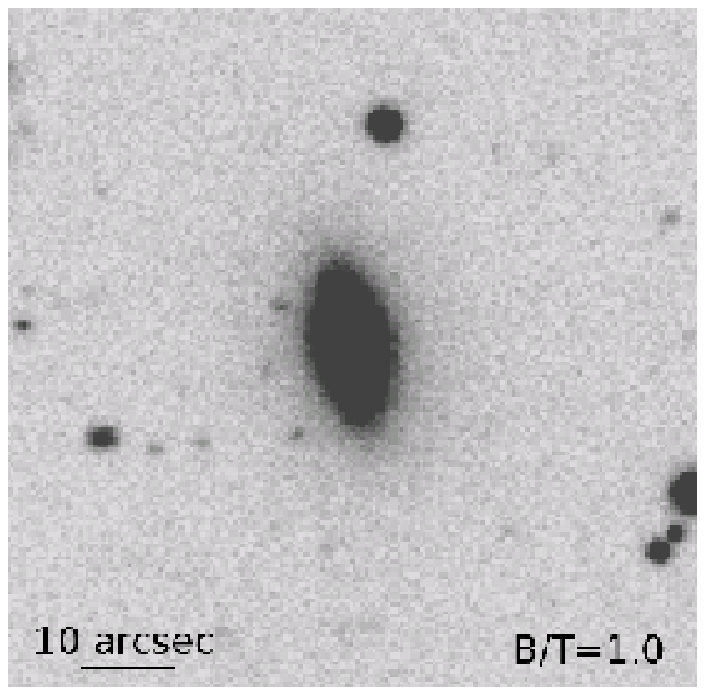}
 \includegraphics[scale=0.70]{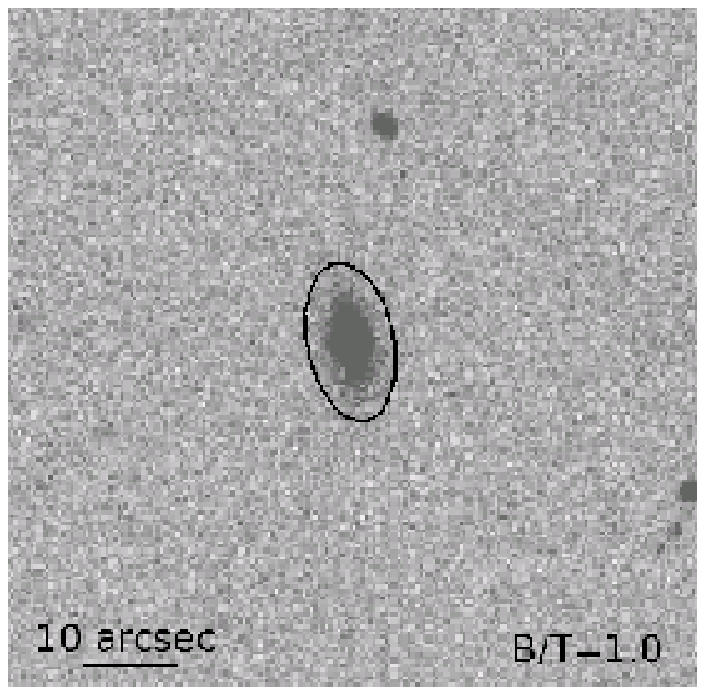}
 \includegraphics[scale=0.42]{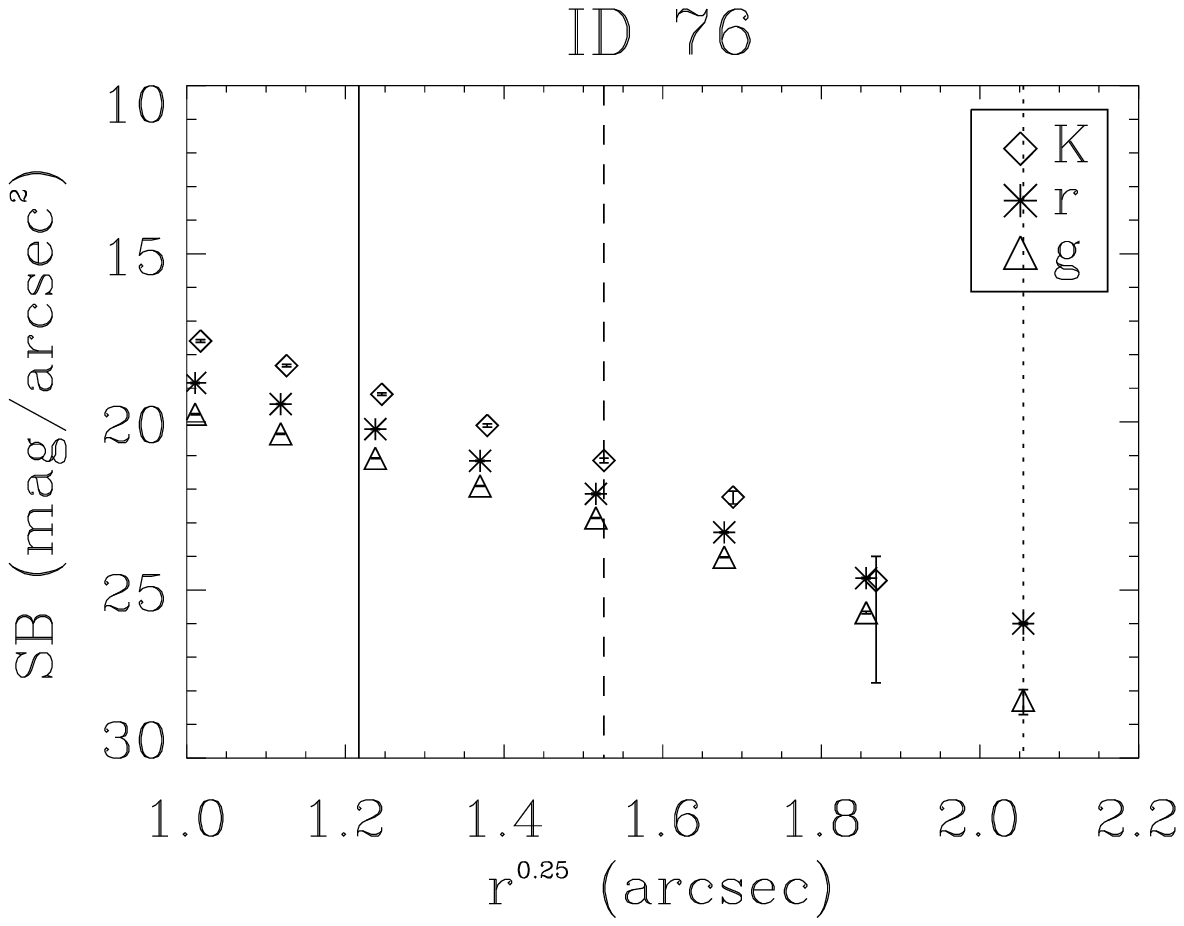}\\
 \includegraphics[scale=0.70]{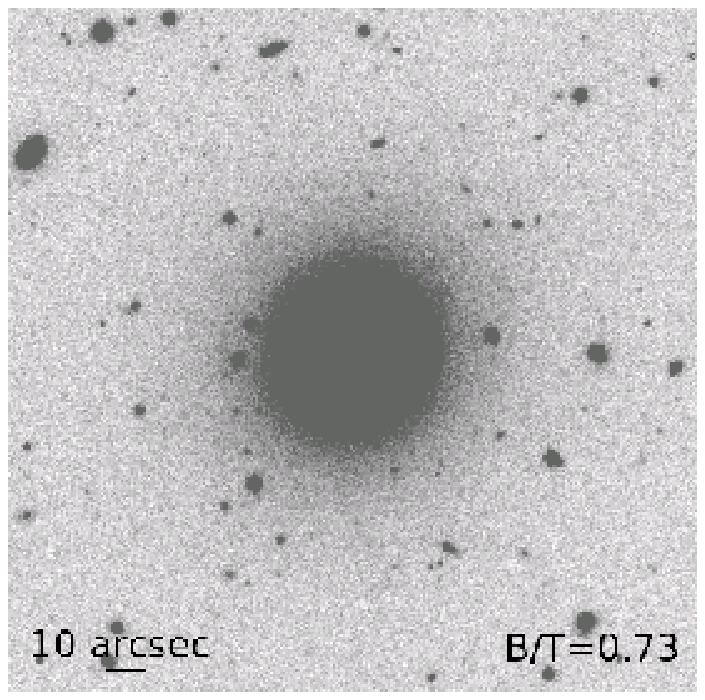}
 \includegraphics[scale=0.70]{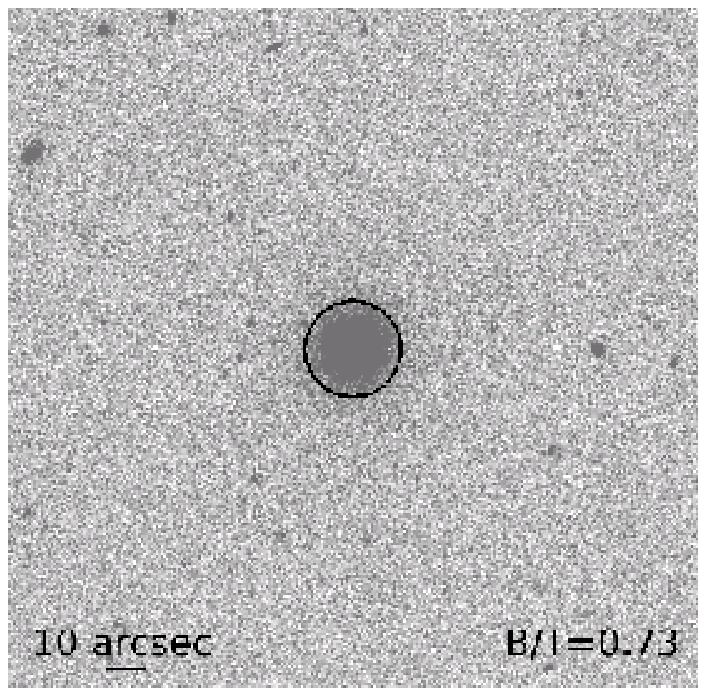}
 \includegraphics[scale=0.42]{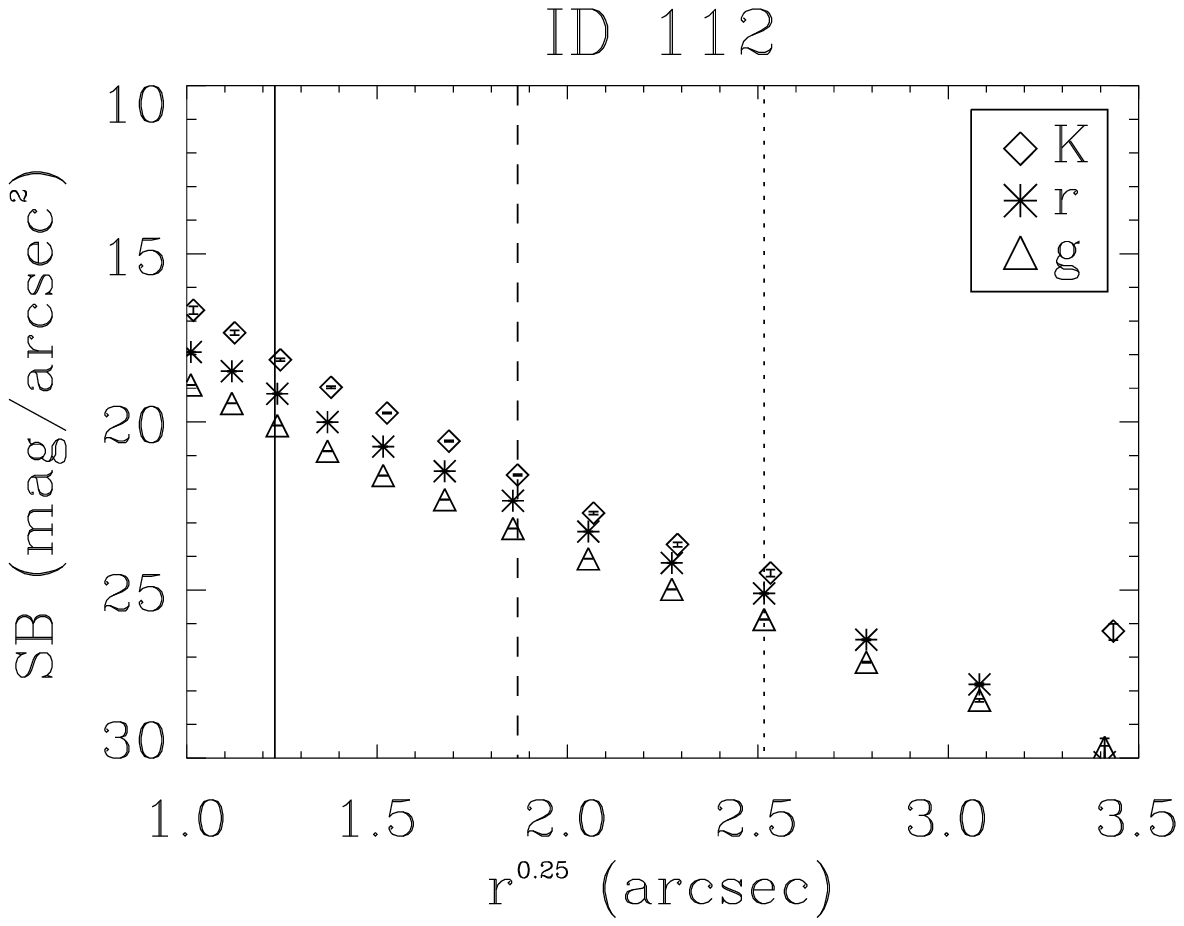}\\
 \includegraphics[scale=0.70]{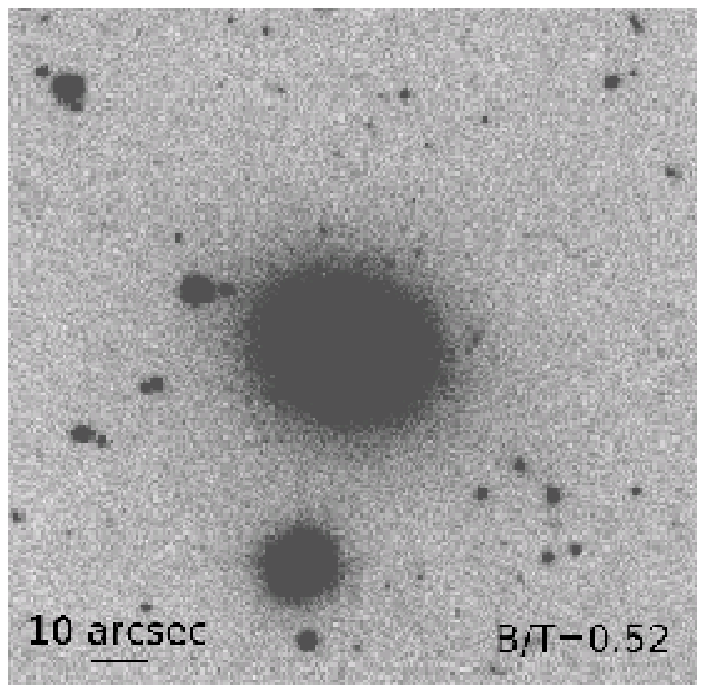}
 \includegraphics[scale=0.70]{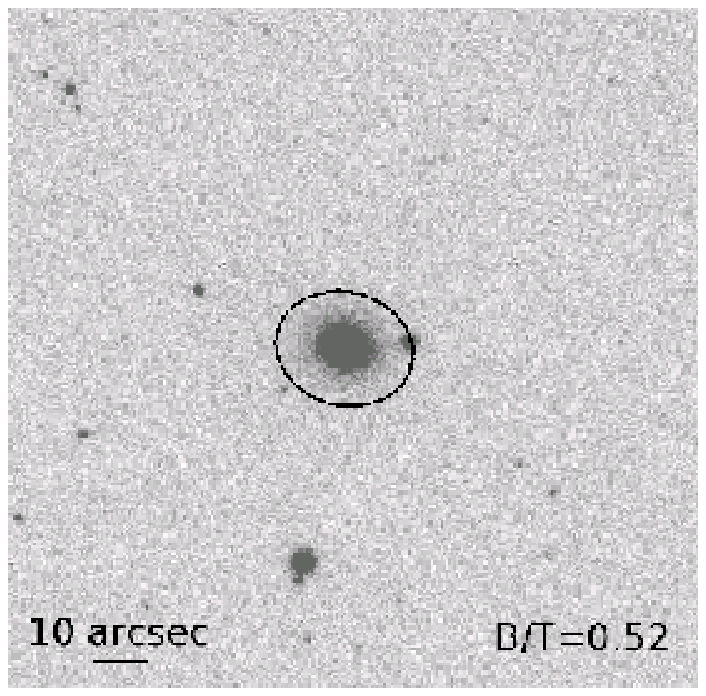}
 \includegraphics[scale=0.42]{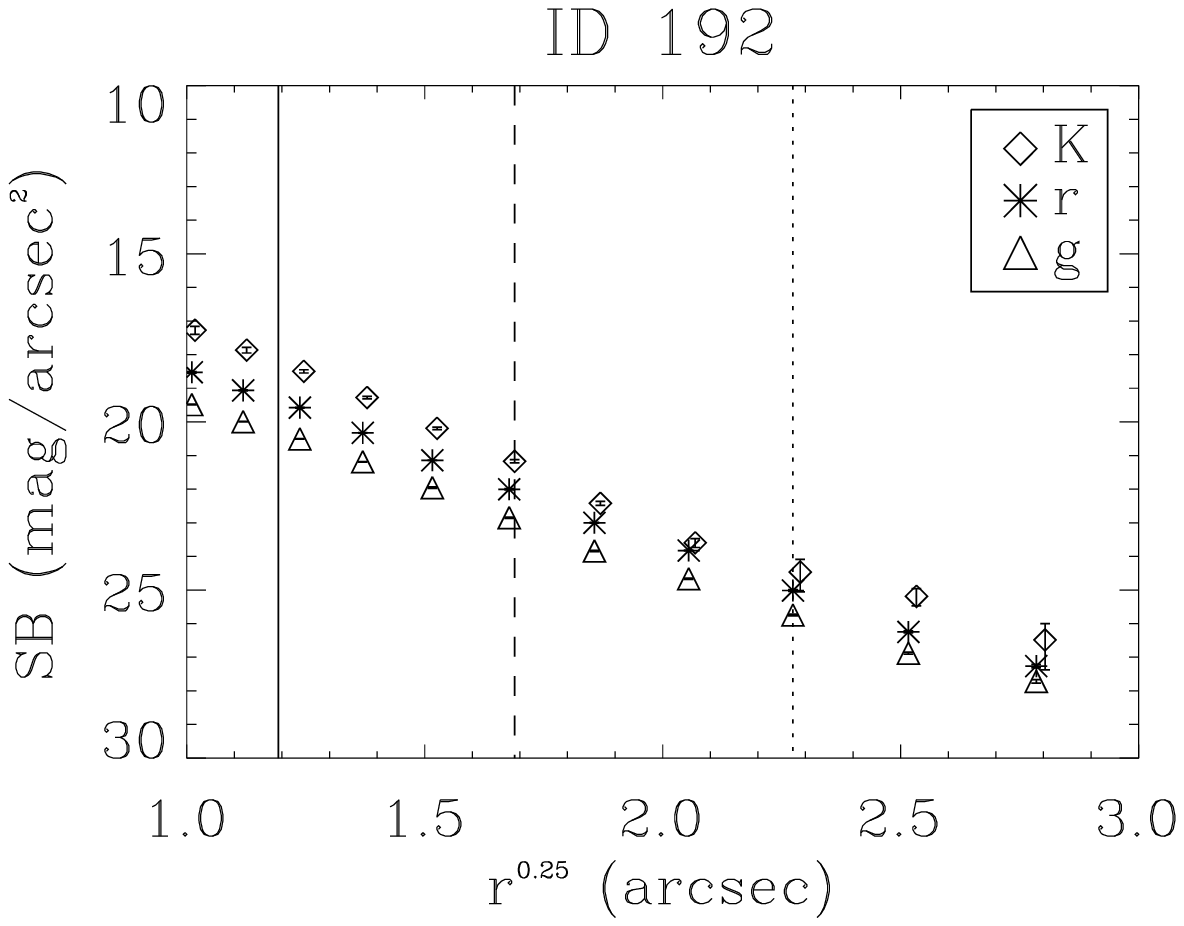}
\caption{Examples of the relaxed elliptical galaxies in $r$-band (left), in $K$-band (middle) and their surface brightness profiles in $g$, $r$ and $K$-bands (right). The scale bar and $B/T$ in the images represent 10 arcsec and bulge to total light ratios. The vertical solid lines and the dashed lines in the plot are the inner and the outer radii for deriving
 $r-K$ and $g-K$ color gradients. The vertical dotted line means the outer boundary for the $g-r$ color gradient fit. The ellipses in the 
$K$-band images indicate the outer boundary for the analysis of color profiles for the $K$-band. 
The ID on top of the SB profile figures indicate the ID of the objects in 
\citet{kav10} as indicated in Table \ref{tab_sp}.}
\label{rel}
\end{figure}

\begin{figure}
\epsscale{1.0}
 \includegraphics[scale=0.70]{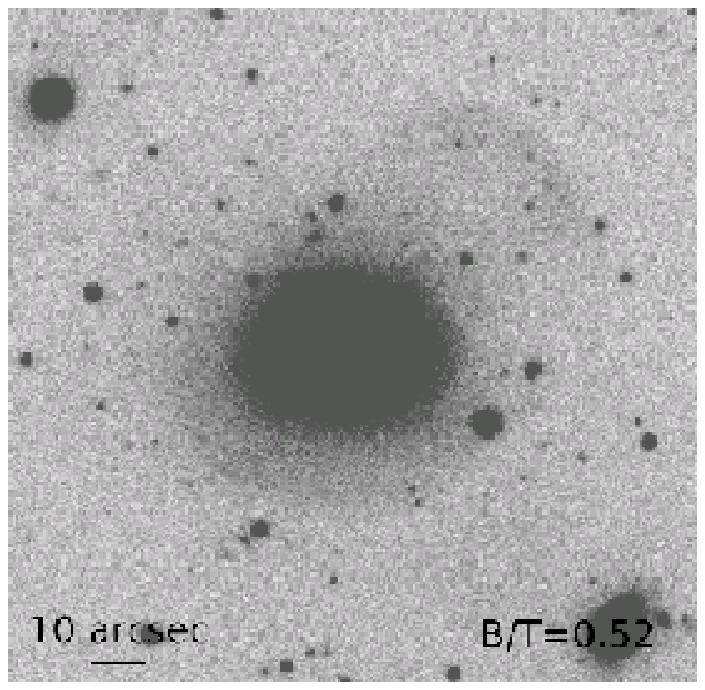}
 \includegraphics[scale=0.70]{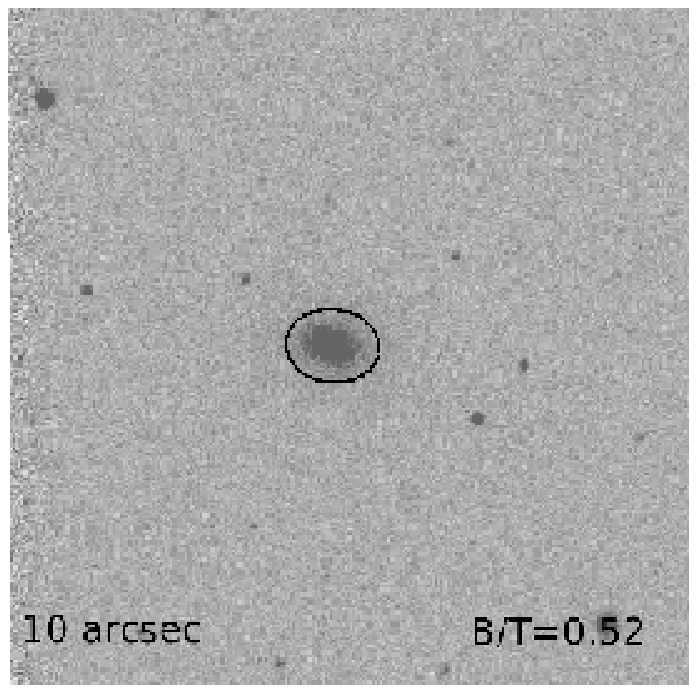}
 \includegraphics[scale=0.45]{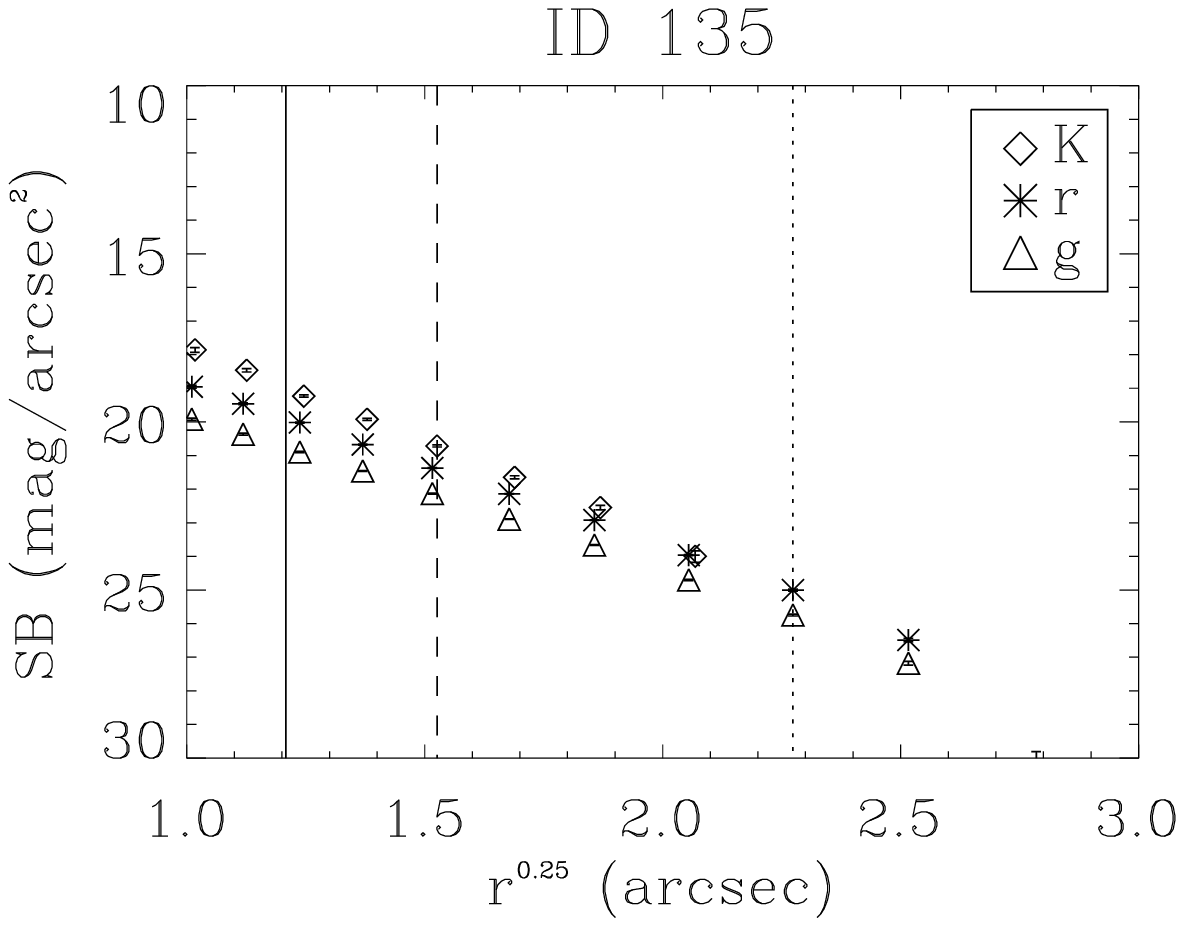}\\
 \includegraphics[scale=0.70]{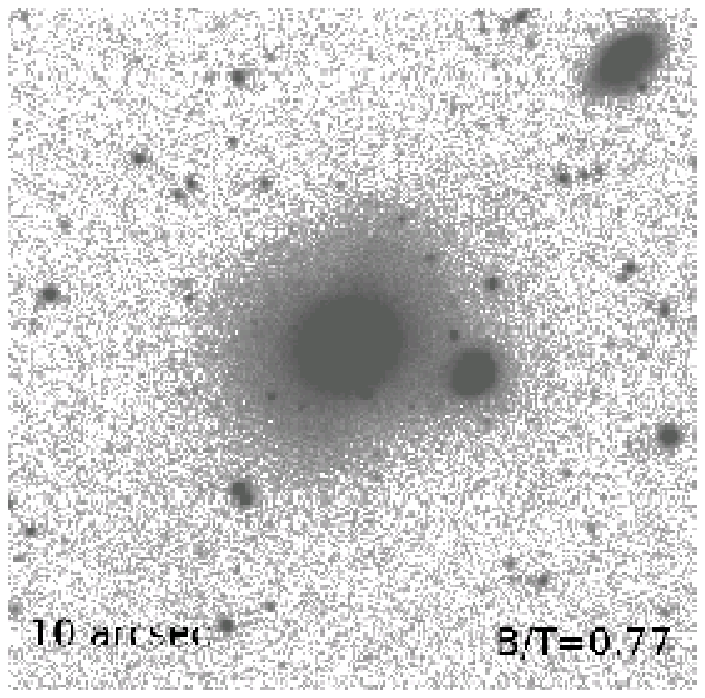}
 \includegraphics[scale=0.70]{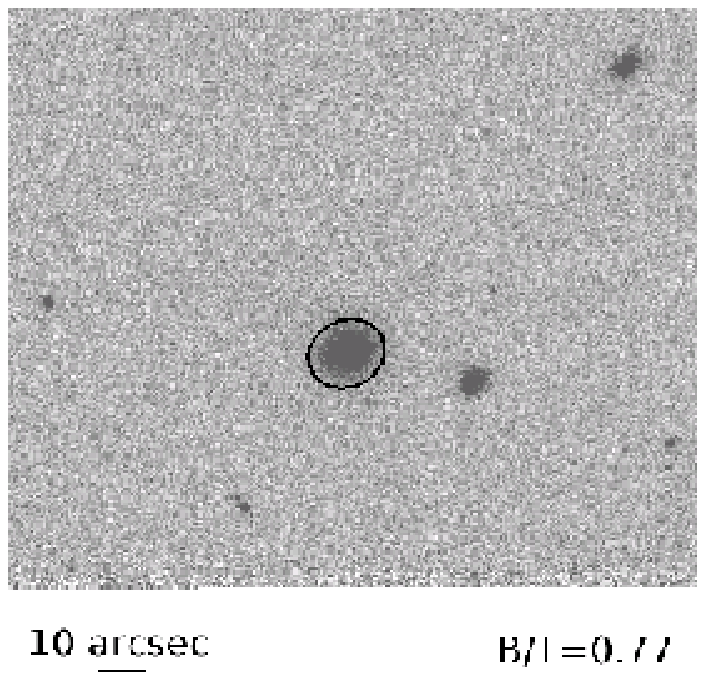}
 \includegraphics[scale=0.45]{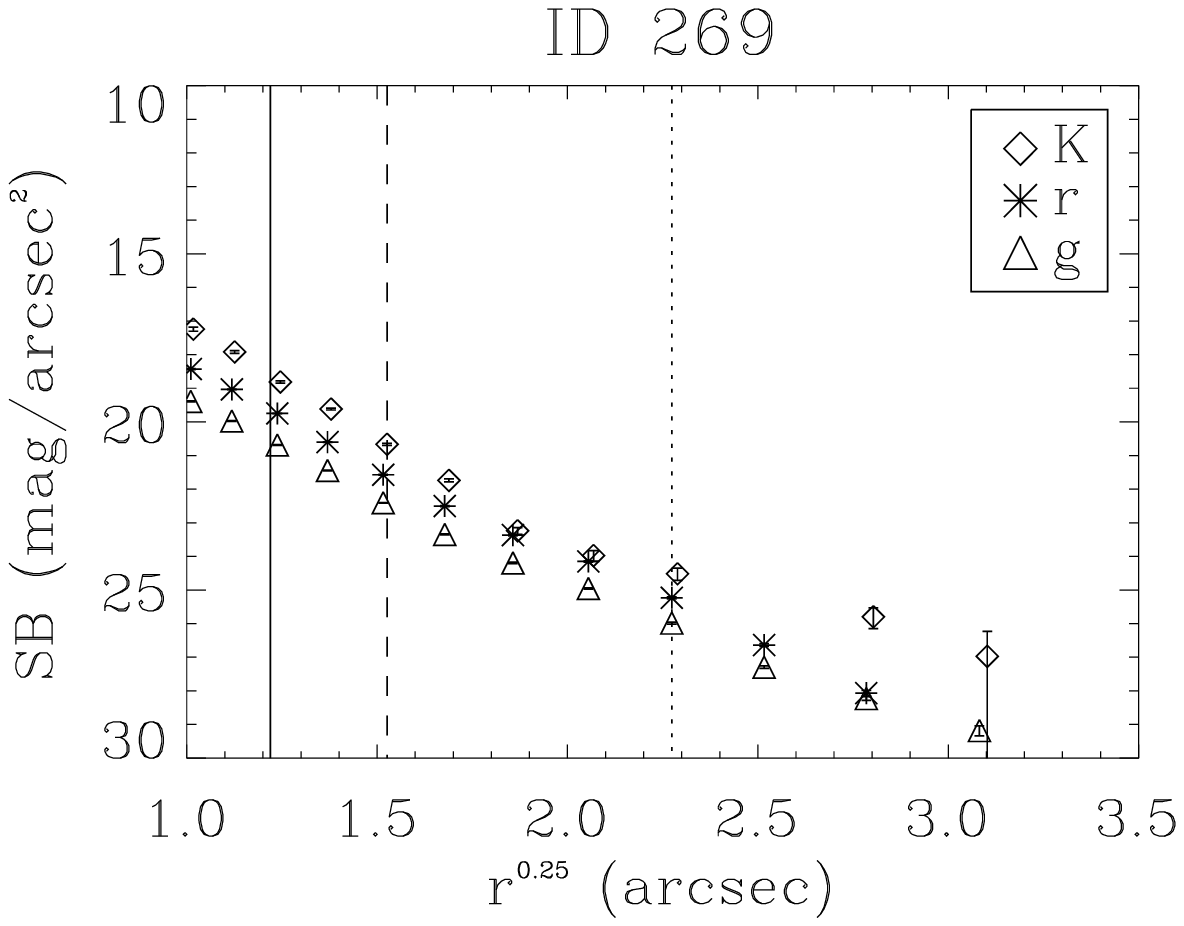}\\
 \includegraphics[scale=0.70]{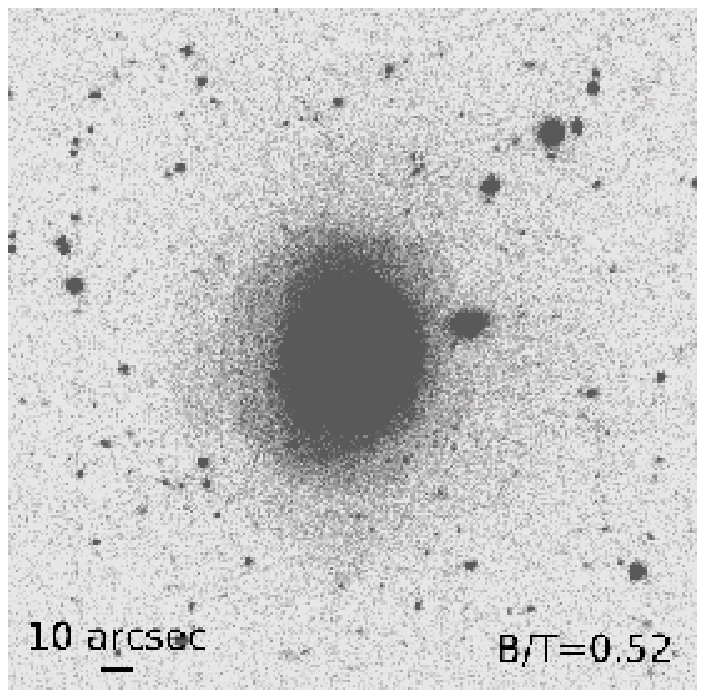}
 \includegraphics[scale=0.70]{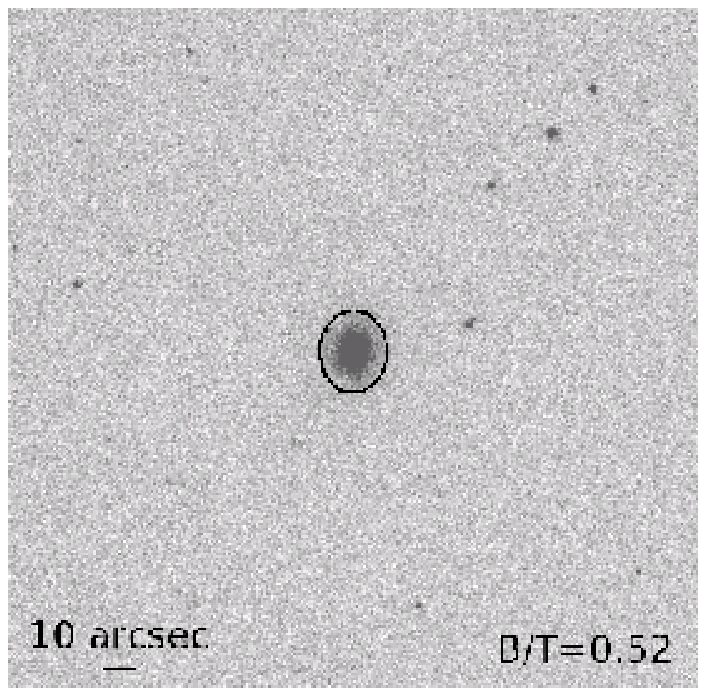}
 \includegraphics[scale=0.45]{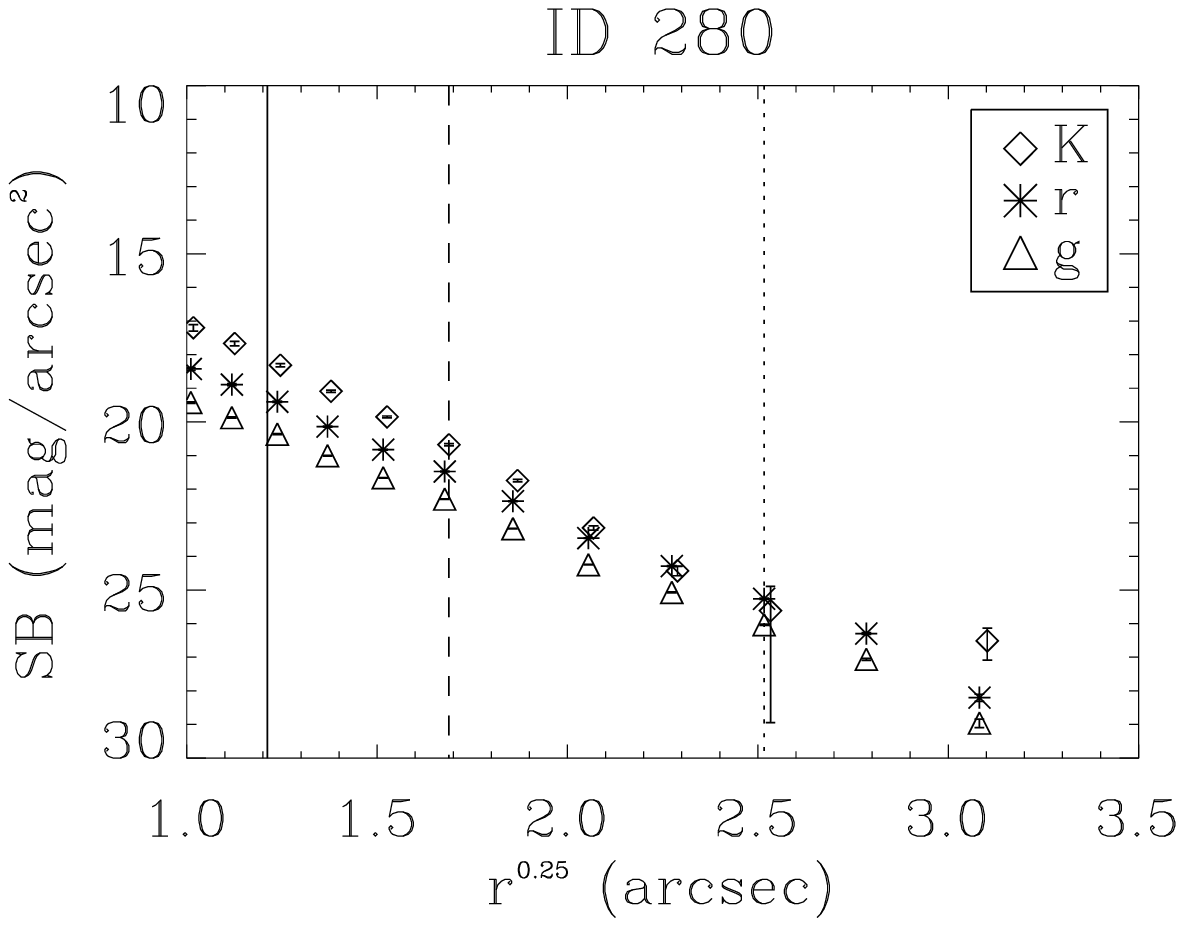}\\
 \includegraphics[scale=0.70]{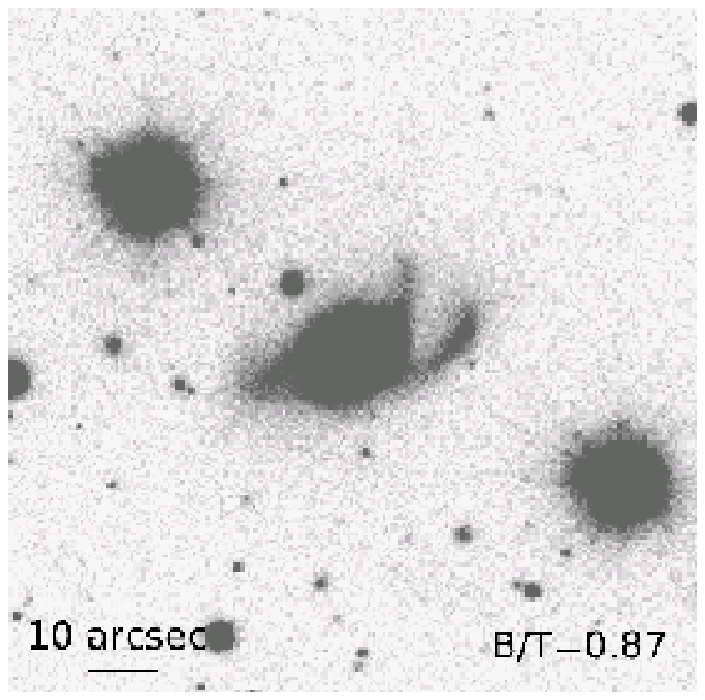}
 \includegraphics[scale=0.70]{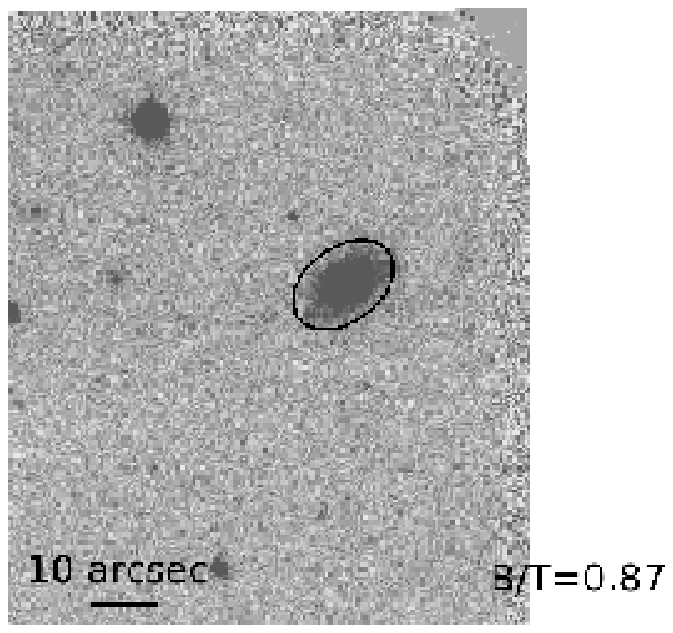}
 \includegraphics[scale=0.45]{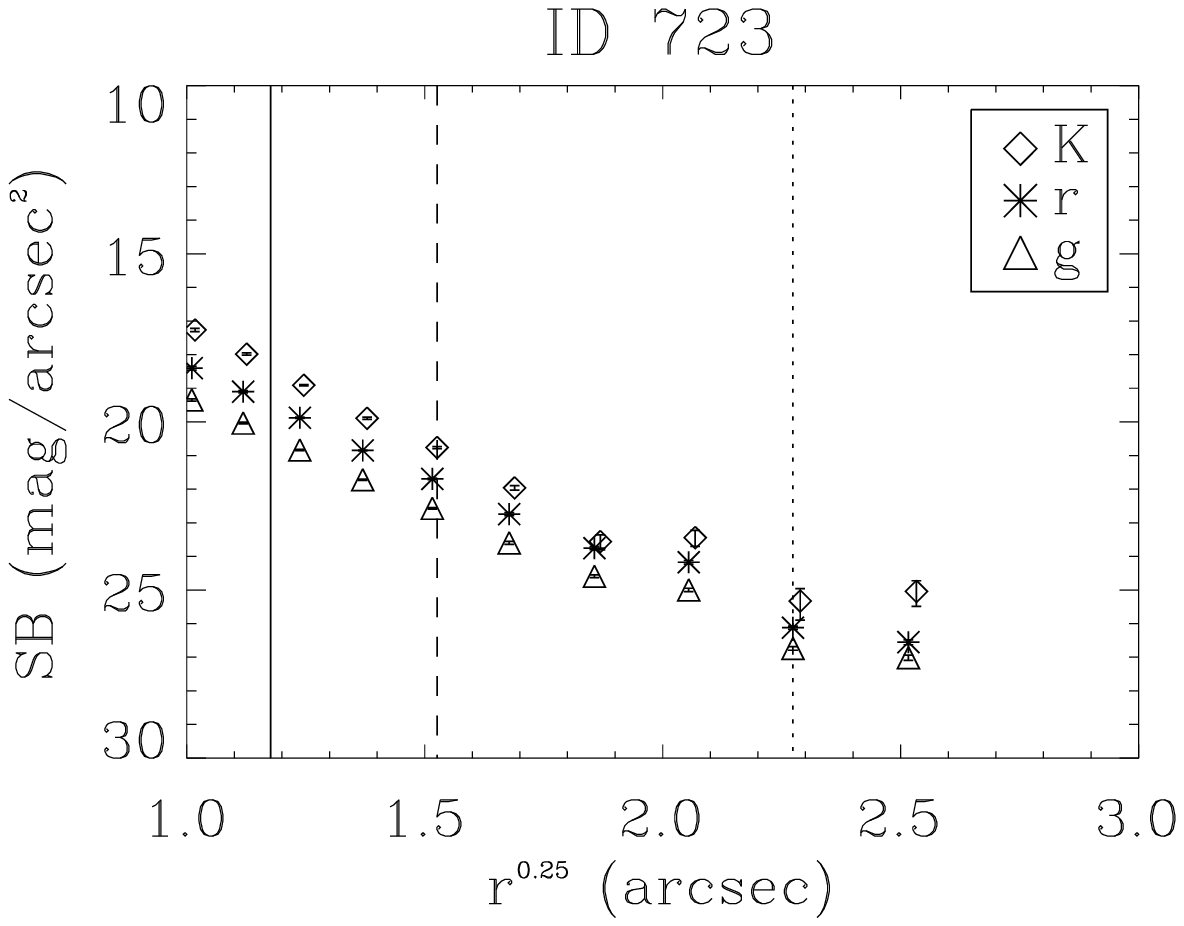}
\caption{Same as Figure \ref{rel} but for tidal-feature ellipticals. Note the faint
tidal features around each elliptical, which are indicative of their past merging 
activities.}
\label{tid}
\end{figure}

\begin{figure}
\epsscale{1.0}
 \includegraphics[scale=0.70]{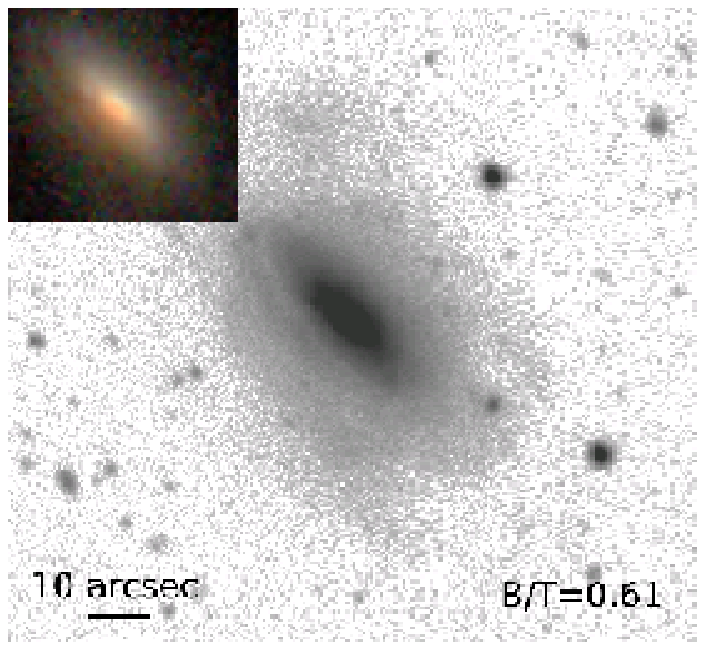}
 \includegraphics[scale=0.70]{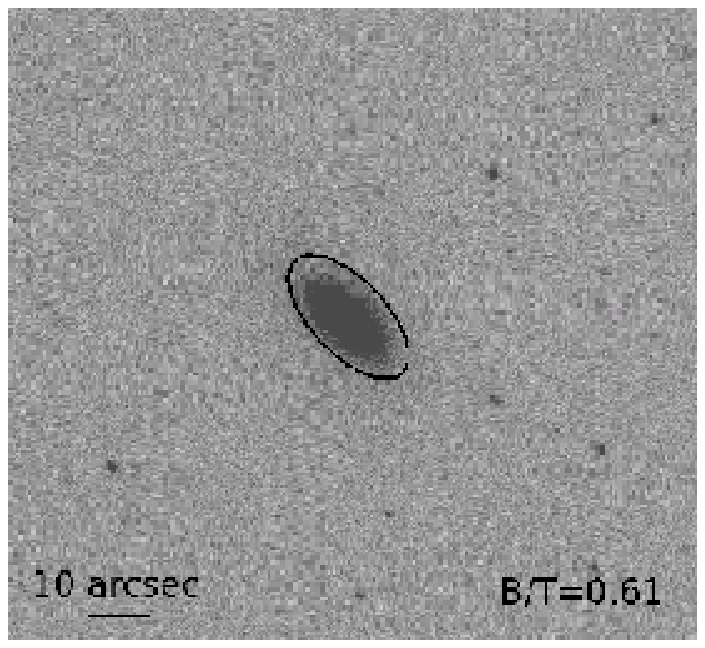}
 \includegraphics[scale=0.45]{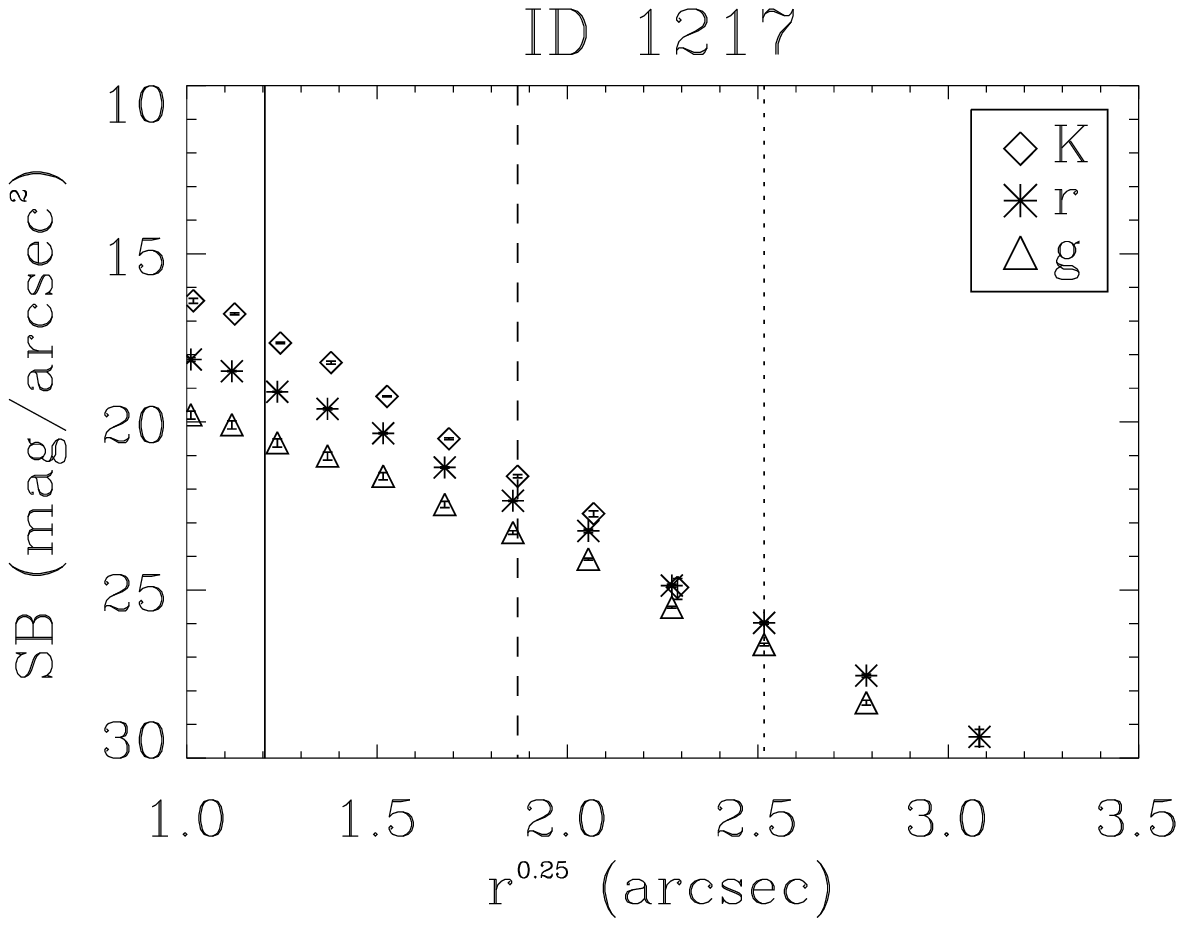}\\
 \includegraphics[scale=0.70]{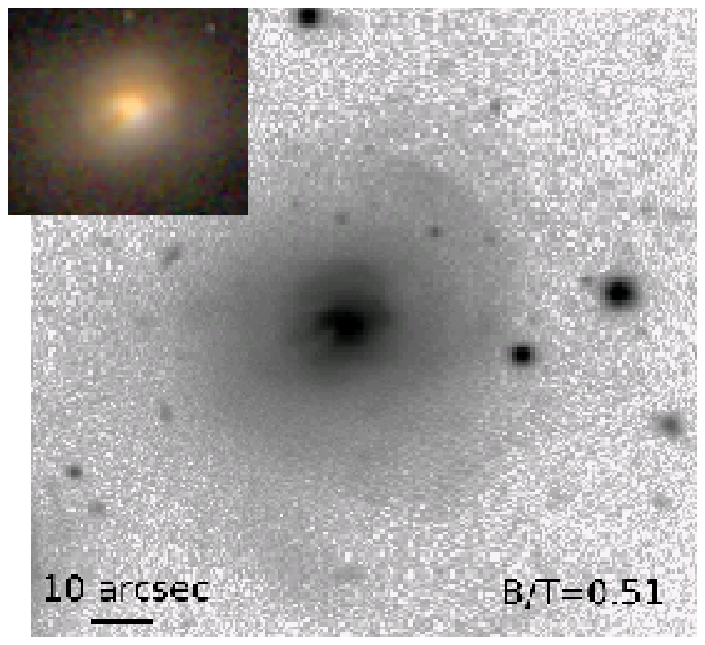}
 \includegraphics[scale=0.70]{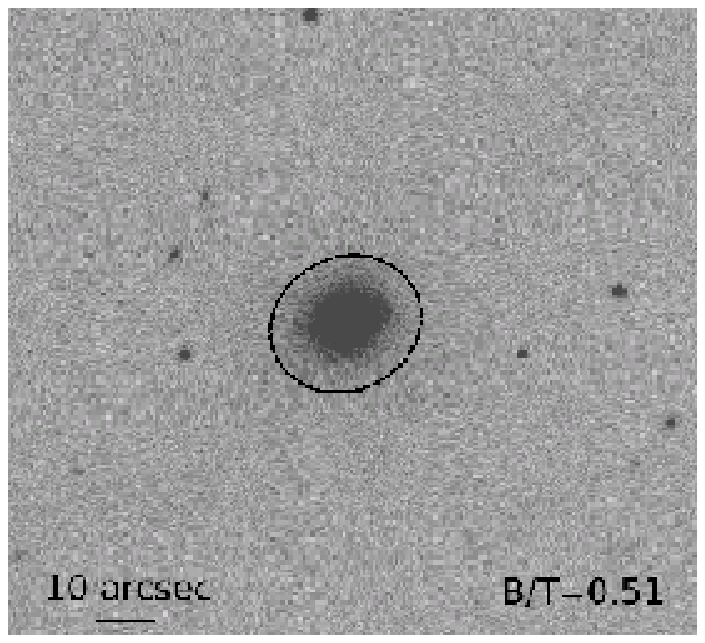}
 \includegraphics[scale=0.45]{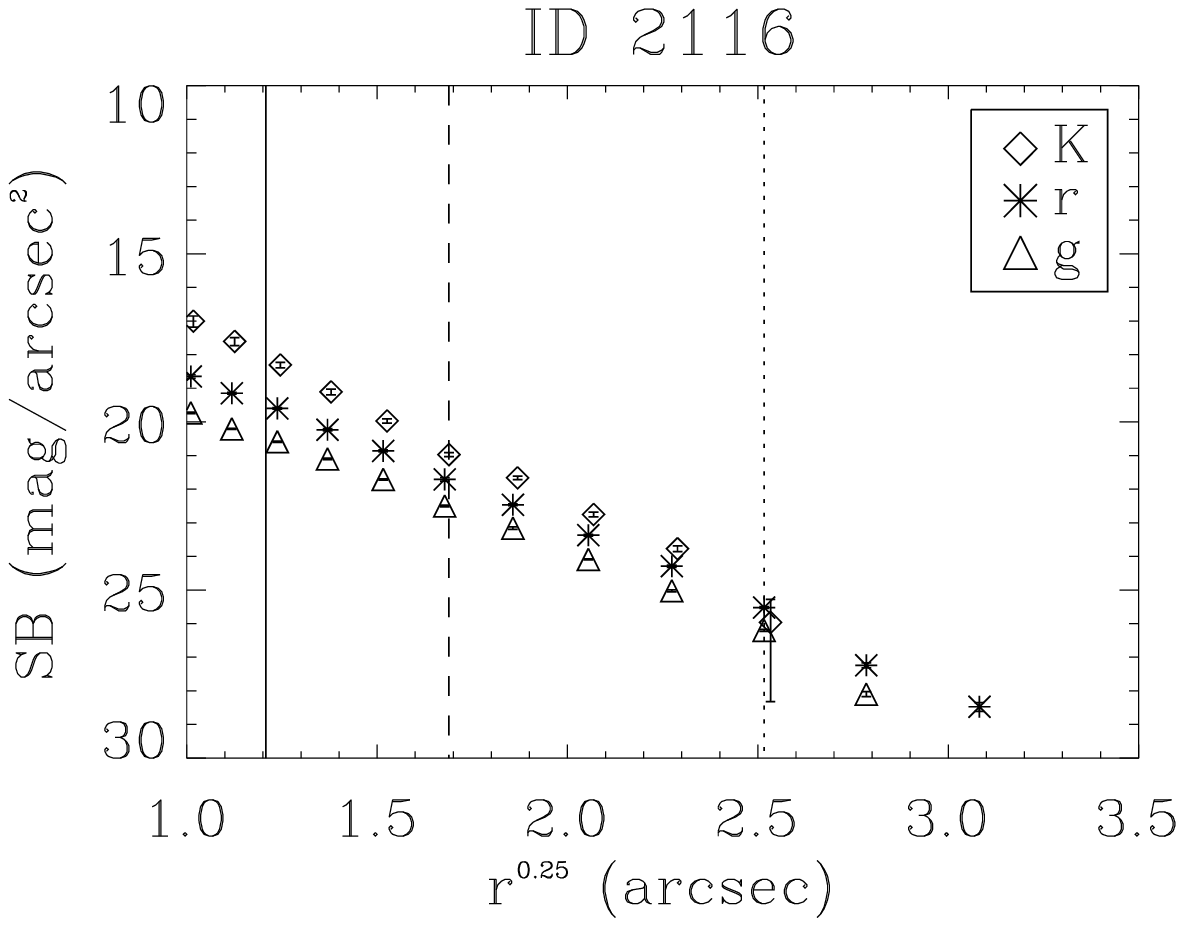}\\
 \includegraphics[scale=0.70]{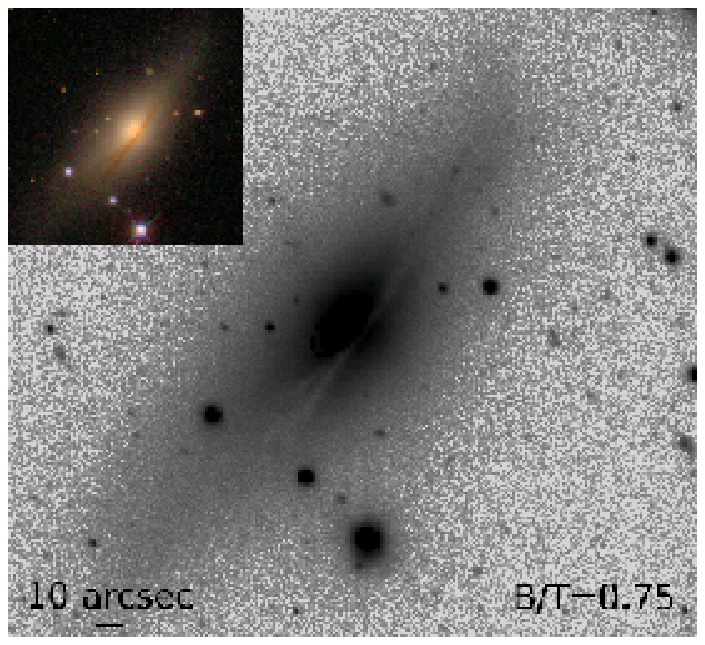}
 \includegraphics[scale=0.70]{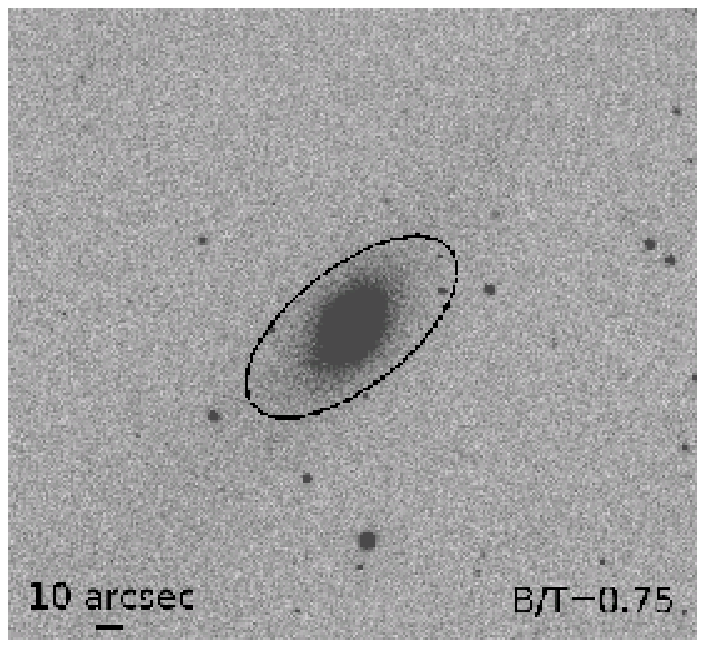}
 \includegraphics[scale=0.45]{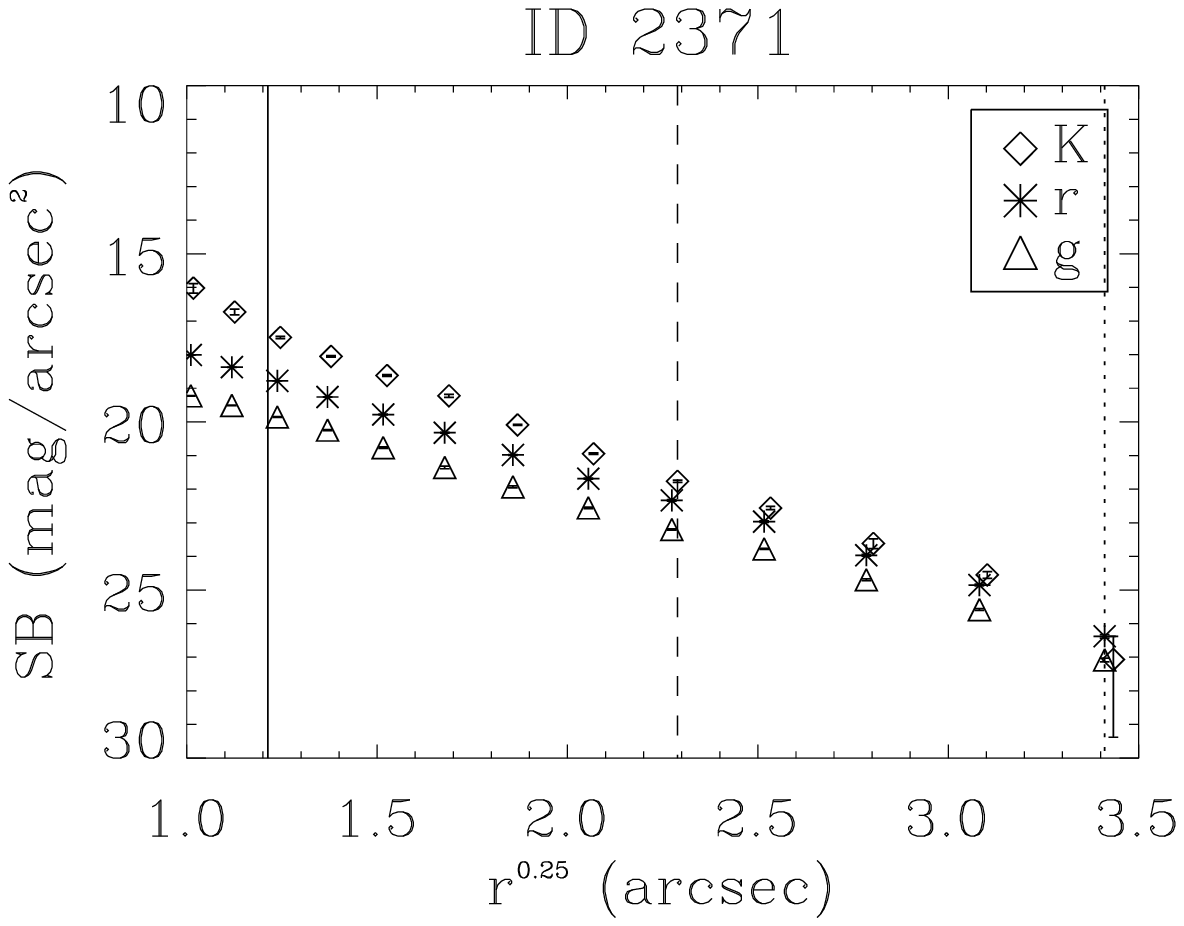}\\
 \includegraphics[scale=0.70]{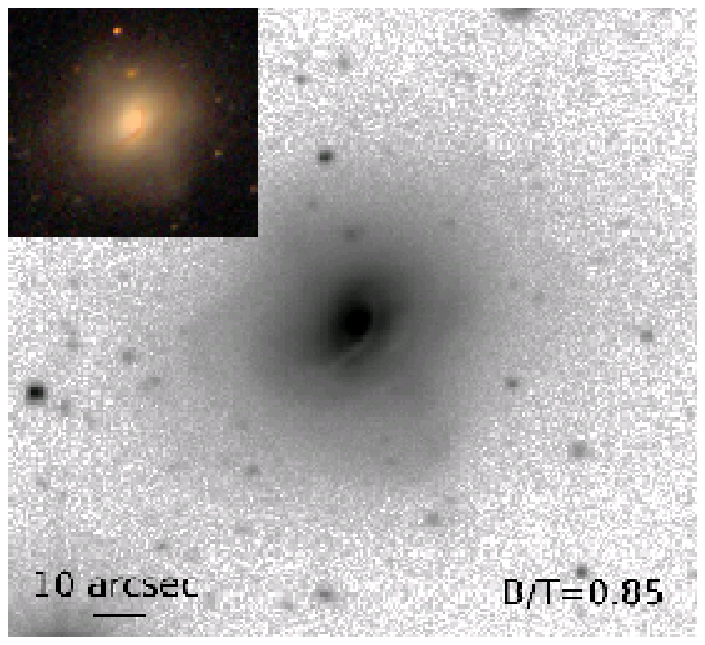}
 \includegraphics[scale=0.70]{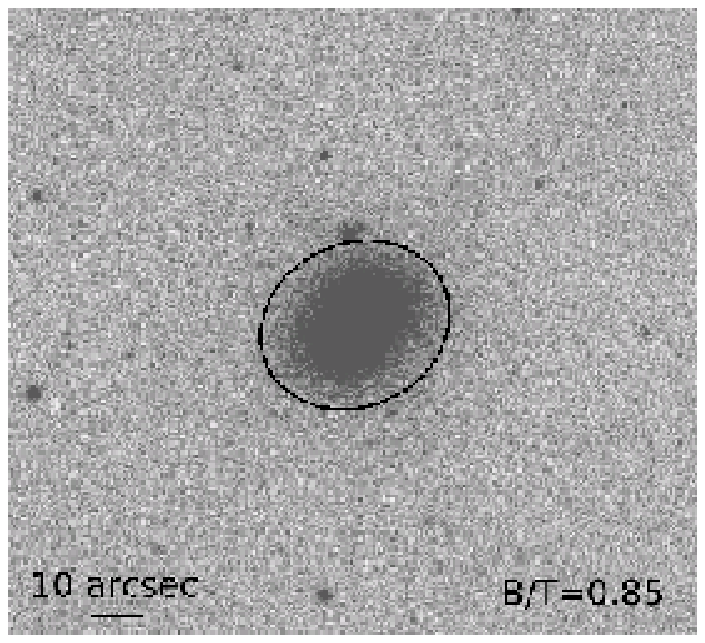}
 \includegraphics[scale=0.45]{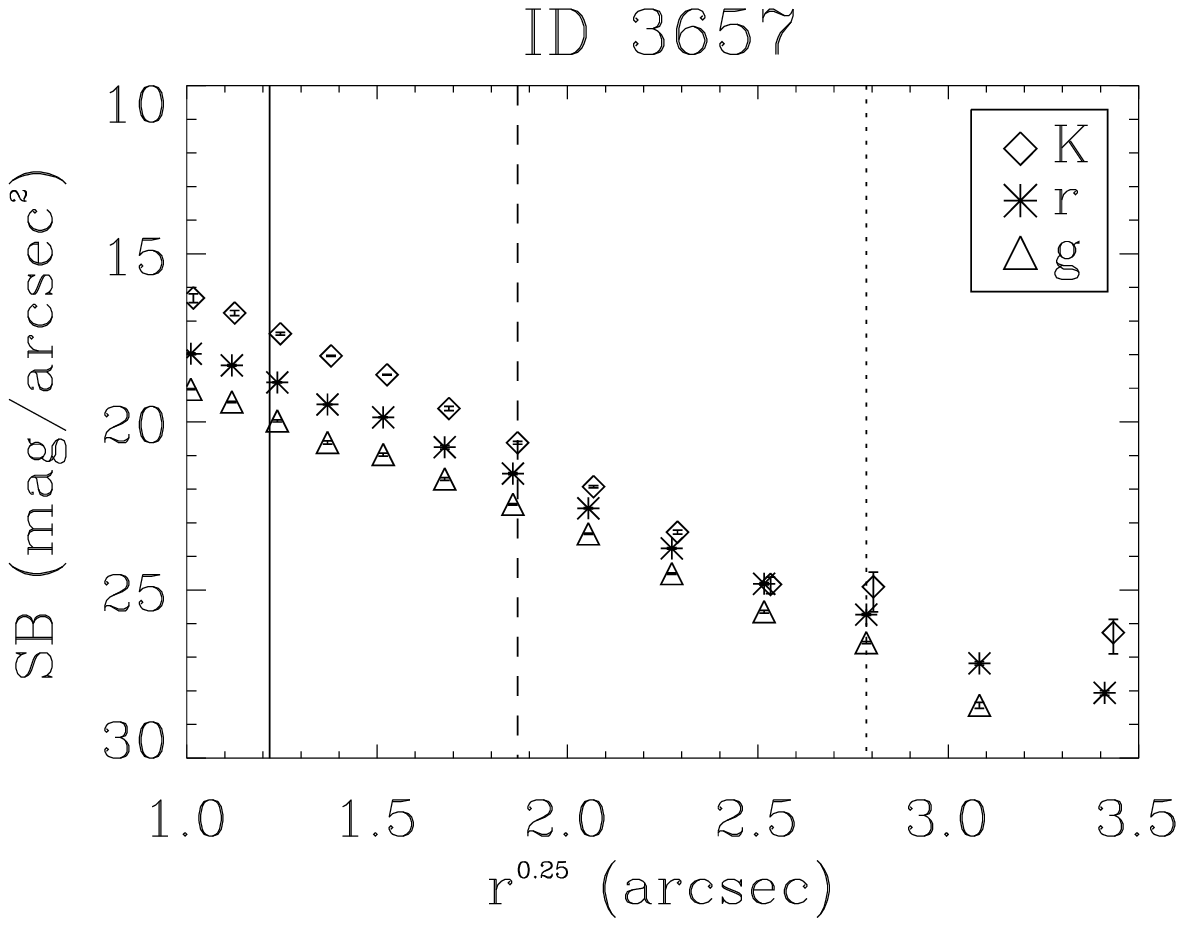}
\caption{Same as Figure \ref{rel} but for dust-feature ellipticals. Note the dust-lanes
in the $r$-band images.}
\label{dust}
\end{figure}

\begin{figure}
\epsscale{1.0}
 \includegraphics[scale=0.70]{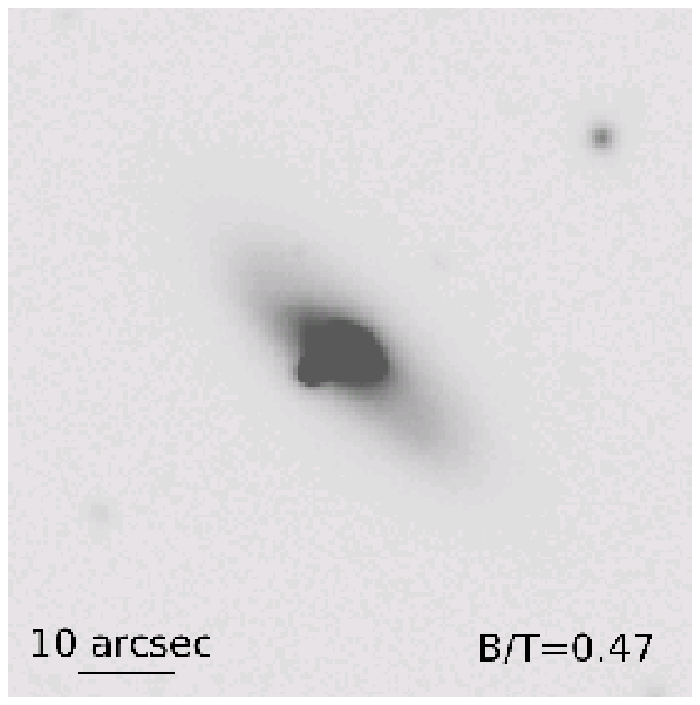}
 \includegraphics[scale=0.70]{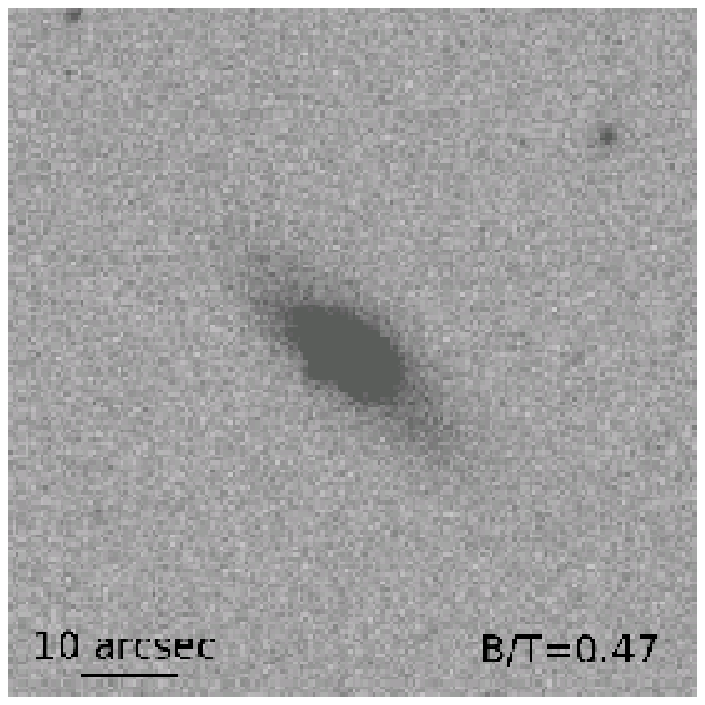}
 \includegraphics[scale=0.45]{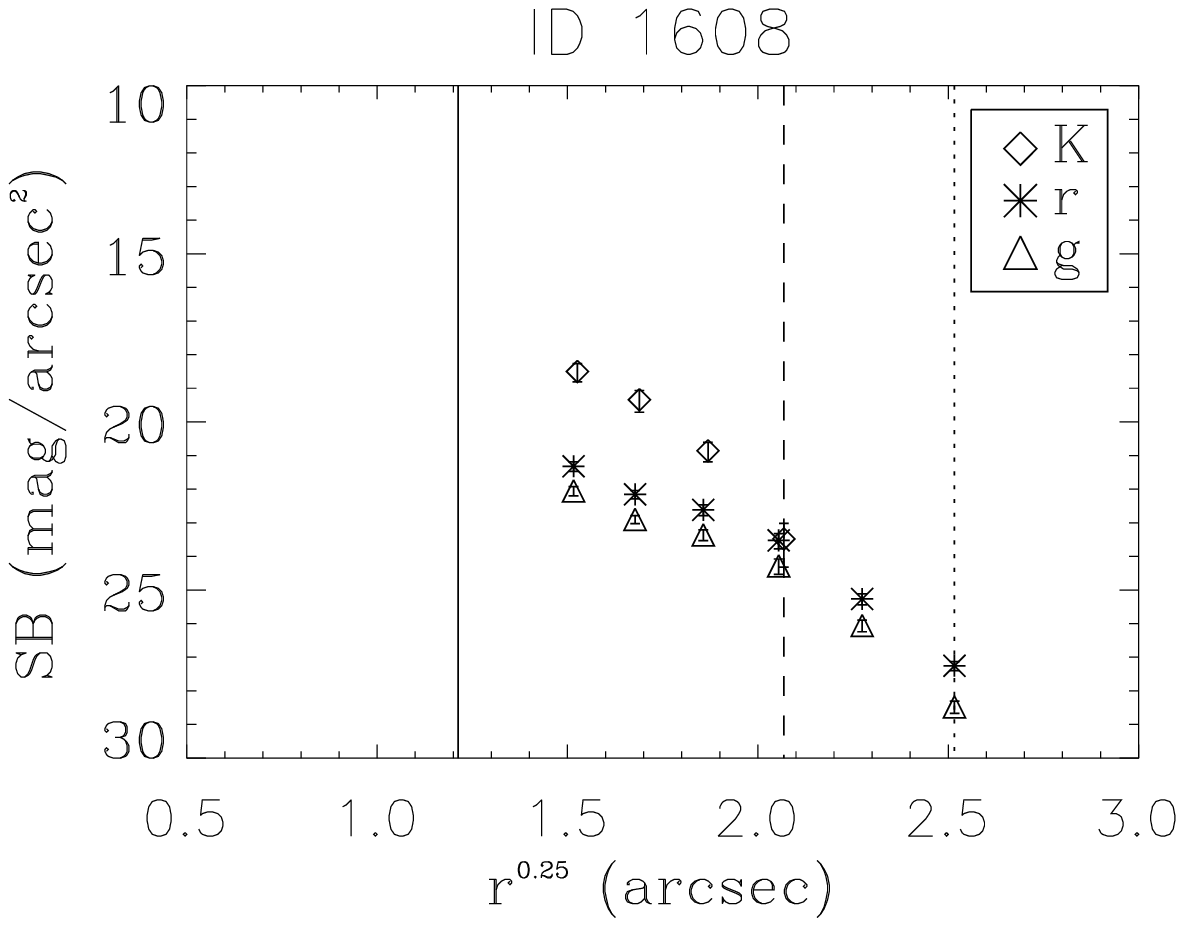}\\
 \includegraphics[scale=0.70]{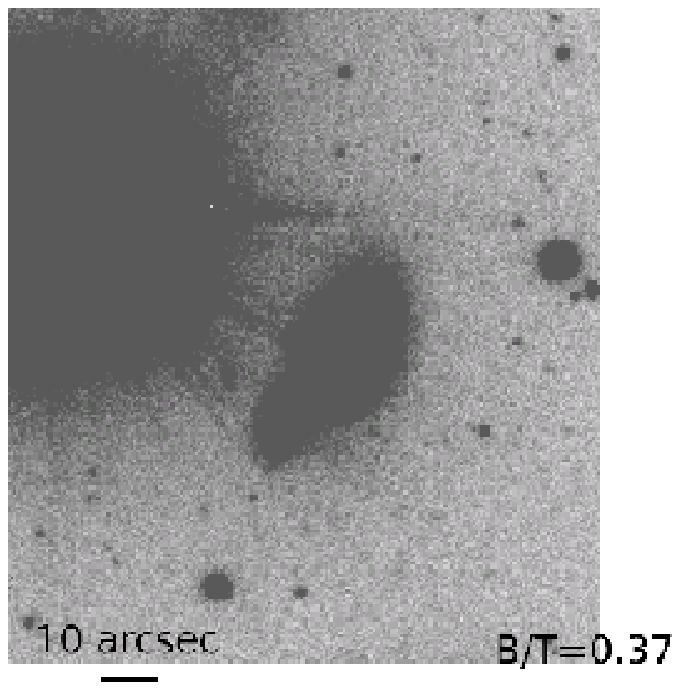}
 \includegraphics[scale=0.70]{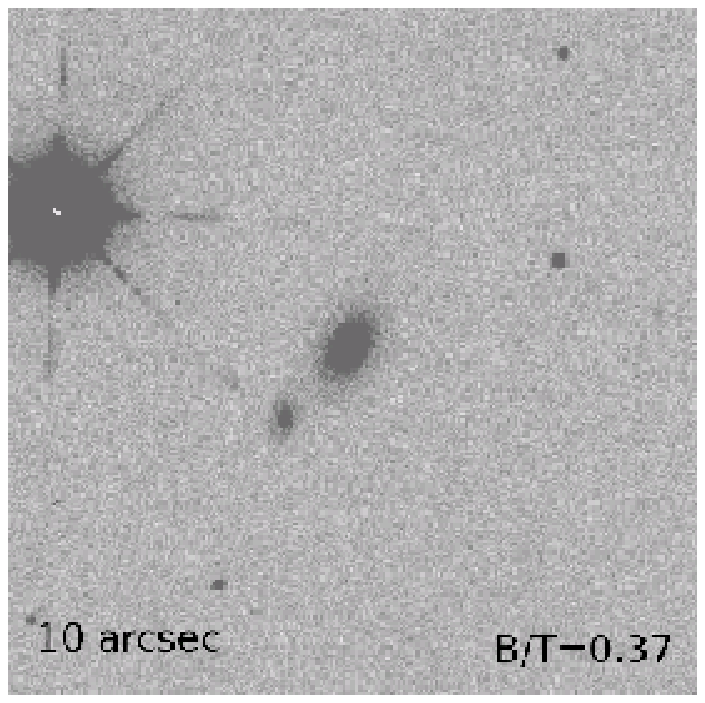}
 \includegraphics[scale=0.45]{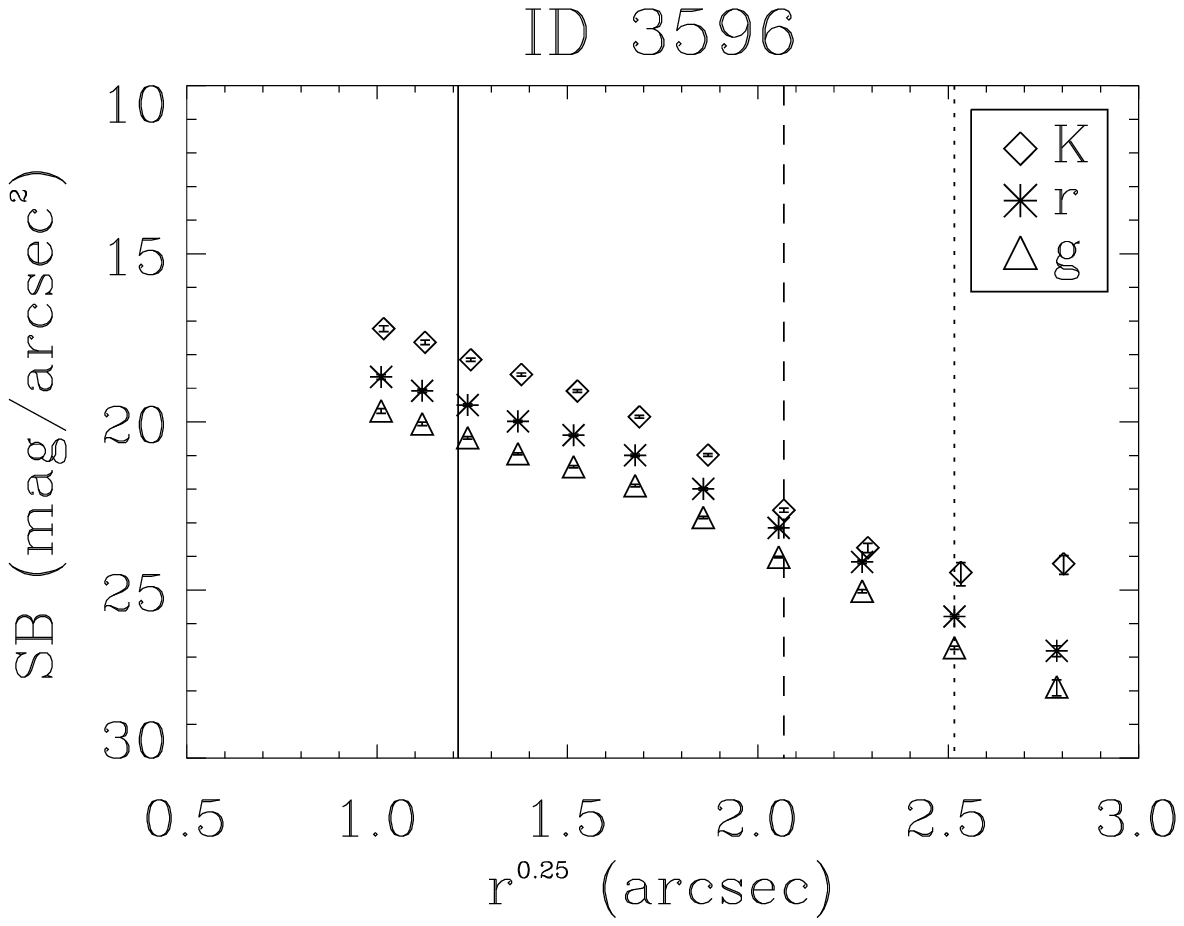}\\
 \includegraphics[scale=0.70]{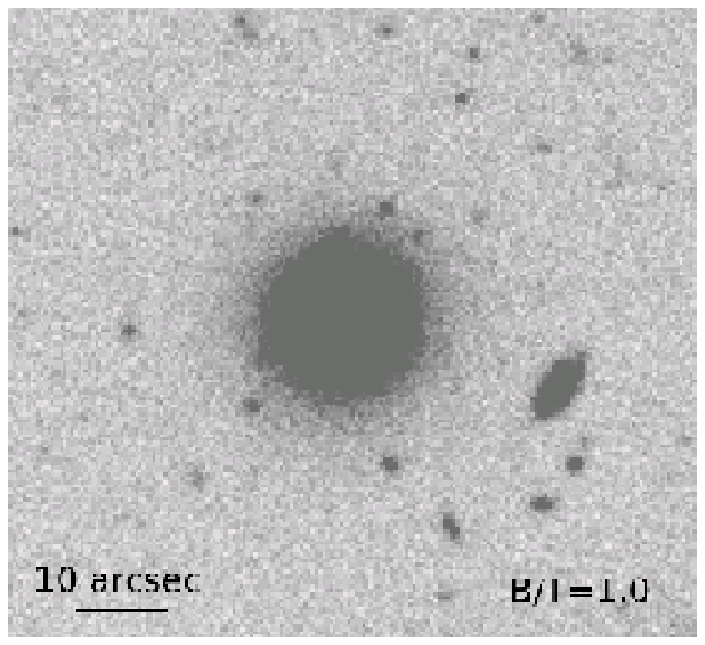}
 \includegraphics[scale=0.70]{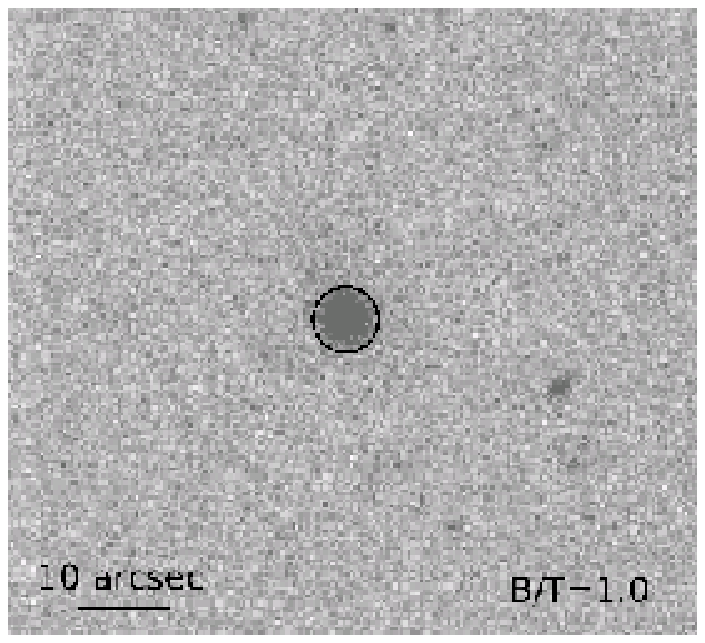}
 \includegraphics[scale=0.45]{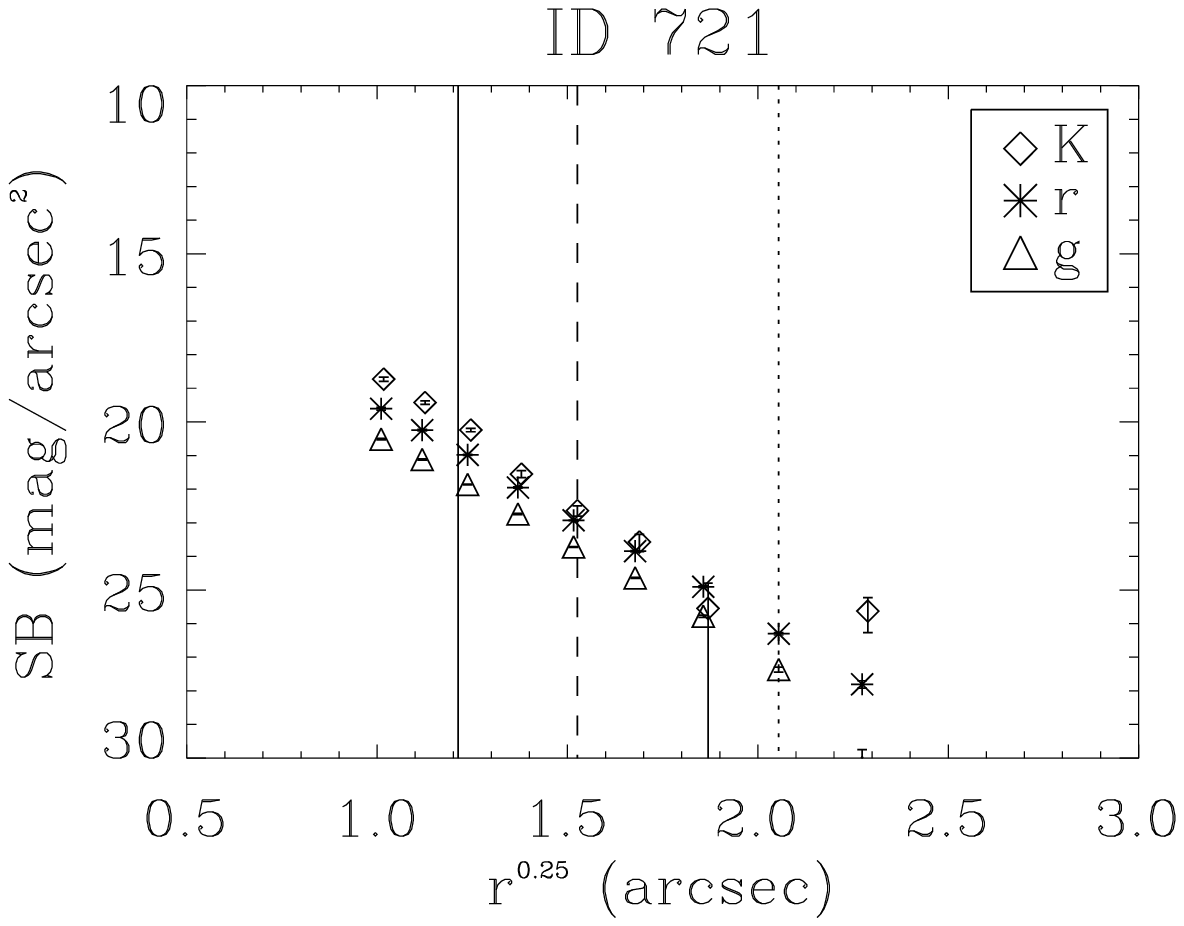}\\
 \includegraphics[scale=0.70]{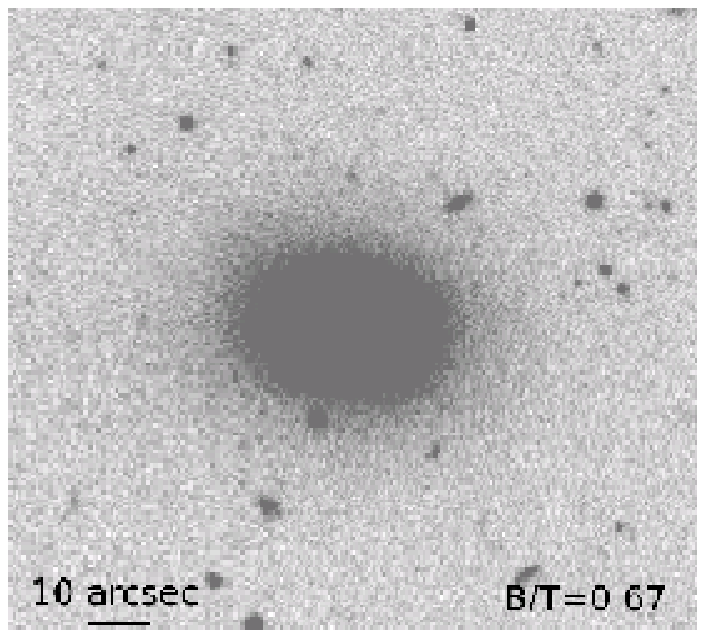}
 \includegraphics[scale=0.70]{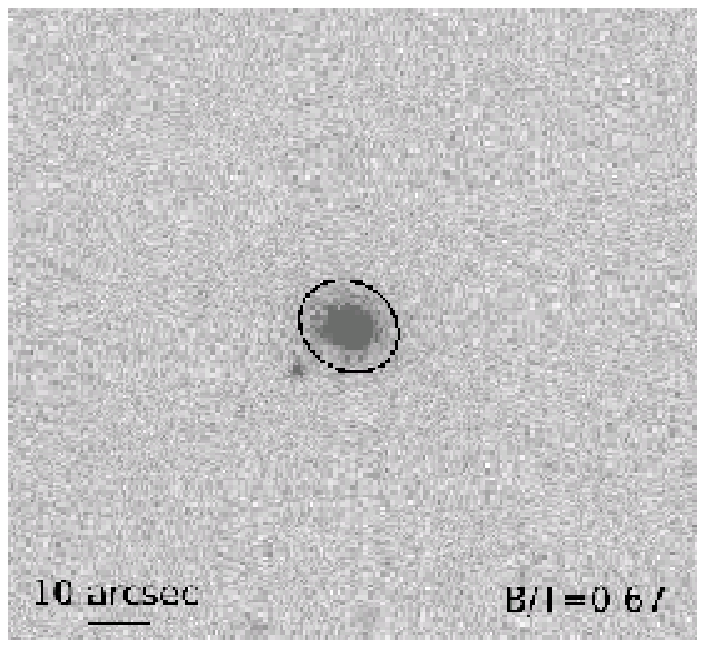}
 \includegraphics[scale=0.45]{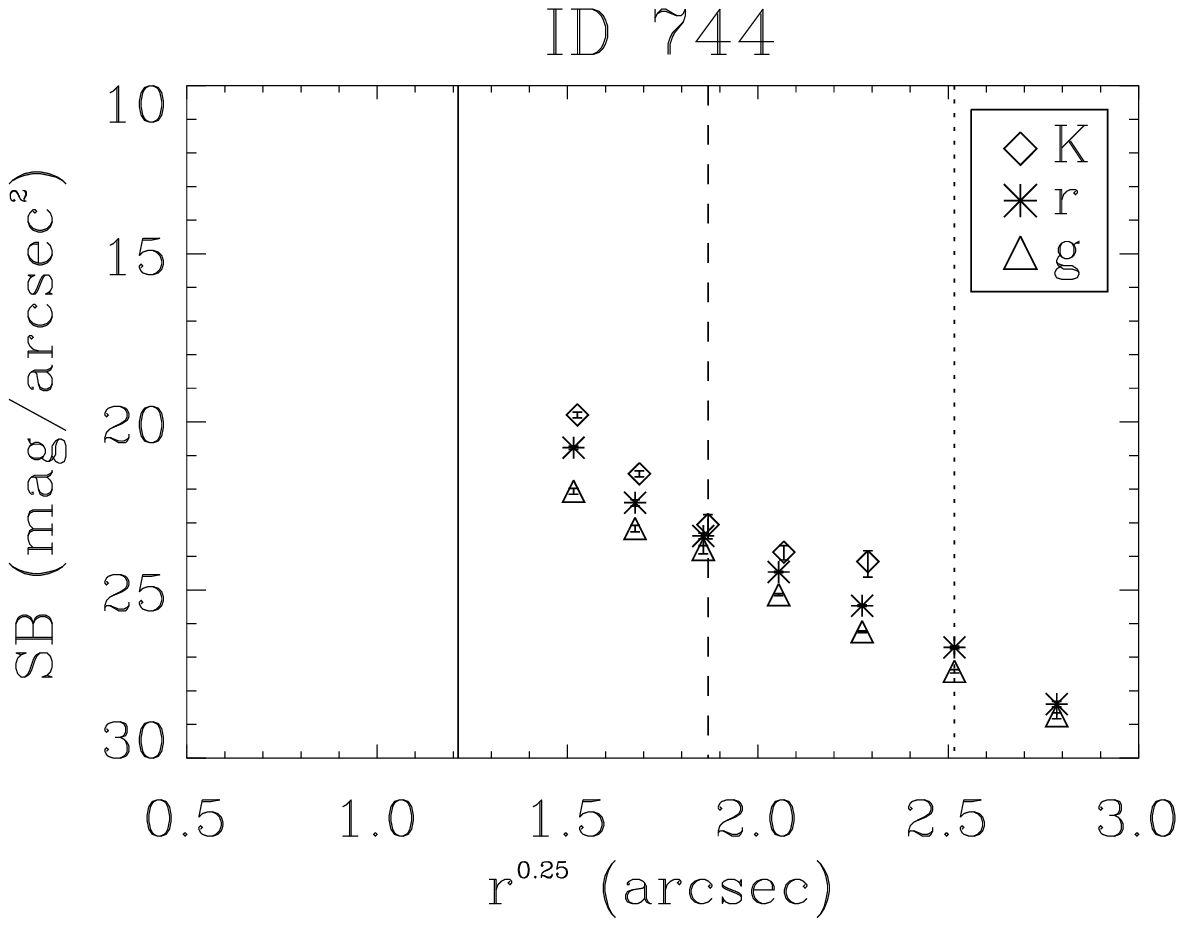}
\caption{Same as Figure \ref{rel} but for the K10 early-types that are excluded from our analysis because of the presence of close neighbors (the top 2 rows) and small sizes (the bottom 2 rows).}
\label{rej}
\end{figure}

\begin{figure}
\epsscale{1.0}
 \includegraphics[scale=0.70]{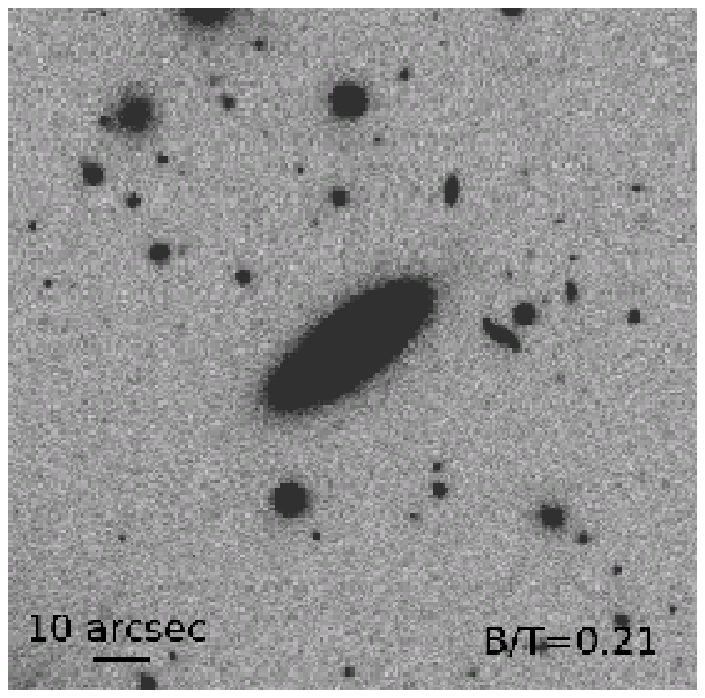}
 \includegraphics[scale=0.70]{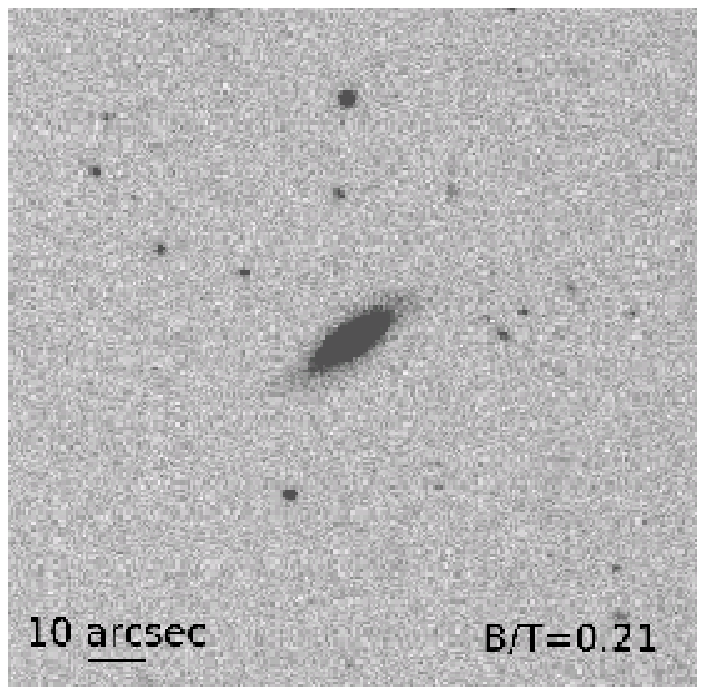}
 \includegraphics[scale=0.45]{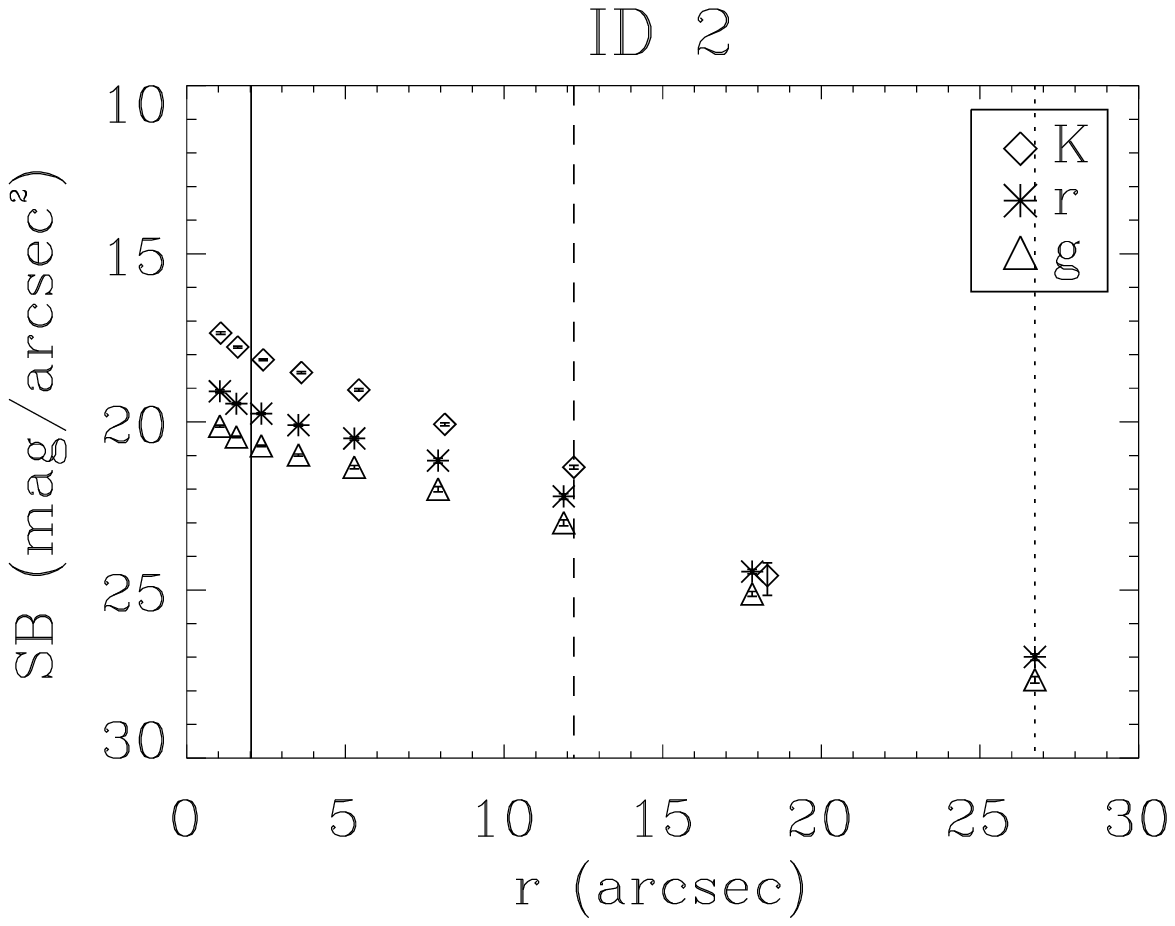}\\
 \includegraphics[scale=0.70]{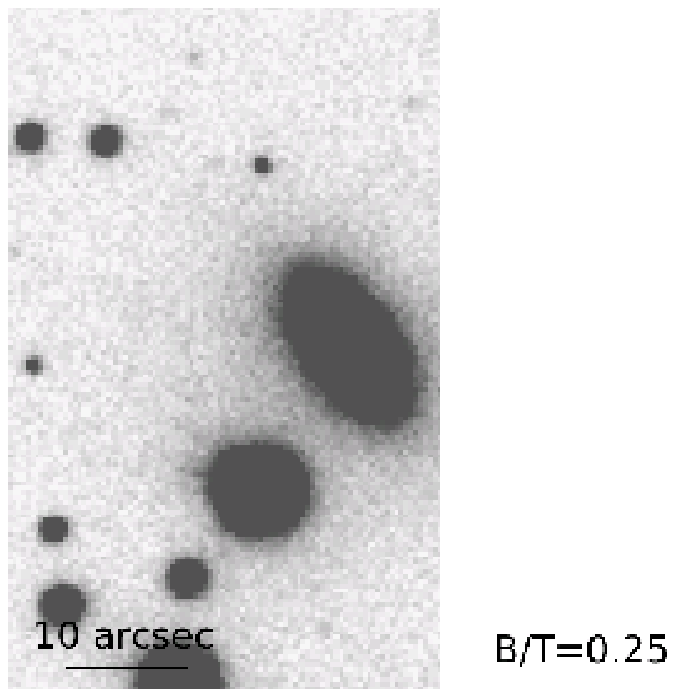}
 \includegraphics[scale=0.70]{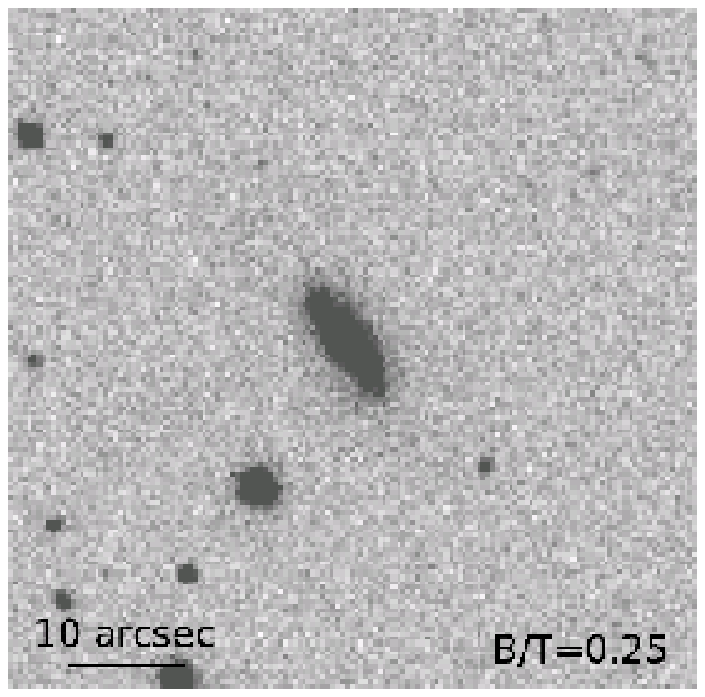}
 \includegraphics[scale=0.45]{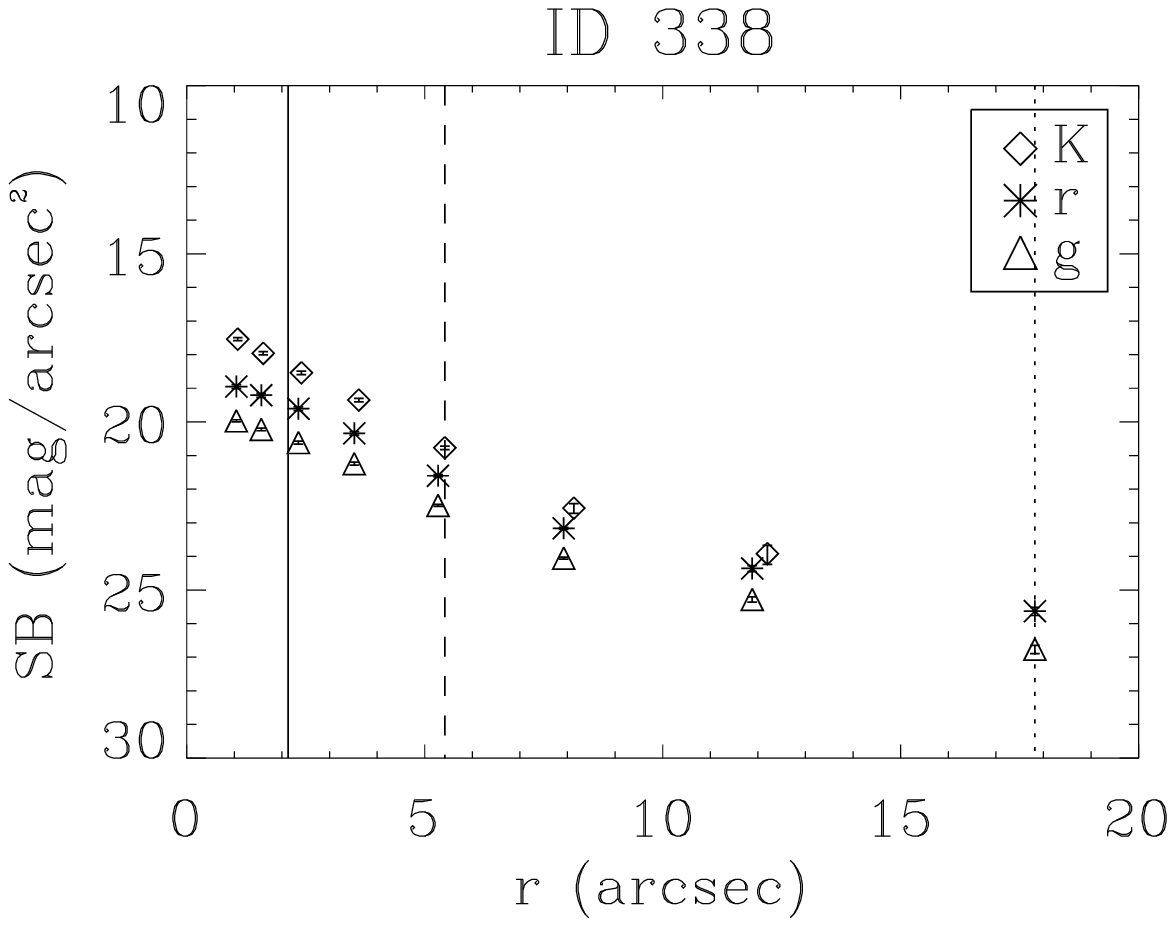}\\
 \includegraphics[scale=0.70]{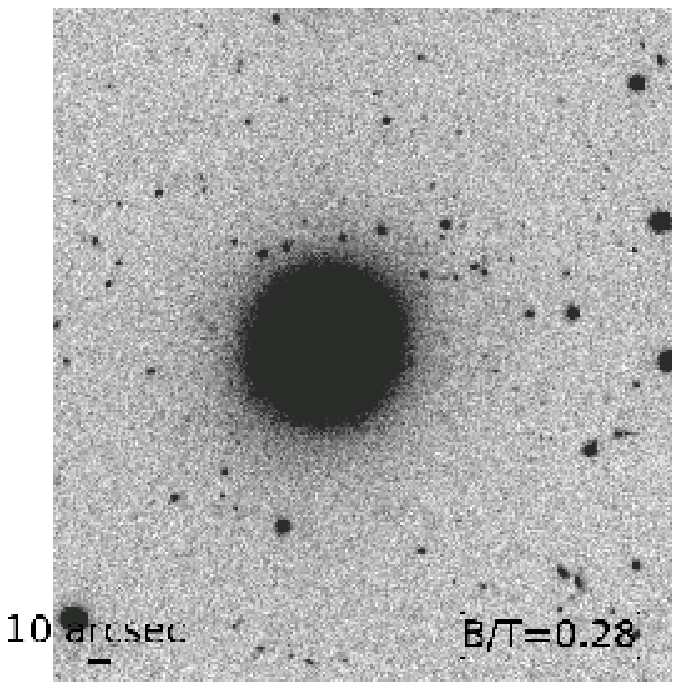}
 \includegraphics[scale=0.70]{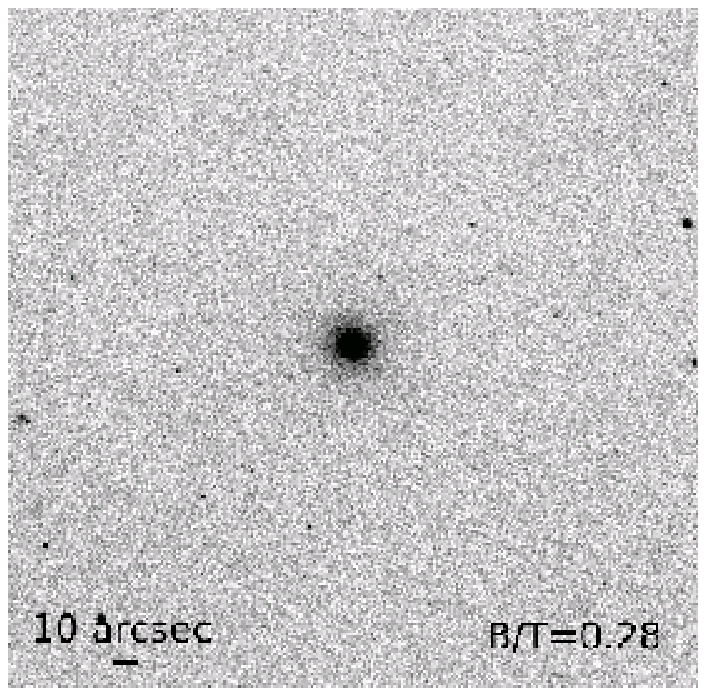}
 \includegraphics[scale=0.45]{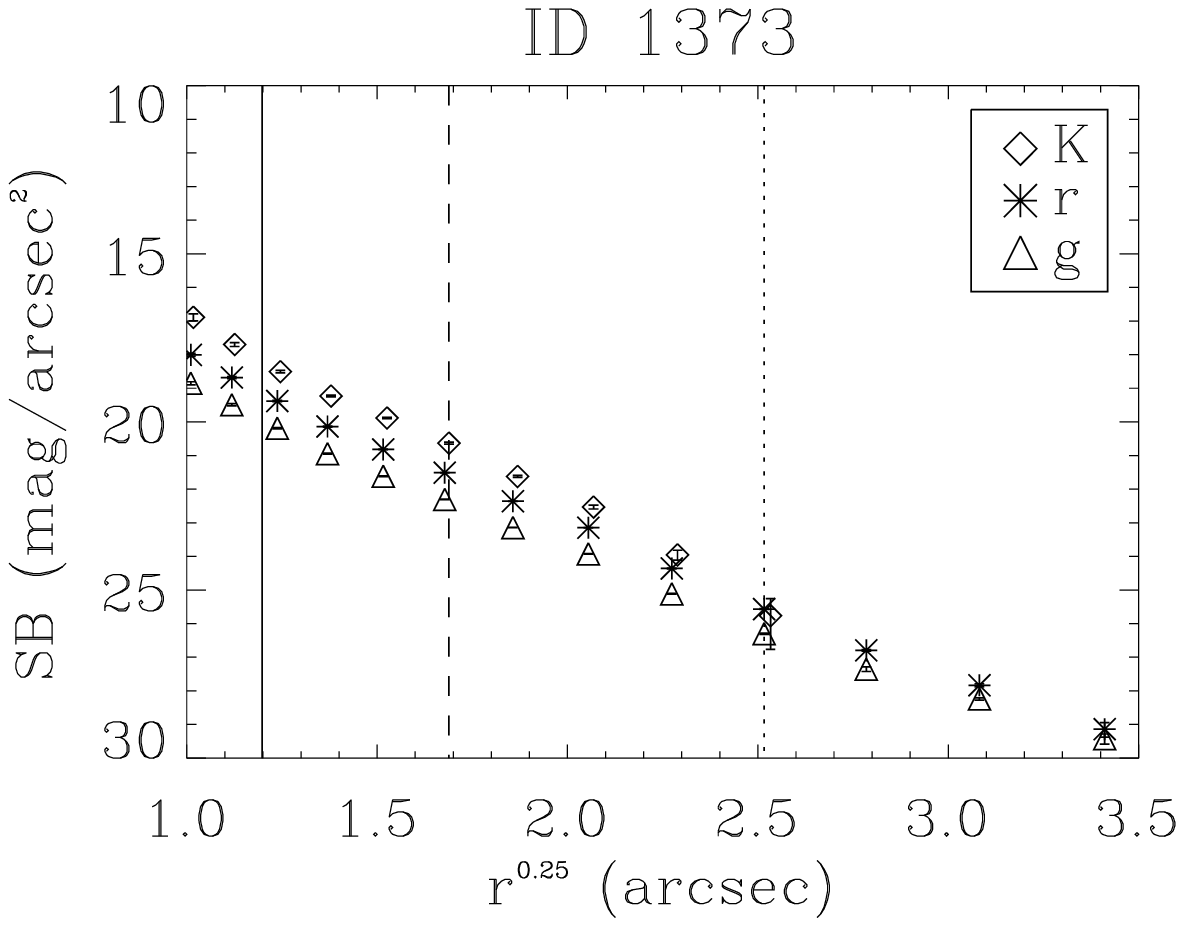}\\
 \includegraphics[scale=0.70]{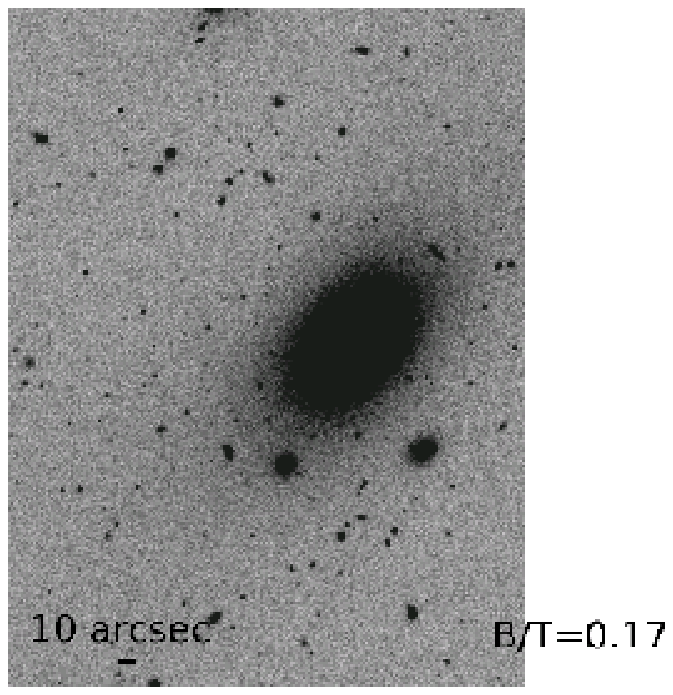}
 \includegraphics[scale=0.70]{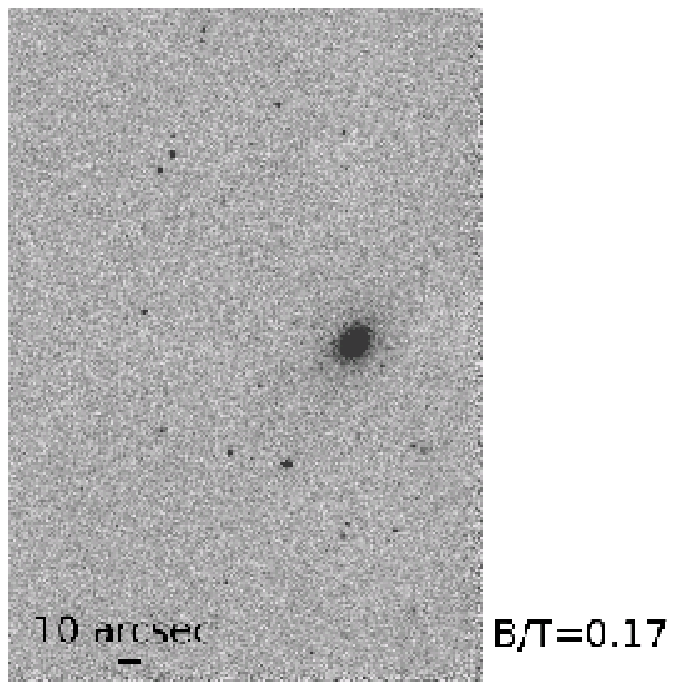}
 \includegraphics[scale=0.45]{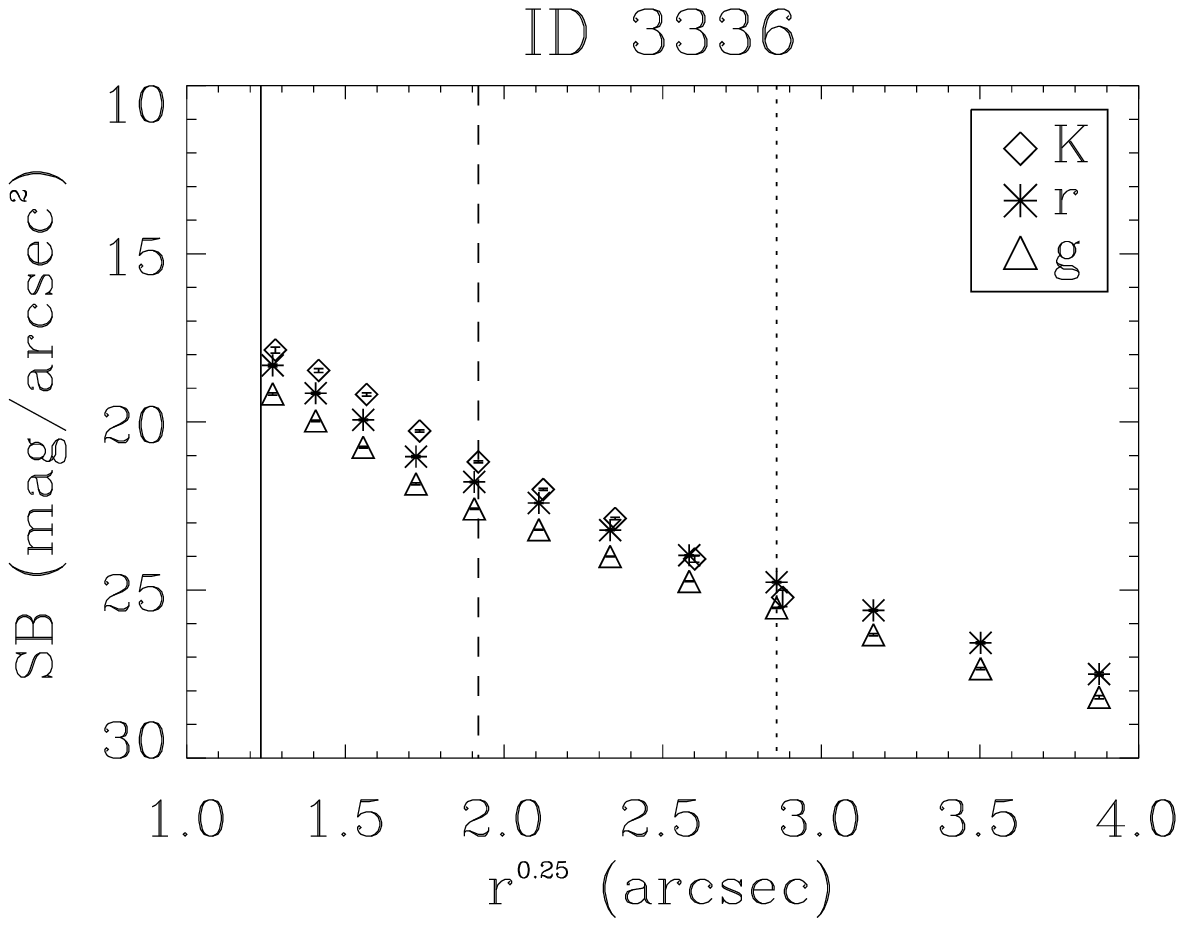}
\caption{Same as Figure \ref{rel} but for the K10 early-types that are excluded from our analysis because of their low ($< 0.5$) $B/T$ ratios.}
\label{disky}
\end{figure}

In order to investigate color/age/metallicity gradients, we derived SB profiles of each elliptical in $g, r$, and $K$ (see Figure \ref{rel} -- \ref{disky}). The first step in the SB profile fitting is to estimate a proper background, 
and to subtract it from the images. This is because, in the outer parts of galaxies where the sky noise is dominant, SB can change substantially depending on the value of the sky background. Therefore, the background subtraction is very important in the analysis of color gradients and the derivation of SB at the outer region of galaxies. Both the SDSS Stripe 82 and the UKIDSS LAS data offer sky background estimates in each image, 
however we find that these values are only approximately accurate. 
Hence, we re-estimated sky background of each image with a growth curve analysis. 
We carried out the growth curve analysis in the following way. The initial guess of the background is obtained from sky value indicated in the FITS image header. The initial guess is 
subtracted from the image, and  the growth curve is constructed using the IRAF ELLIPSE task. We examined
the growth curve out to $\sim 2 \times$ the isophotal radius of the object (typically, 15 -- 40 
arcsec in radius, with isophote at 1.5-$\sigma$ of the sky noise),
 and a small amount of background value was added or subtracted manually until the 
background become flat at such a radius. 
Through this process, we determine the sky background value to an accuracy of $\sim$0.03\%. 

After subtracting the background, we used the IRAF ELLIPSE task to fit the SB profile. We fitted the $r$-band images first in order to use the $r$-band results as a reference to compare with the other band data, because the photometric sensitivity of the $r$-band is highest among the SDSS filters \citep{gun98}. We fixed the center position as the position that is computed by ELLIPSE, and the position angle and the ellipticity are set free. In some cases, elliptical galaxies exhibit a twist of isophotes in position angle and ellipticity as a function of radius, therefore fixing these two parameters to single values often result in a poor fitting of the SB profile with the ELLIPSE model. We set a step size for the radial increase to be 1.5 $\times$ the previous radius of an isophote with the starting radius of 2 pixels (0.8\arcsec). 
The non-linear increase of the isophotal radius allows us to match S/N of the outer isophotes to the inner ones. Once the fitting of the $r$-band images is done, we fitted the $g$-band and the $K$-band images using the best-fit ELLIPSE parameters from the $r$-band image. Since the $r$-band images have a slightly different pixel scale and a position angle in comparison to the $K$-band images (in some cases), we modified the radial steps (in pixel) and the position angle values for the $K$-band ELLIPSE parameters before fitting the $K$-band images to match those of the $r$-band fits. Objects in the vicinity of an elliptical galaxy were masked out by using the positional information from SExtractor \citep{ber96}. Some objects that are not separated from the elliptical galaxy were masked out manually during the ELLIPSE fitting process. Also masked out were dust features in dust-feature type ellipticals. Figure 6 shows examples of the masking of the neighbor objects. Some of the dust-feature type ellipticals have more than 50\% of pixels masked out when the masking was done conservatively (i.e., many pixels outside the dust features were 
also masked). To see if this affected the SB-fitting significantly, 
we varied the fraction of pixels to be masked out by decreasing it to 30\% level and performing SB fittings
to such images.  We did not find a strong change in the derived SB profile through this exercise (See Figure \ref{mask_prof1}).
Finally, SB profiles were constructed along the major axis of each ellipse. Figures \ref{rel} -- \ref{disky} show examples of the SB fitting, together with $r$- and $K$-band images of elliptical galaxies. 
The derived total $K$-band magnitudes and the major-axis effective radii $r_{eff}$ in $K$ and $r$ from the fitting  
are presented in Table \ref{tab_sp}.

\subsection{Color Gradients}

 Color profiles were obtained by subtracting a SB profile in one band from a SB profile in another band. 
As the SDSS and the UKIDSS/LAS images have different pixel scales, we linearly interpolated isophotal radii of 
the $K$-band data in $r^{1/4}$ scale, because elliptical galaxies are known to have nearly linear SB profiles in
logarithmic scale as a function of $r^{1/4}$ (de Vaucouleurs' law). Color gradient slopes were obtained 
by a least square fitting of 
a 1st order polynomial to the color profiles divided by the logarithmic radius scale in units of $r_{eff}$ as
given in Eqs. (\ref{eqn_cg_gr}) -- (\ref{eqn_cg_gk}).  The mpfit of Interactive Data Language (IDL) 
is used for the linear fit \citep{mark09}.

\begin{equation}
\nabla_{g-r}= \frac{\Delta(\mathrm{SB}(g)-\mathrm{SB}(r))}{\Delta \log(r/r_{eff}(r))}
\label{eqn_cg_gr}
\end{equation}

\begin{equation}
\nabla_{r-K}= \frac{\Delta(\mathrm{SB}(r)-\mathrm{SB}(K))}{\Delta \log(r/r_{eff}(K))}
\label{eqn_cg_rk}
\end{equation}

\begin{equation}
\nabla_{g-K}= \frac{\Delta(\mathrm{SB}(g)-\mathrm{SB}(K))}{\Delta \log(r/r_{eff}(K))}
\label{eqn_cg_gk}
\end{equation}

 The effective radius in $K$-band, $r_{eff}(K)$, was used for $\nabla_{r-K}$ and $\nabla_{g-K}$ and the $r_{eff}(r)$ was used for $\nabla_{g-r}$. Magnitude errors of color profiles at each radius were taken as SB profile 
 errors in two bands added in quadrature.
 The inner boundary of the fitting range was set to be 1.5 times the seeing FWHM of the image with the worse seeing, 
and the outer boundary is set to be the location where SB starts to drop below 1.5 times the background noise of 
the image with the shallower depth (see \S\ 4.1). 
Note that we used SDSS images with regular depths (the depth comparable to UKIDSS LAS images) to check if the shallower depth of the UKIDSS LAS image affect the color gradient values. Through such an exercise, we confirm that limiting the fitting range to the depth of the shallower image does not introduce systematic bias in the color gradient measurements although uncertainties in the gradient values increase as a result of the loss of S/N. 
 The error in the color gradient is the formal fitting error returned at the end of
 the least-squares fitting process. 
   Figures \ref{col_prof_lm} -- \ref{col_prof_hm} show examples of the $g-r$, $r-K$, and $g-K$ color profiles 
 in three different mass (luminosity) bins, together with the color gradient values.
   Also, we present in Figure \ref{col_prof_comb} the median $g-r$, $r-K$, and $g-K$ color profiles 
 in three different mass (luminosity) bins, overlaid on the color gradient profiles of all the 
 ellipticals belonging to each mass bin.

\subsection{Age and Metallicity Gradients}

 We derived age and metallicity gradients by using a $g-r$ and $r-K$ color-color diagram. Figure \ref{ssp}
 demonstrates the procedure used for deriving the age and the metallicity gradients, showing three example cases.
 We constructed an age-metallicity grid on the $g-r$ and $r-K$ diagram using a simple stellar population (SSP) model
of \citet[hereafter BC03]{bc03}, The Salpeter initial mass function (IMF) and the 
 Padova 1994 stellar evolutionary track were used for the SSP model. 
 To derive the age and the metallicity gradients, locations of the points on a color profile are 
 plotted on the color-color diagram. By doing a bilinear interpolation on the age-metallicity grid,
 we derived the age and the metallicity for each radial point. 
 Then, the age and the metallicity gradients were calculated in a similar way to the color gradients 
 using the following equations. 
 The errors for the age and the metallicity measurements were determined 
 by a drawing 1-$\sigma$ error box over each data point on the grid, 
 and taking the largest/smallest age/metallicity values at corners of the error box to be the upper and 
 lower 1-$\sigma$ limits. 

\begin{equation}
\nabla_{t}= \frac{\Delta \log(Age)}{\Delta \log(r/r_{eff}(K))}
\label{eqn_ag}
\end{equation}

\begin{equation}
\nabla_{Z}= \frac{\Delta \log(Z)}{\Delta \log(r/r_{eff}(K))}
\label{eqn_zg}
\end{equation}

  Finally, we discuss how the derived age/metallicity gradients can be affected by the choice of 
 the star formation history and the IMF.
  For example, a small amount of star formation may make the underlying stellar population look to be younger and more metal than they really are, 
  if they are analyzed by an SSP model (Li et al. 2007). 
  Figure \ref{csp} shows the age-metallicity grid in $g-r$ versus $r-K$ color-color space for three models incorporating different star formation histories (exponentially decaying star formation history and a constant burst with a burst duration of 1 Gyr) and a different IMF.
  As shown in Figure \ref{csp}, extending the duration of the star formation activity 
 (e.g. by adopting a longer burst period in a burst model, or $\tau$, the exponential time scale of an
 exponentially decaying star formation model) would broaden the age intervals in the age-metallicity grid 
 in the $g-r$ direction due to an enhanced contribution of younger stars to the bluer part of the spectrum. 
   The age-metallicity grid expands in the $r-K$ direction too, but to a lesser extent than in $g-r$.
  On the other hand, the extended star formation activity has little effect in the metallicity grid
  in the color-color space. The net effect is that the extended star formation can make 
  the age gradient substantially shallower (by a factor of a few or more in some cases) than in the analysis 
  with the SSP model, while keeping the metallicity gradient results nearly intact. 
  
   This point is made clear in Figure \ref{grad_t_Z_comp}, where we compare the age/metallicity gradients 
  of our ellipticals calculated from different SF history models against those from
  the SSP model. We find that the average of the age gradient flattens by 50\% going from the SSP model
  to the exponentially decaying SF model with $\tau = 1$ Gyr ($\nabla_{t} = 0.31$ to $\nabla_{t} = 0.17$),
  much less so in the average $\nabla_{Z}$ values.
  The effect of the IMF is mild in the color-color grid as discussed in \citep{ren06,han12,man12}. 
  A bottom-heavy IMF \citep{van10,van11} would deform the color-color grid with respect to the Salpeter IMF prediction, but by a very small amount ($<0.1$ mag in color) since low mass stars have little contribution
  to the integrated light of elliptical galaxies. An example is given in Figure \ref{grad_t_Z_comp} for the 
  Chabrier IMF \citep{cha03}, showing a non-significant change in the gradient values.
   Therefore, we consider that the effect of the IMF is negligible in the analysis of the age/metallicity gradient. Figures \ref{age_z_prof_lm} -- \ref{age_z_prof_hm} show examples of age and metallicity profiles in three different mass (luminosity) bins, together with the gradient values.

\section{COLOR, METALLICITY, AND AGE GRADIENTS}

\subsection{Color Gradient}

  Figure \ref{hist_col} shows the marginal distributions of color gradients of different
  types of elliptical galaxies. Also, Table \ref{tab_col_grad} summarizes the median 
  color gradients of each type of elliptical galaxies.
  The overall median color gradient steepens from 
  $\nabla_{g-r} = -0.06$ to $\nabla_{g-K} = -0.33$ as we increase 
  the wavelength base line as we expected. 
   The overall color gradients are in good agreement with previous 
  measurements made by \citet{lab10}.

     Among different types of ellipticals, the color gradient 
     median values are very similar between the relaxed ellipticals and
  the tidal-feature ellipticals. 
   A K-S test of the color gradient distribution
  confirms that the relaxed types and the tidal feature types
  share a nearly identical distribution of color gradients 
  in all the three colors (Table \ref{tab_ks}). This is similar to the result of \citet{kim12} who
  find no strong distinctions between the fundamental planes of relaxed and tidal feature 
  ellipticals.
   On the other hand, the median color gradient of 
  the dust-feature type ellipticals is markedly steeper than
  the other two types by twice to three times
  (Table \ref{tab_col_grad}). A K-S test gives less than 5 \% probability that 
  the color gradient distributions of the 
  dust-feature ellipticals and the relaxed ellipticals are 
  identical (Table \ref{tab_ks}). 
   Note that we masked out the dust-features
  when fitting the SB profiles of the dust-feature ellipticals, 
  in order to minimize the dust extinction affecting 
  color gradient measurements. Therefore, 
  we suggest that a difference in the formation and the evolution
  process is the origin for the discrepancy, rather than
  the dust extinction.
   
   The fact that the relaxed type ellipticals and the tidal-feature type ellipticals 
  share a similar color gradient distribution means that they share a similar
  merging history. The steep color gradient in the dust-feature types
  is in agreement with the expected steep color gradient from a  monolithic
  collapse or a dissipative major merger with a high star formation
  efficiency. Therefore, we suspect
  that the dust-feature type ellipticals are the products of merging between
 gas-rich galaxies where the remaining gas producing the dust 
  features.

  In Figure \ref{kmag_col}, we present the $K$-band absolute magnitude
 ($M_{K}$) versus the color gradients in three different colors. 
  Since $M_{K}$ is closely related to the stellar mass ($\mathrm{M}_{*}$),
 we also mark the stellar mass on the x-axis, where the
 stellar mass is computed assuming a mass-to-light ratio of 1.
  For a $K$-band mass-to-light ratio of 1, the absolute magnitudes of
  $M_{K}=-22$ and $-23$ mags correspond
 to  the stellar mass of $10^{10.88}$ and $10^{11.28}\,\mathrm{M_{\odot}}$
 respectively \citep{dro04}. 
  Different types of elliptical galaxies are plotted with different symbols,
 and the median, the error of the median, and the rms dispersion 
 of the color gradients are indicated
 at each absolute magnitude bin, allowing us to understand 
 how the color gradient changes as a function of the absolute magnitudes
 (or the stellar masses). At the bottom of each figure, we indicate
  median errors in the color gradient measurements.

   If dry merging mixes up stellar population in each galaxy, 
   a naive expectation is that the color gradients become shallower and the dispersion
   becomes smaller toward the brighter magnitude. 
   However, Figure \ref{kmag_col} shows that the median values of the color gradients 
  are nearly constant as a function of absolute magnitude, 
  somewhat in contrary to the naive expectation (possibly with an exception at
 the brightest end). On the other hand, the dispersion of the color gradients appears  
  to decrease at the brighter end, consistent with the expectation. 
   A similar trend is evident in Figure \ref{col_prof_comb} too. However,
 the steady decrease in the scatter is illusionary and reflects the larger
 measurement errors as discussed below.

   To examine this trend more closely, we plot the intrinsic 
  scatter of the color gradients as a function of $K$-band absolute 
  magnitude in Figure \ref{scat_col}. Note that the measurement errors add noises
  to the intrinsic scatter, therefore we subtracted 
  the contribution of the measurement errors from the observed scatter 
  to obtain the intrinsic scatter as described below. Note that
  the subtraction of the measurement errors is more important at the faint
  end than at the bright end since faint objects have large color 
  gradient errors due to the lack of data points to fit for their SB.
   We assumed that the color gradients are distributed in Gaussian
  with its mean value equal to the measured median value of the 
  color gradient with a certain intrinsic dispersion.
   Then, we constructed the measurement error distribution for
  actual measurement errors.
   We took a median value of the measurement errors (at each
   absolute magnitude bin), and subtracted
  it from the observed rms scatter to obtain an initial guess for 
   the intrinsic scatter.
   In the next step, we randomly picked color gradients from the normal
   distribution of the color gradients, assigning measurement errors
   to each value by randomly picking a measurement error from
   the measurement error distribution that we constructed earlier.
   The median value and the rms scatter are  
  computed from this simulated set of color gradients. The rms 
   scatter was compared with the observed rms scatter, and the value of the 
   intrinsic scatter was adjusted if the rms scatter from the simulation
   does not match the observed rms scatter.
   Finally, after iterations of the above process,
   we determined the intrinsic scatter to be the intrinsic
   scatter in the simulation where the rms scatter of the simulation
   is identical to the observed rms scatter. 
  
    Error of the intrinsic scatter of color gradients, $\sigma_{x}$, 
   was estimated by propagating 
   the color gradient measurement errors in the following way.
    Here, $x$ is the intrinsic scatter, $N$ is the number of galaxies 
   in a certain magnitude bin, $a_{i}$ is the color gradient value
   of the i'th galaxy, $\sigma_{a_{i}}$ is the measurement 
   error of the color gradient of the i'th galaxy, 
   and $\mu$ is the average of the color gradients. 
   The values of the scatters are presented in Table \ref{tab_scat} 
   and shown in Figure \ref{scat_col}. 
  Although the estimates of the error values are mathematically sound, 
  we caution that the the error estimates do not include potential systematic effects 
  discussed in Section 3.3, and may be under- or over-estimated due to 
  small number statistics in some of the bins.

\begin{equation}
x = \sqrt{\frac{\left( \sum_{i=1}^N a_{i}^{2} \right) }{N} - \left( \frac{\sum_{i=1}^N a_{i}}{N}\right) ^2}
\label{eqn_scat_err1}
\end{equation}

\begin{equation}
{\sigma_{x}}^{2} = \sum_{i=1}^N {\sigma_{a_{i}}}^2 \left( \frac{\partial x}{\partial a_{i}} \right) ^2 ~,~
\frac{\partial x}{\partial a_{i}} = \frac{a_{i}-\mu}{Nx} 
\label{eqn_scat_err2}
\end{equation}

\begin{equation}
\sigma_{x} = \sqrt{ \sum_{i=1}^N {\sigma_{a_{i}}} ^2 \left( \frac{a_{i}-\mu}{Nx} \right) ^2}  
\label{eqn_scat_err3}
\end{equation}
             
    In Figure \ref{scat_col}, we find that the steady decrease of the rms scatter
   toward the massive end disappears for the optical-NIR color gradients when the
   intrinsic scatter is considered. The decrease in the intrinsic scatter is only
   visible at the most massive end at a few $\sigma$ level. The implication of the
   result will be discussed later, in relation to the merging activity (\S 5).

\subsection{Age and Metallicity Gradients}  
     
  Our analysis (\S\ 3.3) allows us to derive
  age and metallicity gradients. The results of the age and 
  the metallicity gradient are described in this section.
  
  Figure \ref{hist_tz} shows the marginal distribution of the age and the metallicity 
  gradients. As was the case for the color gradients, the relaxed type
  ellipticals and the tidal-feature type ellipticals share a nearly identical
  distribution, with a positive age gradient ($\nabla_{t} \simeq 0.2$) 
  and a negative metallicity gradient ($\nabla_{Z} \simeq -0.4$).  
    The dust-feature type ellipticals have small age gradients (except for
  two outliers), and significantly steep metallicity gradients 
  of $\nabla_{Z} \simeq -0.8$.
    The very steep metallicity gradient of dust-feature type ellipticals
  supports the idea that they are the products of 
  wet merging events. 
  
    The derived mean $\nabla_{Z}$ is steeper than $\nabla_{Z}$ values derived
  from studies that are based on the line indices 
  ($\nabla_{Z} \sim -0.2$ -- $-0.3$; \citealt{oga05};
  \citealt{spo10}; \citealt{raw10}; \citealt{san07}; citealt{ani07}),
  but is in agreement with the values in \citet{lab10} and \citet{tor12}   
  that are based on color profile analysis like our study. The origin 
  of the discrepancy is unclear although a slightly positive age 
  gradient in the color gradient analysis steepens the metallicity 
  gradient but not to the extent that it resolves the difference. The line indices
  analysis that derive the flatter $\nabla_{Z}$ assume $\nabla_{t} = 0$ or find 
  null $\nabla_{t}$. While the line indices studies are limited at $r<r_{eff}$,
  the color gradient studies are done out to a few $\times~r_{eff}$, so this may
  be the cause of the discrepancy.

  The absolute values of the gradients are
  quantities that are subject to the analysis methods 
  (e.g., the adopted stellar population synthesis models, see Li et al. 2007) 
  so that we caution readers not 
  to pay too much attention to the absolute values but concentrating on 
  the relative changes in the gradients.

  In Figure \ref{kmag_az}, we show the $M_{K}$ versus the age and the metallicity gradients.
  With the thick points, we indicate the median $\nabla_{t}$ and $\nabla_{Z}$ values
  and their errors. On these points, the thin error bars show the rms dispersion. 
  The thick error bars represent the errors of the median, which are derived by 
  dividing the rms dispersion by $\sqrt{N}$ where $N$ is the number of objects
  in the corresponding bin.
   Like the color gradients, the age and the metallicity gradients do not show 
  a significant change as a function of $M_{K}$.
  One exception is the brightest bin (corresponding to the stellar mass of 
  $\sim 5 \times 10^{11} \, \mathrm{M_{\odot}}$) where both $\nabla_{t}$ and 
  $\nabla_{Z}$ start to flatten. However this trend at the brightest bin should 
  be taken with a caution since only six objects are included in the bin.

  Just like the color gradients, this result is consistent with a similar analysis
  performed by \citet{lab10}. On the other hand, \citet{lab10} do not show  
  a result at the very massive end of $\sim 5 \times 10^{11} \, \mathrm{M_{\odot}}$,
  so the flattening of the age and the metallicity gradients at the 
  brightest bin is not noticed in their work.  
  The flattening of the age and the metallicity gradients at the brightest
  end is interesting
  since this is exactly what is expected if these ellipticals underwent
  frequent dry merging. Obviously, more number statistics at the bright
  end is desired to make a stronger case for this.
  
   Similarly to the color gradients, the scatters in the age and the color gradients
  appear to decrease at the brightest magnitude bin.
   As a closer examination of this trend, we show in Figure \ref{scat_az} and Table \ref{tab_scat}
  the intrinsic scatter of the age and the metallicity gradients as a function
  of the absolute magnitude.
   The intrinsic scatter in these gradients are obtained in an iterative process
   identical to the method we employed for the intrinsic scatter of color 
   gradients (see the previous subsection). Errors of the intrinsic scatters
   are estimated using Eqs. (\ref{eqn_scat_err1}) -- (\ref{eqn_scat_err3}), but here
   replacing the color gradients with $\nabla_{t}$ or $\nabla_{Z}$.

   The derived intrinsic rms scatter is about 0.3 at a lower mass range of  
 $10^{10.7} \, \mathrm{M_{\odot}} < \mathrm{M} < 10^{11.4} \, \mathrm{M_{\odot}}$,
 which is roughly in agreement with the scatter found in \citet{oga05} at a similar mass range.
   Like the scatters of color gradients, the scatters of these quantities
  stay nearly constant out to $M_{K} \sim -23.40$ or $\mathrm{M} \sim 10^{11.4} \, 
  \mathrm{M_{\odot}}$, but decreases at the most massive end by nearly a factor
  of two per a factor of two change in mass (or luminosity).
   The reduction in the scatter is dominated by the relaxed and the tidal-feature type 
  ellipticals. On the other hand, the reduction in the scatter is inversed for
  the tidal-feature type ellipticals, although the larger errors in the scatter
  (due to the smaller number of galaxies available) for these types of objects 
  make it difficult to draw a firm conclusion.

  Overall, the two evidences -- the possible flattening of the age and the metallicity gradients
  and the decrease in the scatter of these gradients at the most massive end 
  -- are in agreement with the expectation of what would happen under dry merging, and
  this point will be discussed in more detail below.

\section{IMPLICATION ON DRY MERGING}

\subsection{Model}

 In the previous section, we saw that the scatters in the color/age/metallicity gradients
 decrease at the brightest $M_{K}$. 
 Here, we present a simple model that calculates how the metallicity 
 gradient would change as a function of absolute magnitude and mass if major dry 
 merging is the dominant mechanism for the growth of elliptical galaxies at
 the massive end.
 
  To do so, we use a simulation result from \citet{dim09} where
  they perform simulations of major dry merging of
 elliptical galaxies with various initial conditions,
 and present the metallicity gradients of pre-mergers and merger products. 
  Borrowing their results, 
  we establish a relation between the average metallicity gradients of
  two galaxies before merging, and the metallicity gradient of the merger
 product (Figure \ref{dmerg}). We find that there is a good correlation between
 the average of the metallicity gradients of pre-merging galaxies 
 ($[\nabla_{Z}(gal1) + \nabla_{Z}(gal2)]/2$) and
 the metallicity gradient of the merger product ($\nabla_{Z}(merger)$) as
 
 \begin{equation}
 \nabla_{Z}(merger) = 0.55 \times \frac{\nabla_{Z}(gal1) + \nabla_{Z}(gal2)}{2} - 0.02.
 \label{eq_dmerg} 
 \end{equation}
 
   The relation shows that the metallicity gradient flattens on average when two
 galaxies merge, with the resulting $\nabla_{Z}$ being
 about 60 \% of the average value of the $\nabla_{Z}$'s of the two pre-merging galaxies.
  The correlation is reasonably tight, with the largest rms dispersion of
  0.03 in $\nabla_{Z}(merger)$. This dispersion is sufficiently smaller than
  the reduction in the metallicity gradient value (which is order of 0.1), therefore
  we ignore the dispersion in Eq. (\ref{eq_dmerg}) in the following discussion. 
 
  Two consequences are expected from Eq. (\ref{eq_dmerg}).
  One is that the metallicity gradient flattens gradually as the masses of
 ellipticals increase in a dry merging dominated evolution.
  Roughly speaking, there should be a decrease in the metallicity gradient by
 a factor of 0.55, over a mass doubling scale (0.3 dex in log(M)).
  Another consequence is the decrease of the scatter in the color and 
 the stellar population gradients at a high mass end. Suppose that
 we have a variety of the stellar population gradients for pre-merging
 ellipticals. If we randomly merge galaxies with different 
 metallicity gradients, the merging averages the metallicity gradient values
 and remove galaxies with extreme metallicity gradients. Since the merger 
 product will be more massive than each of its predecessors, we expect
 that the major dry merging would reduce the scatter in the color gradients at high
 mass end.
 
   To see how the mean $\nabla_{Z}$ and its intrinsic scatter would evolve
  under dry merging, we performed a simple simulation. Here, we assumed an initial
  distribution of $\nabla_{Z}$ in Gaussian with its mean value at 
  $\nabla_{Z} = -0.4$ and the intrinsic scatter at $\sigma_{\nabla_{Z}} = 0.3$.
  These values correspond roughly to the observed mean $\nabla_{Z}$ and its scatter at 
  $\mathrm{M} = 10^{10.9} \mathrm{M_{\odot}}$. We assume that a certain fraction,
  $f_{merg}$, of galaxies in a mass bin underwent dry merging. This free parameter
  $f_{merg}$ will control the amount of change in the mean metallicity gradient and
  its scatter.
 
   Figure \ref{dm_frac} shows the result of this simulation.
   The metallicity gradient and its scatter change the most 
   when $f_{merg}$ is the largest. At $f_{merg}=1$, the mean 
   metallicity gradient and its scatter changes by a factor of
   0.6 and 0.4 respectively, per each mass doubling. We find that
   this trend is independent of the initial mean $\nabla_{Z}$ and
   $\sigma_{\nabla_{Z}}$ values. 
   
   Based on the simulation, we derive the following relations at
   $f_{merg}=1$:  
   
   \begin{equation}
   \Delta\,\mathrm{log}(\langle \nabla_{Z} \rangle) = -0.74 \Delta\,\mathrm{log}(\mathrm{M}),
   \end{equation}
   
   and
   
   \begin{equation}
   \Delta\,\mathrm{log}(\sigma_{\nabla_{Z}}) = -1.3 \Delta\,\mathrm{log}(\mathrm{M}).
   \end{equation}
  
   Here, $\Delta\,\mathrm{log}(\langle \nabla_{Z} \rangle)$, $\Delta\,\mathrm{log}(\sigma_{\nabla_{Z}})$, 
   and $\Delta\,\mathrm{log}(\mathrm{M})$ are the logarithmic changes in the mean 
   metallicity gradient,
   the intrinsic scatter of the metallicity gradient, and the stellar mass, respectively.
   
   The above relations show that the change in $\sigma_{\nabla_{Z}}$ is more prominent
   than the change in $\langle \nabla_{Z} \rangle$ as a result of dry merging by 
   a factor of two.

  \subsection{Is Major Dry Merging Important?}
  
  With the theoretical expectations from major dry merging at hand, we now
  proceed to interpret our observational results about $\nabla_{Z}$. Figure \ref{scat_sim}
  shows the comparison of the observed change in $\nabla_{Z}$ and 
  $\sigma_{\nabla_{Z}}$ with a model prediction. In the figure, the error
  in $\sigma_{\nabla_{Z}}$ is estimated in the same way as the errors in
  the scatter of the color gradient values as described in Section 4.2 and
  using Eq. (8). The model plot assumes no major dry merging at 
  $\mathrm{M} < 10^{11.3} \, \mathrm{M_{\odot}}$, $f_{merg} = 0.3$ at 
  $\mathrm{M} = 10^{11.4} \, \mathrm{M_{\odot}}$, and $f_{merg} = 0.9$ at 
  $\mathrm{M} = 10^{11.6} \, \mathrm{M_{\odot}}$.

  As we have shown earlier, the reduction in $\sigma_{\nabla_{Z}}$
  is observed at the massive end ($> 10^{11.4} \, \mathrm{M_{\odot}}$). The major dry merging can provide
  an explanation for this result. The observed reduction in the scatter is about
  0.1 over the mass increase of 0.4 dex, and according to Figure \ref{dm_frac}, 
  this can be achieved by assuming that 70 -- 100 \% of ellipticals at the higher
  mass bin have experienced major dry merging while the rest went through
  the formation mechanisms similar to those at the lower mass bins 
  (i.e., they possess similar $\nabla_{Z}$ and $\nabla_{t}$ distributions to those
  at the lower mass bin). The possible formation and the evolution mechanisms
  at the lower mass bin include the wet merging, the minor merging, and
  the monolithic collapse (see discussions below).

   According to our model, this amount of major dry merging needs to flatten 
  $\nabla_{Z}$ by 0.15 over a mass doubling scale. This value is rather small
  compared to the change in the scatter, but should be detectable with a 
  sufficient number of objects. At the moment, we find such a trend, but with 
  a rather large error bar. In comparison to other works, our results are in 
  agreement with \citet{roc10} where they find a reduced scatter in color 
  gradients of BCGs. Also note that some other studies report a flattening
  of the metallicity gradient at the massive end \citep[e.g.,][]{spo10}.

   While we find a supporting evidence for major dry merging at the high 
  mass end, major dry merging is probably not a main 
  mechanism driving the evolution of lower mass ellipticals with 
  $\mathrm{M} < 10^{11.4} \, \mathrm{M_{\odot}}$. Between 
  $\mathrm{M} = 10^{10.7} \, \mathrm{M_{\odot}}$ and 
  $\mathrm{M} = 10^{11.4} \, \mathrm{M_{\odot}}$, the $\sigma_{\nabla_{Z}}$ values 
  stay nearly constant. The same trend is true for the color and the age gradients.
  Clearly, the trend is not consistent with the expectation from
  the major dry merging model we presented in this paper. 
  In order to explain the null change in
  the metallicity gradients, $f_{merg}$ should be less than $\sim 20\%$.  The effect 
  of major dry merging seems negligible in the evolution of ellipticals at this
  mass range. The origin of the large scatter in $\nabla_{Z}$ at this mass range 
  is unclear, although there are plenty of possibilities responsible for
  the dispersion. Minor dry merging could steepen the metallicity gradient,
  rather then flattening it, by distributing the merged stellar remnants 
  in the outer region not the inner region \citep{ber11}. 
   Dissipative collapse could create a variety of gradients depending on
  the star formation efficiencies how the gravitational potential is shaped 
  for each object \citep{pip10}. Minor merging
  with a gas-rich galaxy could place younger generation stars at central or 
  outer regions, making galaxies blue at certain areas and complicating 
  the analysis of the result \citep[e.g.,][]{im01}.

 \section{SUMMARY}
 
   We examined the color, the age, and the metallicity gradients of
  of 204 nearby ($z < 0.05$), massive ($M_{r} < −20.5$ mag, or
 $\mathrm{M} > 10^{10.6} \, \mathrm{M_{\odot}}$) elliptical galaxies 
 (early-type galaxies with $B/T > 0.5$), divided into three morphological types
 -- the relaxed,  the tidal-feature, and the dust-feature types. The subdivision of
 elliptical galaxies was possible by using the SDSS Stripe 82 images 
 which goes about 2 mag deeper than regular SDSS images, and the derivation of the age
 and the metallicity gradients was done by adding $K$-band data of the UKIDSS
 LAS to the SDSS data which enabled us to break the age-metallicity degeneracy.
  No significant difference is found between relaxed and
 tidal-feature types in the gradient properties. However, the dust-feature type
 ellipticals have steeper color gradients which seem to originate from wet merging
 rather than the dust extinction effect. The gradient values of elliptical galaxies
 are found to be nearly constant over the explored mass range, but we find that the
 intrinsic scatter of $\nabla_{Z}$ decreases at $\mathrm{M} > 10^{11.4} \, \mathrm{M_{\odot}}$
 by about 0.1.

      To interpret our results, we constructed a simple model that describes  
  changes in $\nabla_{Z}$ and $\sigma_{\nabla_{Z}}$ as a function of mass under 
  major dry merging. The model predicts that  the mean $\nabla_{Z}$ flattens, and
  $\sigma_{\nabla_{Z}}$ decreases as mass of an elliptical galaxy doubles as a result
  of major dry merging. 
    
     From the analysis of the change in $\nabla_{Z}$ and its scatter, we draw the 
  following conclusions, with a caveat that the result is based on a rather small
  number of objects at the massive end ($\sim 10$). First, the observed reduction
  in the intrinsic scatter in the $\mathrm{M}$ versus $\nabla_{Z}$ relation
  supports the idea that major dry merging has acted on a significant number of ellipticals 
  at the very massive end ($\mathrm{M} \sim 10^{11.6} \, \mathrm{M_{\odot}}$).  
  Second, if the reduction of the scatter in $\nabla_{Z}$ is indeed caused by
  major dry mergers, $\nabla_{Z}$ should flatten at the higher mass.
  The result from our analysis of the SDSS/UKIDSS-LAS ellipticals is consistent
  with this scenario at the massive end, but not conclusive.
  Third, at the lower mass regime ($\mathrm{M} < 10^{11.4} \, \mathrm{M_{\odot}}$), 
  the major dry merging does not appear to be the dominant process in the evolution
  of ellipticals. At such a mass range, the mean $\nabla_{Z}$ is nearly constant, 
  and the scatter in the $\mathrm{M_{*}}$ versus $\nabla_{Z}$ relation is large and
  nearly constant ($\sim 0.4$) too, implying that various physical mechanisms are
  responsible for the large scatter.

\acknowledgments

 This work was supported by the Creative Initiative program, No.
2010-0000712, of the National Research Foundation of Korea (NRFK) funded by the Korean
government (MEST). We thank our CEOU/SNU colleagues for useful discussion.

This paper uses the data taken with the United Kingdom Infrared Telescope (UKIRT) which
is operated by the Joint Astronomy Centre on behalf of the Science and Technology
Facilities and from the Sloan Digital Sky Survey project.

\clearpage

\clearpage

\clearpage

\clearpage

\clearpage

\clearpage

\begin{figure}
\epsscale{1.0}
\plotone{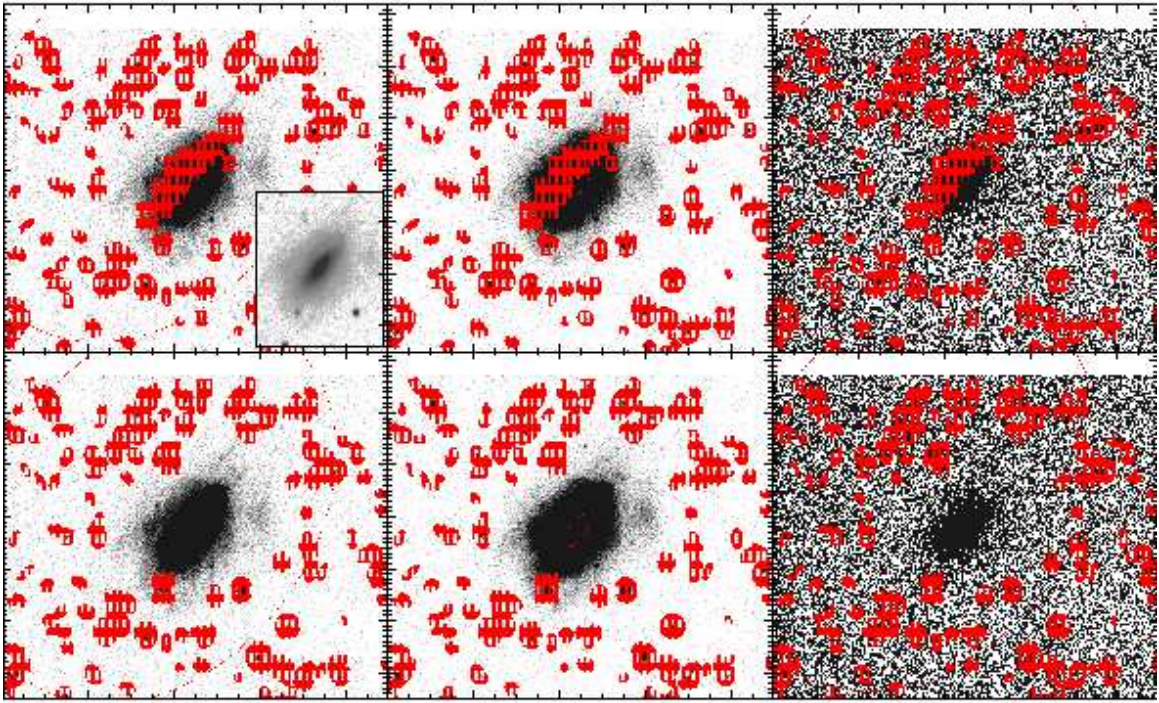}
\caption{Example of masking processes during ELLIPSE fitting of elliptical galaxies (with dust lane). Top panels represent $g$, $r$ and $K$-band images when dust lane is masked out from left to right, and the bottom panels represent the same but without masking of the
dust lane.}
\label{mask}
\end{figure}

\clearpage

\begin{figure}
\epsscale{1.0}
\plotone{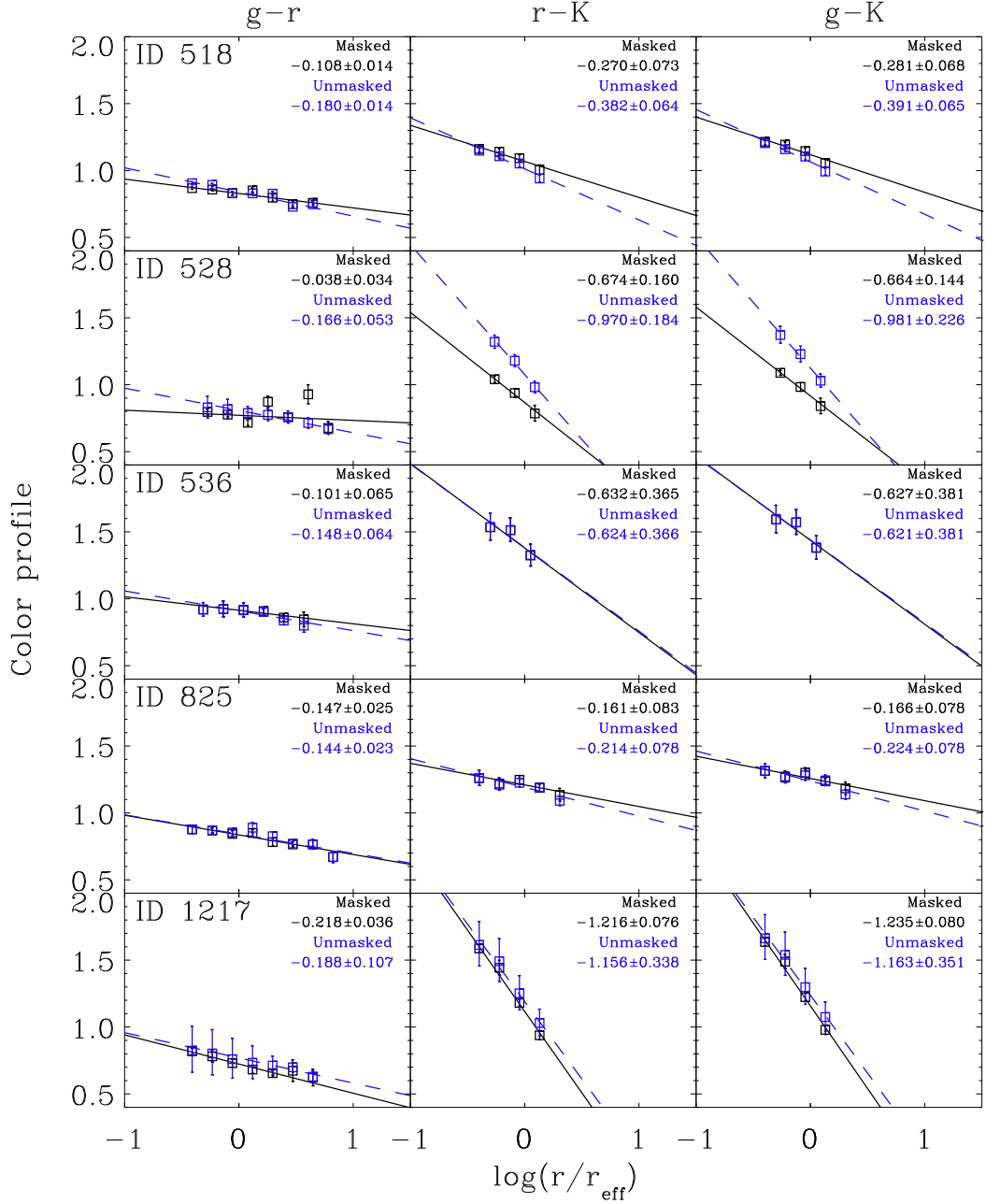}
\caption{Color profiles of dust-feature elliptical galaxies in $g-r$, $r-K$ and $g-K$ from left to right. Black squares and solid lines are their color profiles after masking out dust lanes during the ELLIPSE fit. The blue squares and the dashed lines represent the cases
without masking dust lanes.}
\label{mask_prof1}
\end{figure}

\clearpage

\begin{figure}
\epsscale{1.0}
\plotone{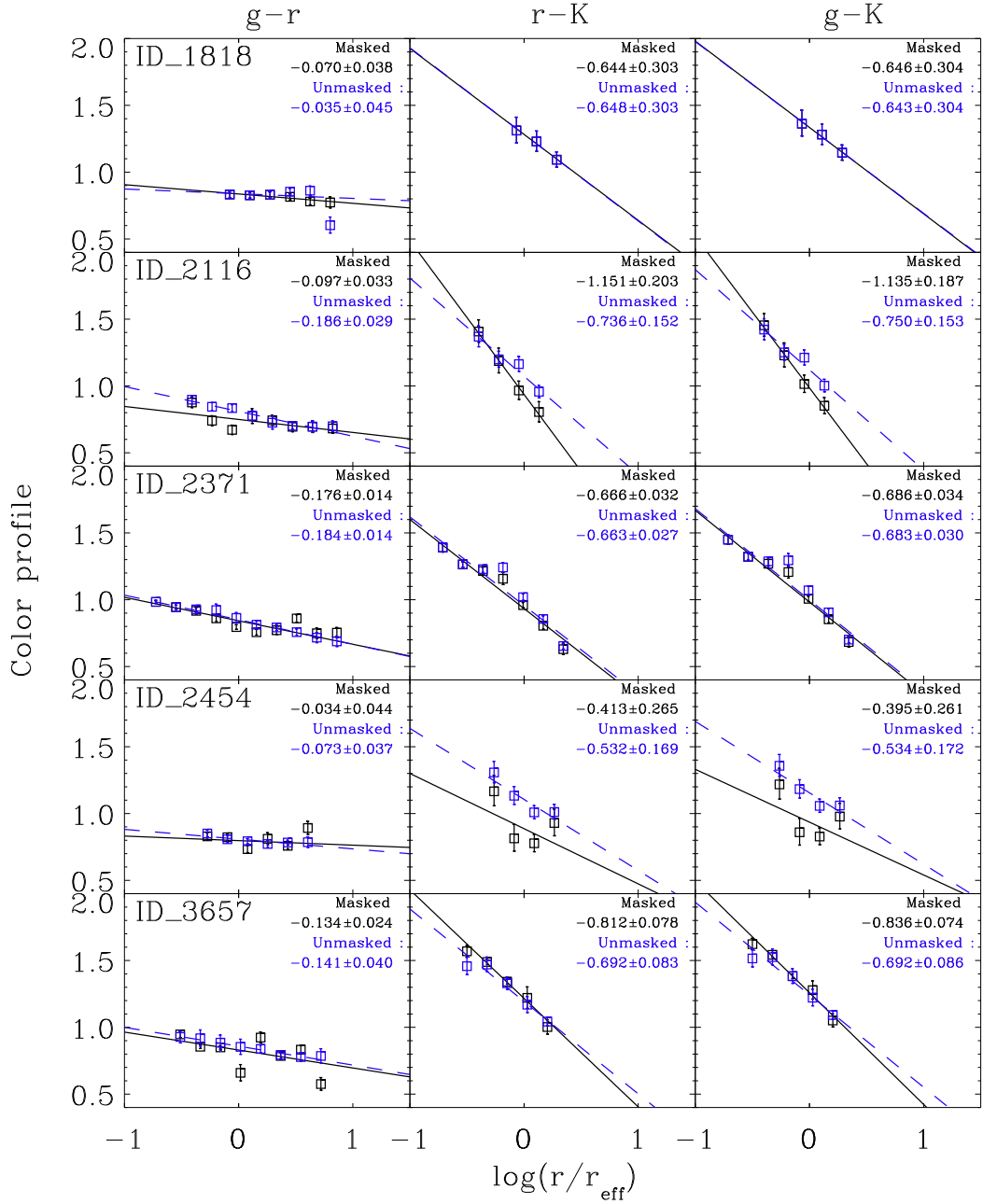}
\caption{Figure \ref{mask_prof1}, continued.}
\label{mask_prof2}
\end{figure}

\clearpage

\begin{figure}
\epsscale{1.0}
\plotone{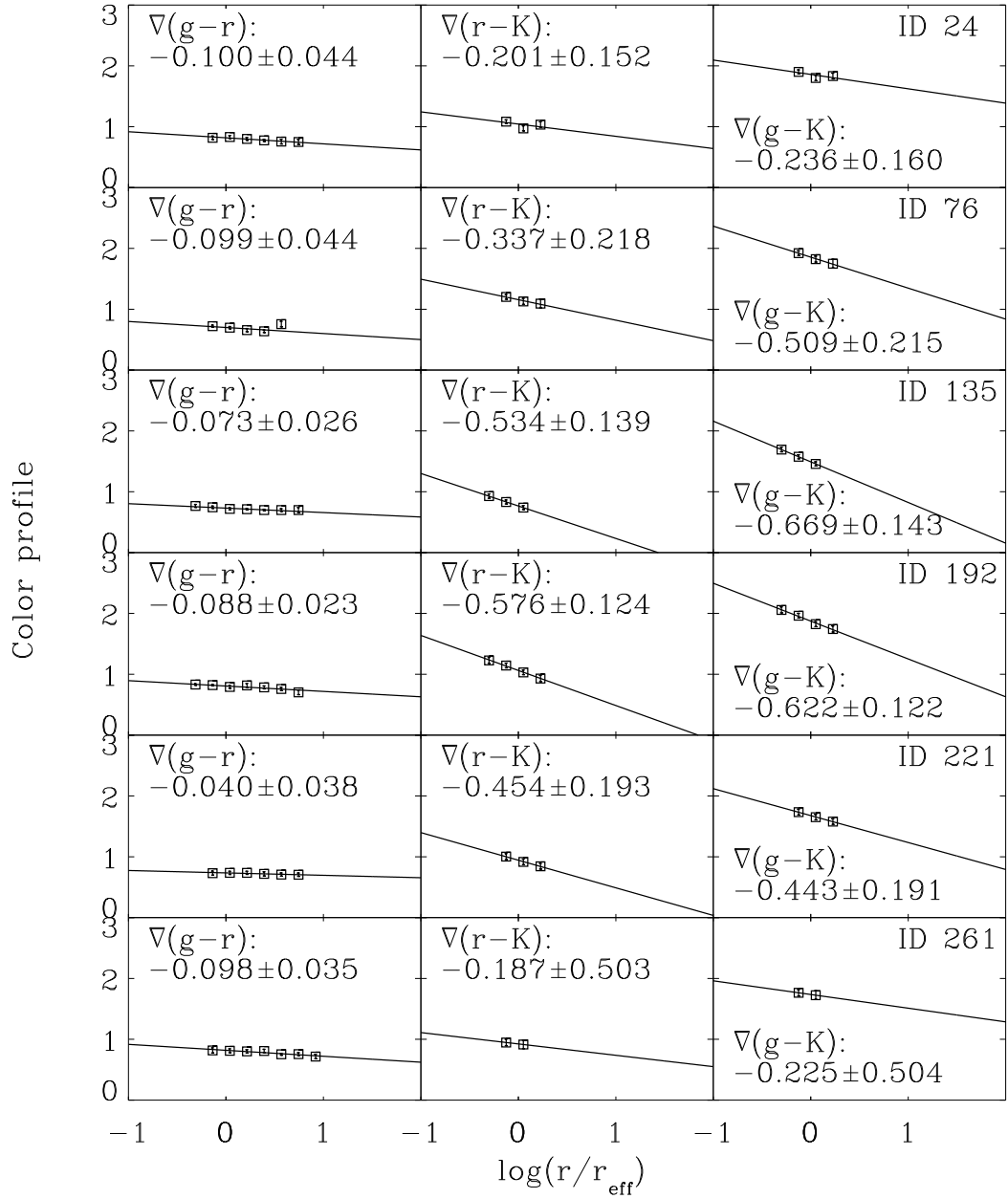}
\caption{Color profiles and their color gradient values of elliptical galaxies in low-mass regime ($10^{10.6} \, \mathrm{M_{\odot}} < \mathrm{M_{*}} < 10^{10.88} \, \mathrm{M_{\odot}}$). The best-fit lines are indicated for the color gradients too 
(the solid line).}
\label{col_prof_lm}
\end{figure}

\clearpage

\begin{figure}
\epsscale{1.0}
\plotone{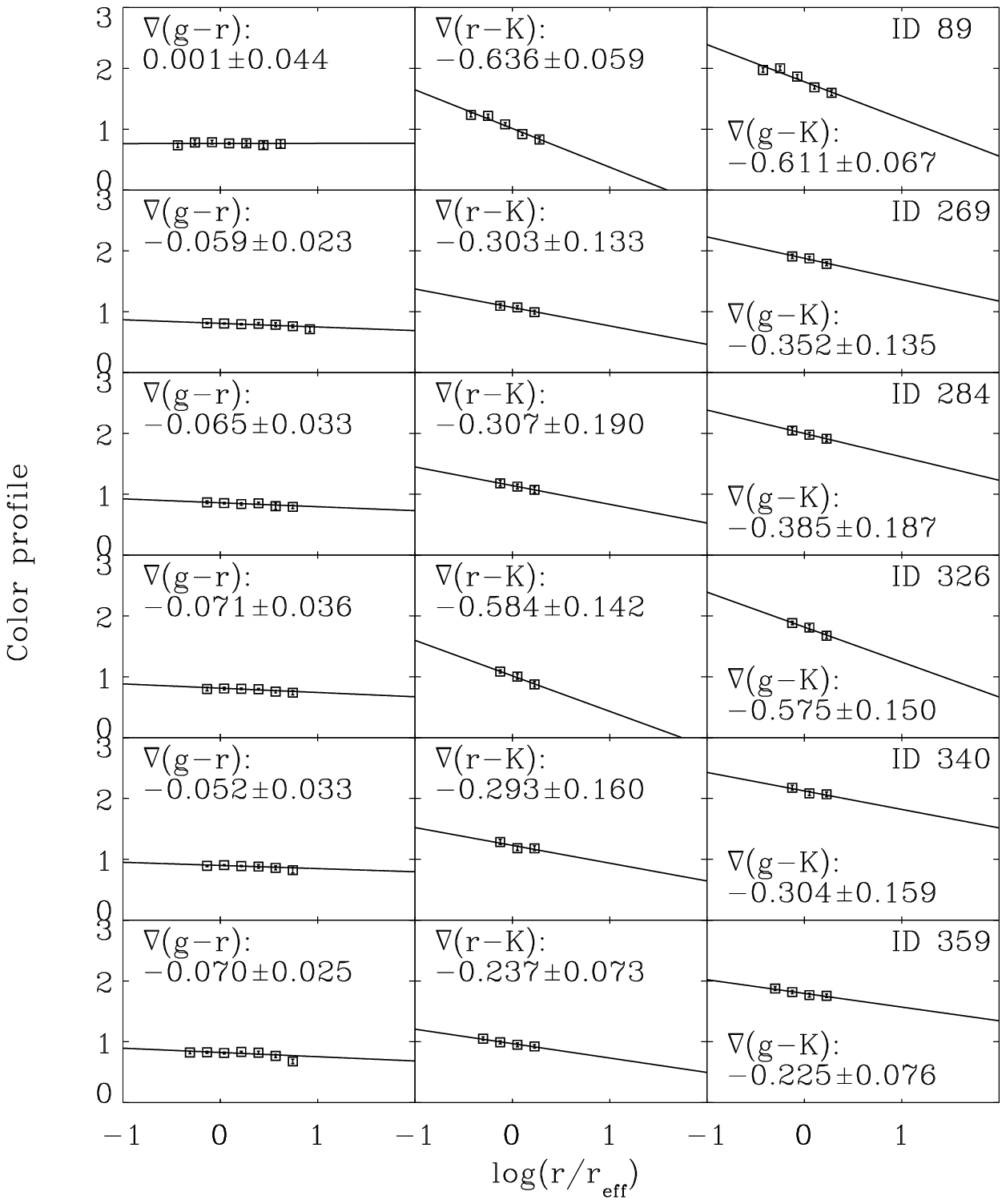}
\caption{Same as Figure \ref{col_prof_lm} but for elliptical galaxies in the intermediate mass regime ($10^{10.88} \, \mathrm{M_{\odot}} < \mathrm{M_{*}} < 10^{11.28} \, \mathrm{M_{\odot}}$).}
\label{col_prof_im}
\end{figure}

\clearpage

\begin{figure}
\epsscale{1.0}
\plotone{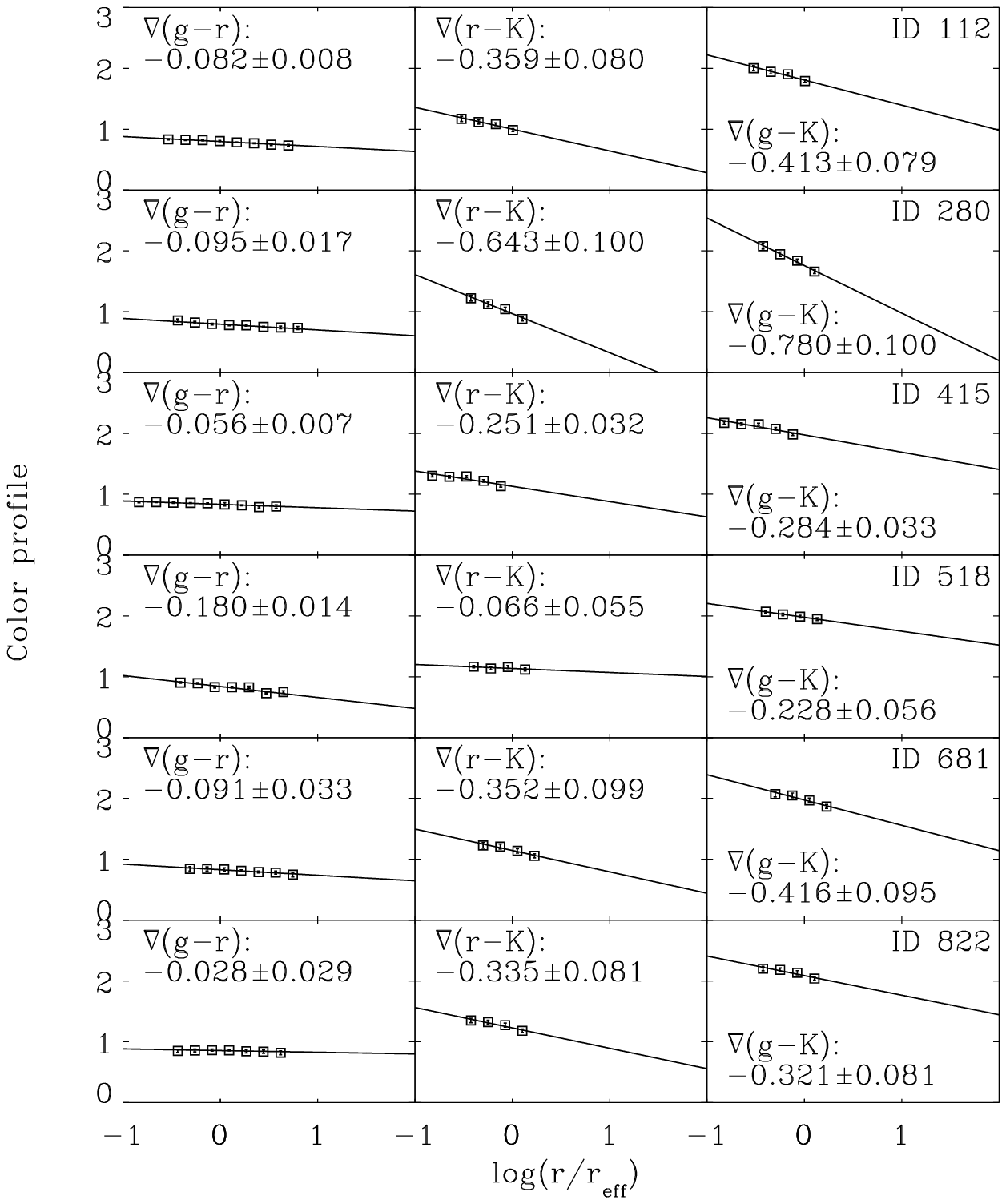}
\caption{Same as Figure \ref{col_prof_lm} but for elliptical galaxies in high-mass regime
($\mathrm{M_{*}} > 10^{11.28} \, \mathrm{M_{\odot}}$).}
\label{col_prof_hm}
\end{figure}

\clearpage

\begin{figure}
\epsscale{1.0}
\plotone{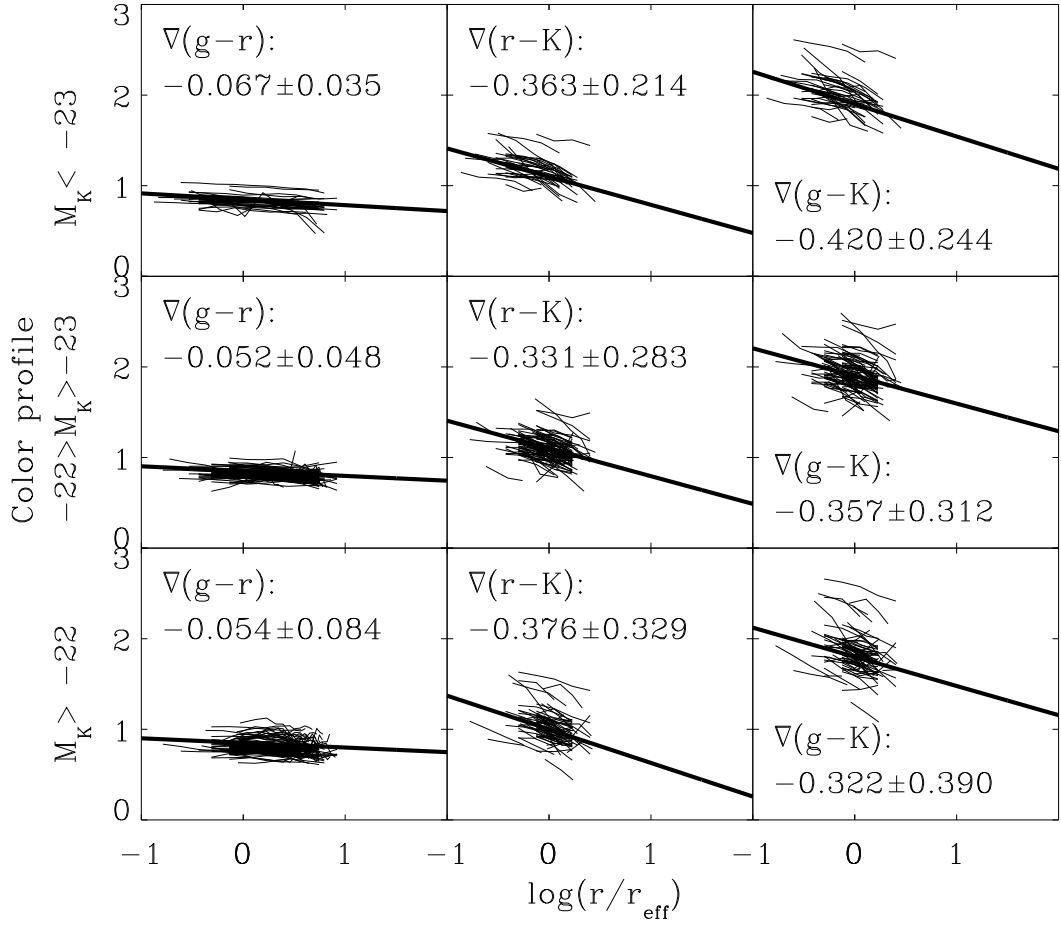}
\caption{Color ($g-r$, $r-K$ and $g-K$ from left to right) profiles and median color gradients (solid line) of elliptical galaxies in high-mass regime
($\mathrm{M_{*}} > 10^{11.28} \, \mathrm{M_{\odot}}$), intermediate mass regime ($10^{10.88} \, \mathrm{M_{\odot}} < \mathrm{M_{*}} < 10^{11.28} \, \mathrm{M_{\odot}}$) and low-mass regime ($10^{10.6} \, \mathrm{M_{\odot}} < \mathrm{M_{*}} < 10^{10.88} \, \mathrm{M_{\odot}}$) from top to bottom.}
\label{col_prof_comb}
\end{figure}

\clearpage

\begin{figure}
\epsscale{1.0}
 \includegraphics[scale=0.6]{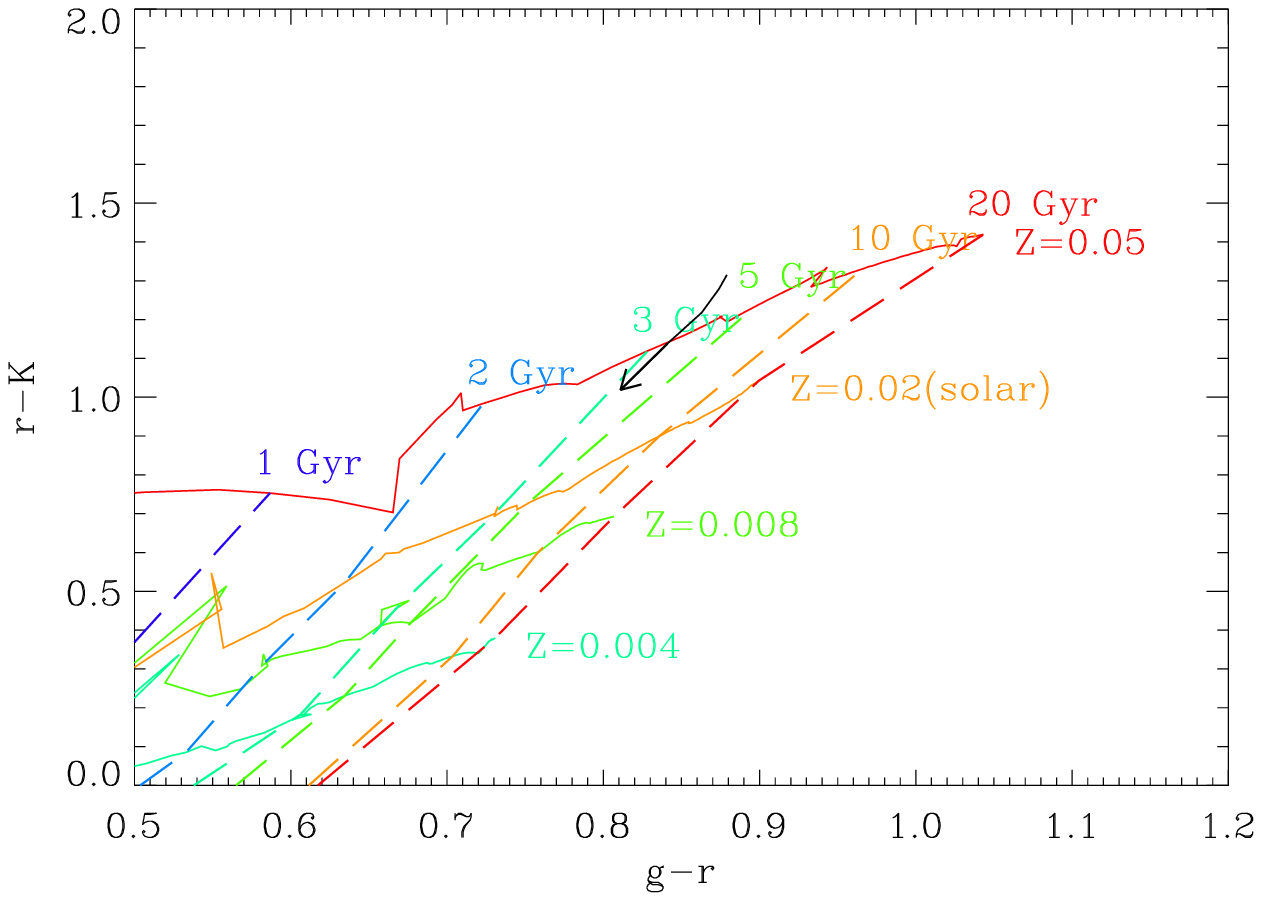}
 \includegraphics[scale=0.7]{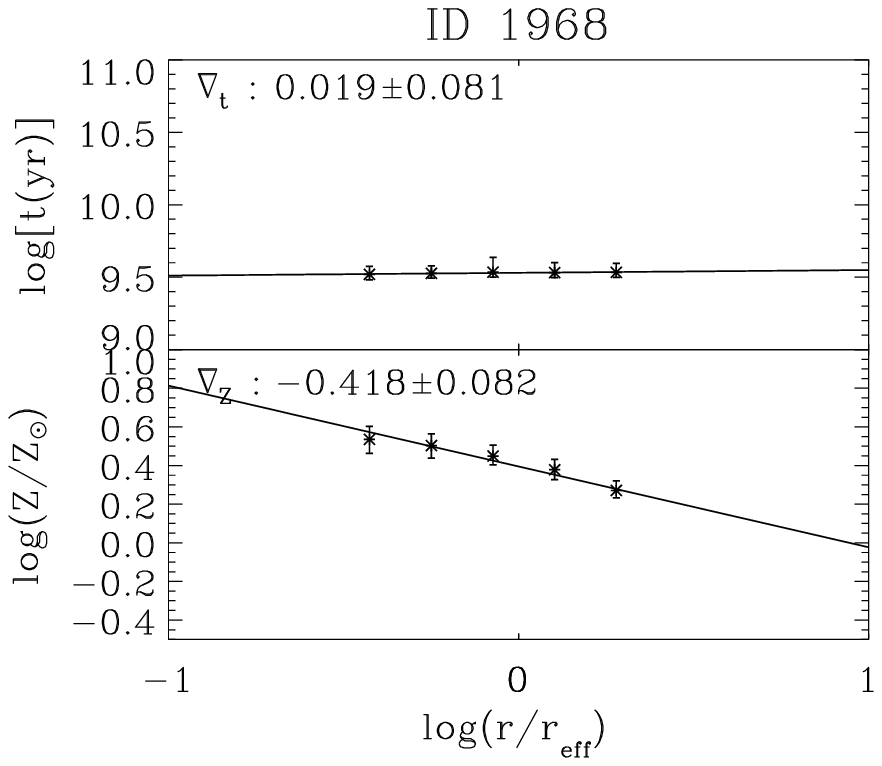}\\
 \includegraphics[scale=0.6]{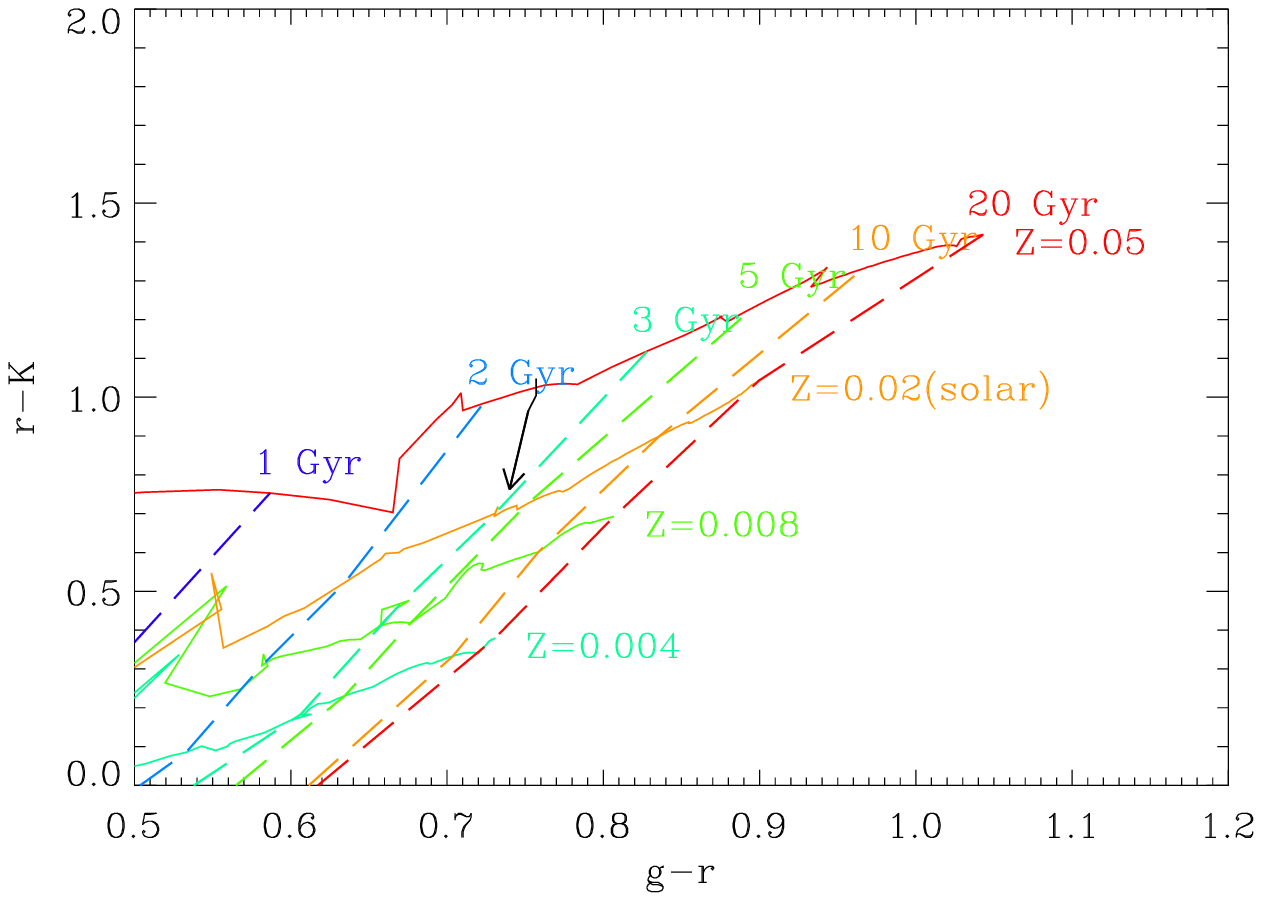}
 \includegraphics[scale=0.7]{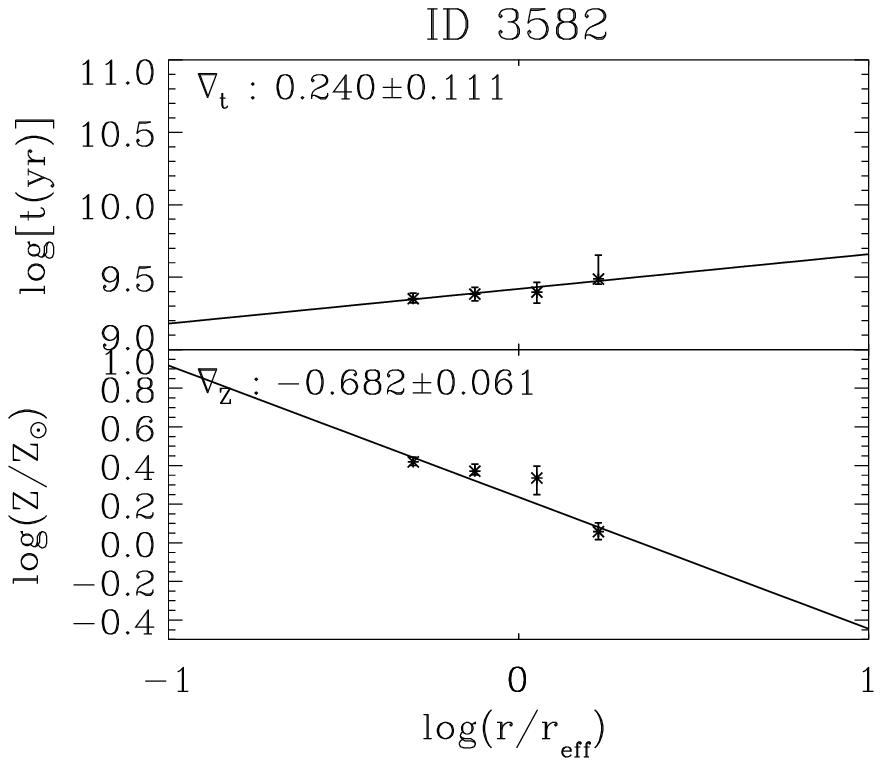}\\
 \includegraphics[scale=0.6]{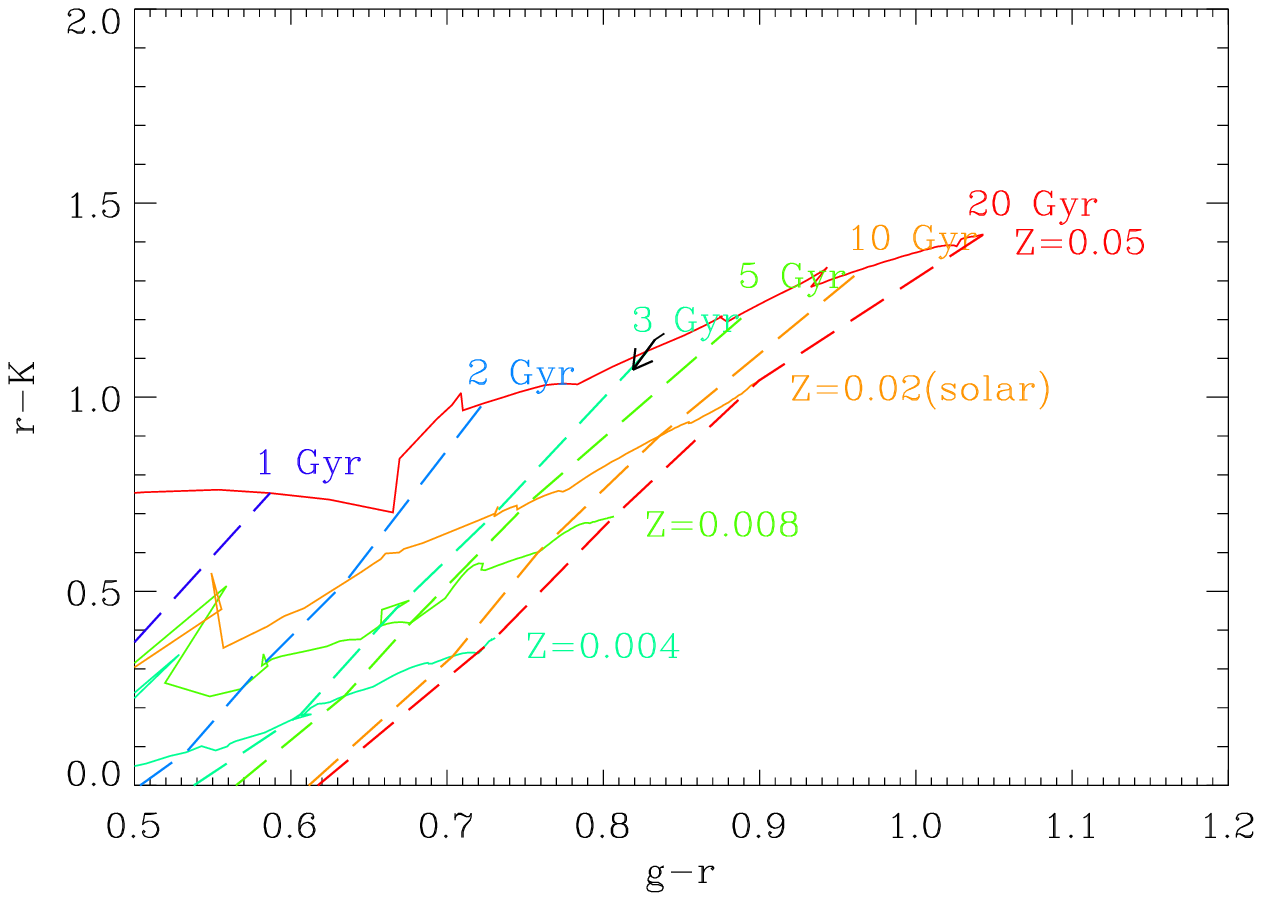}
 \includegraphics[scale=0.7]{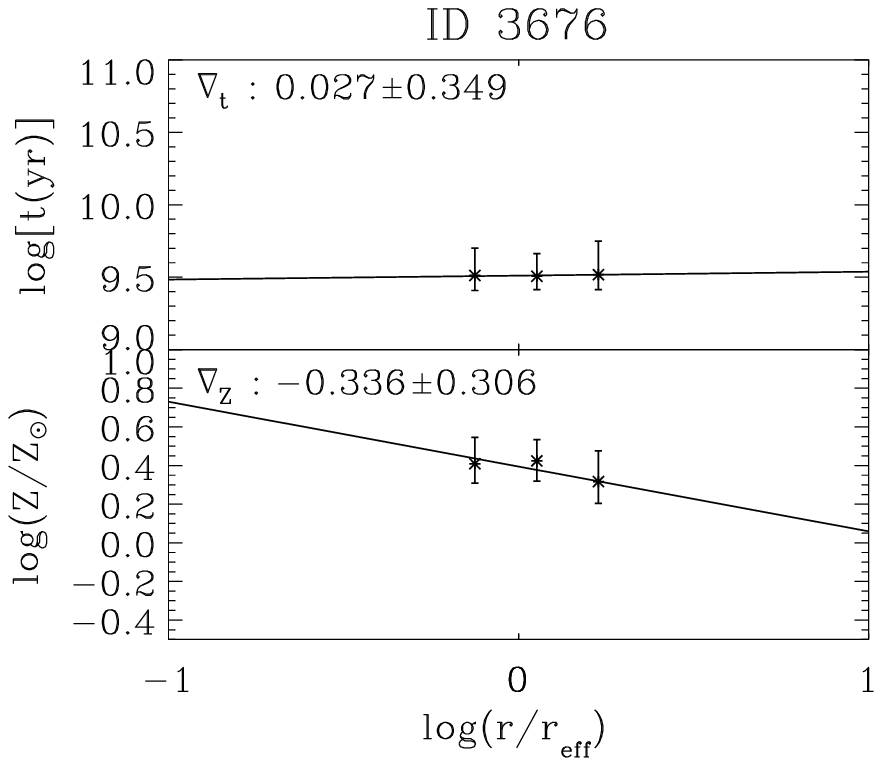}\\
\caption{Examples showing how the age and the metallicity gradients are derived. 
Left: Radial changes of age and metallicitiy (black arrows) plotted in the $g-r$ and $r-K$ diagram. The age and the metallicity grids are created with BC03 SSP model in the manuscript. The arrow head corresponds to the outermost radius. 
Right: The corresponding age and metallicity profiles. Also indicated are the gradient 
values.}
\label{ssp}
\end{figure}

\clearpage

\begin{figure}
\epsscale{1.0}
 \includegraphics[scale=0.6]{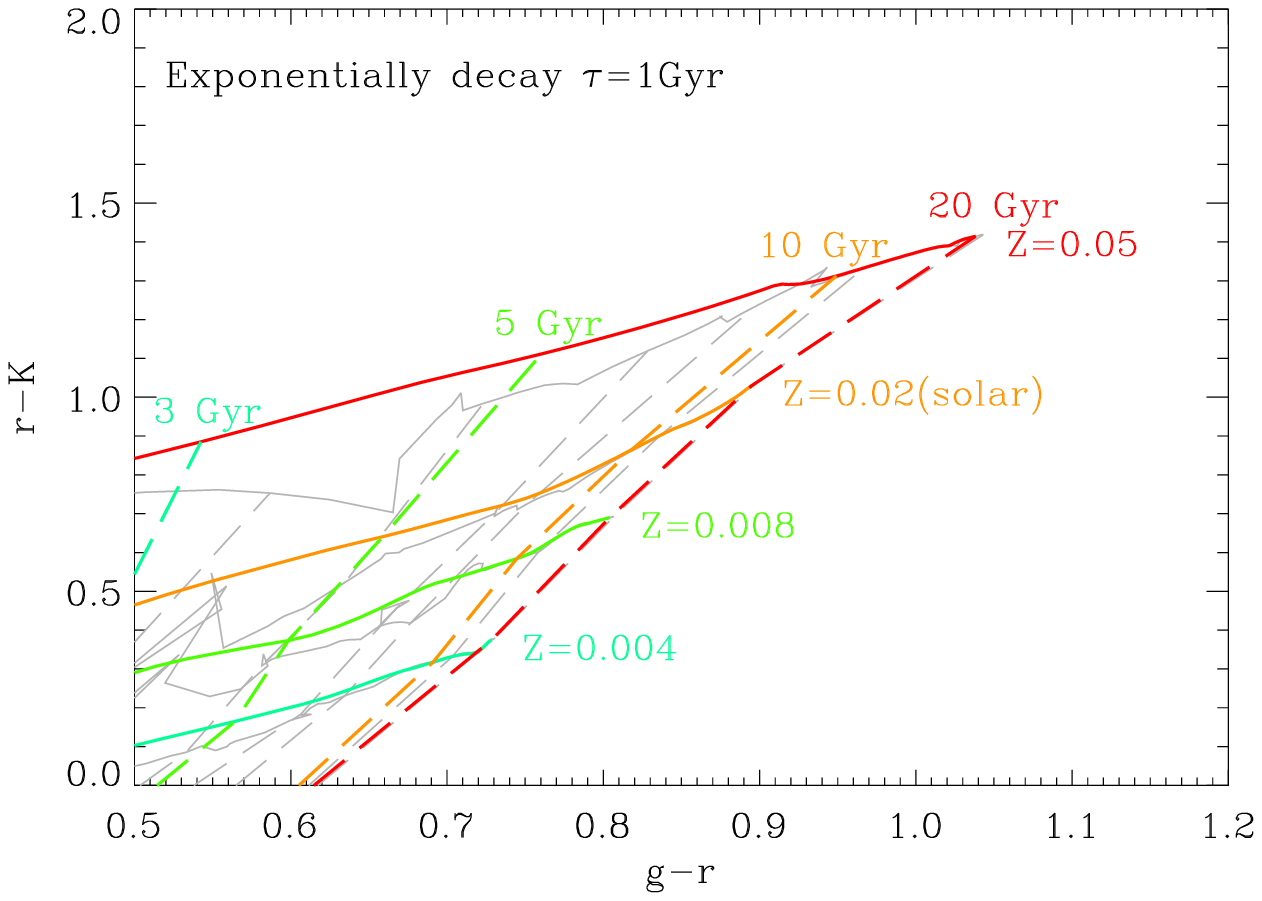}
 \includegraphics[scale=0.6]{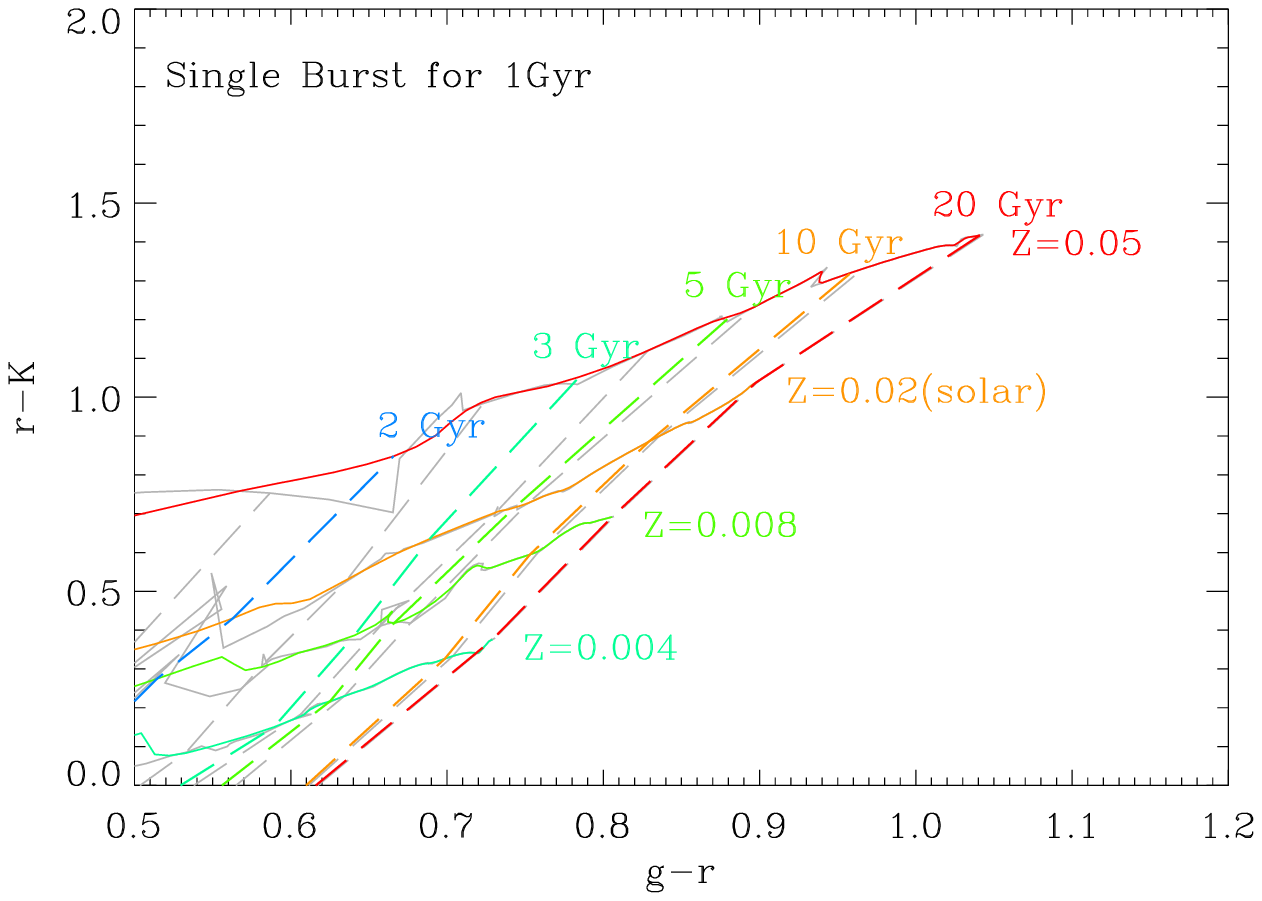}\\
 \includegraphics[scale=0.6]{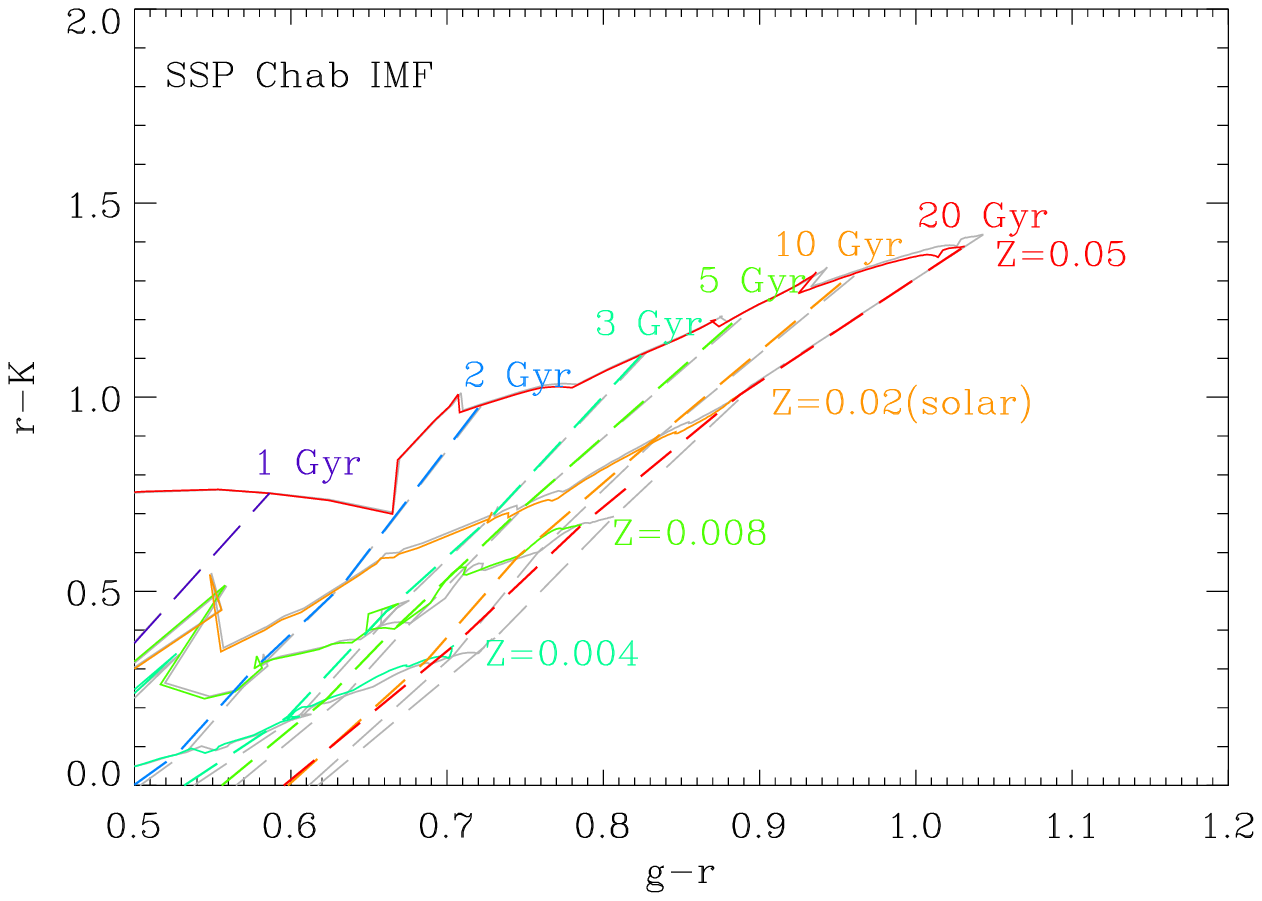}
\caption{Same color color grid as left column of Figure \ref{ssp} but for BC03 composite stellar population model of exponential decreasing star formation rate ($\tau =$ 1 Gyr)(upper left), a single burst expanding 1 Gyr (upper right) and same SSP but with Chabrier IMF (lower right).}
\label{csp}
\end{figure}

\clearpage

\begin{figure}
\epsscale{0.8}
\plotone{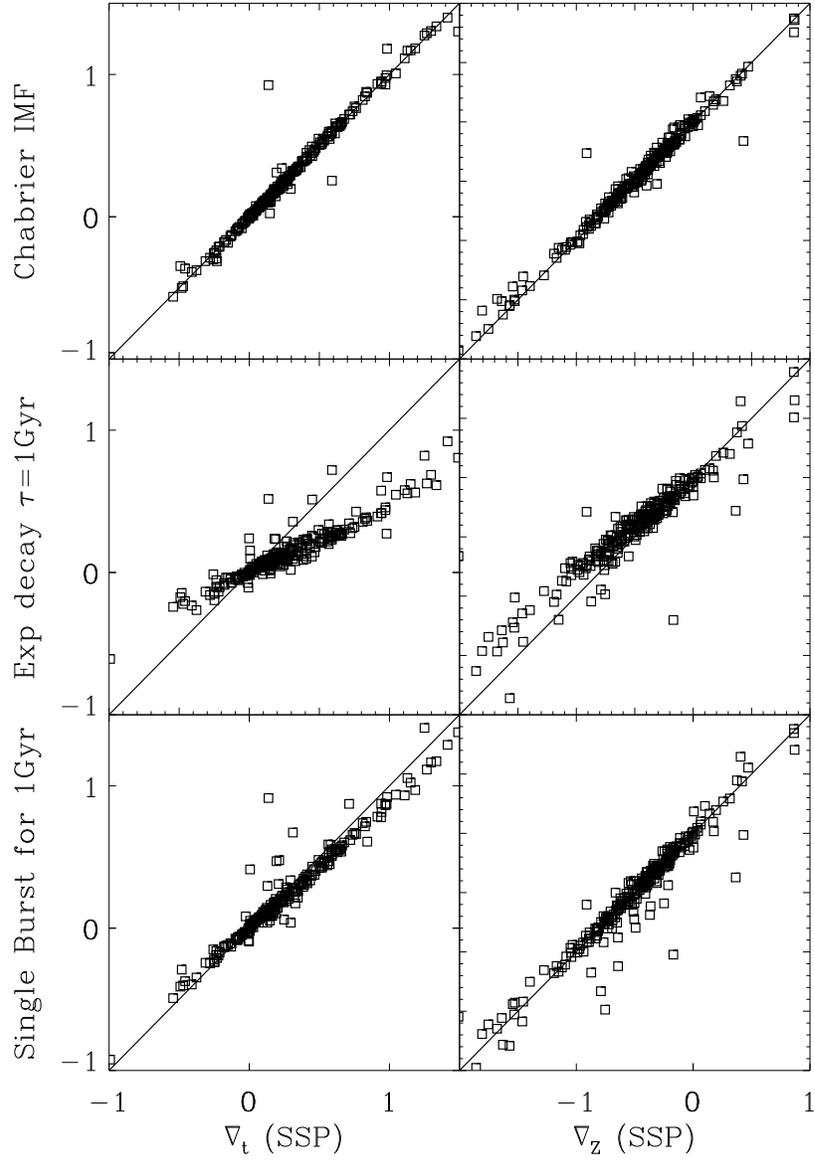}
\caption{Comparison of the gradient values derived by using a different IMF 
(Chabrier IMF), and different star formation histories (an exponentially decaying SFR
with $\tau = 1$ Gyr, and a single burst model with 1 Gyr burst duration). We find that the age 
gradient become shallower when an extended star formation is added, but the change in the
metallicity gradient values is negligible.} 
\label{grad_t_Z_comp}
\end{figure}

\clearpage

\begin{figure}
\epsscale{1.0}
 \includegraphics[scale=1]{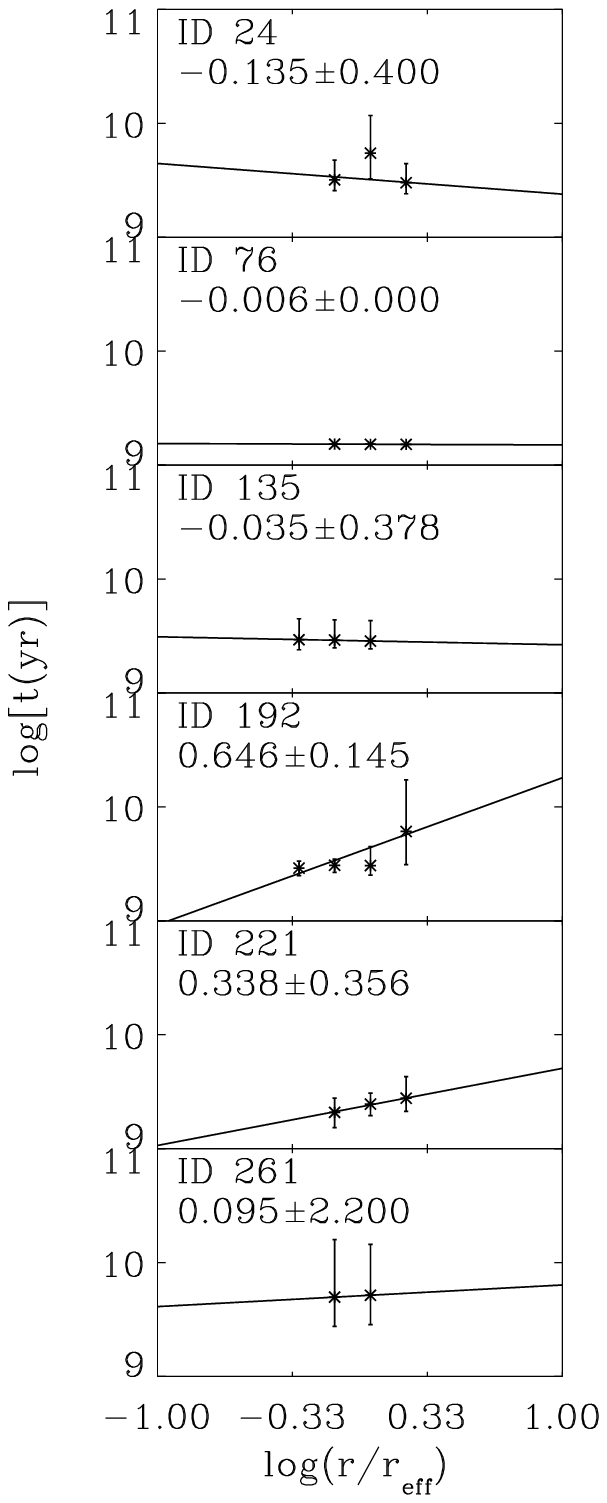}
 \includegraphics[scale=1]{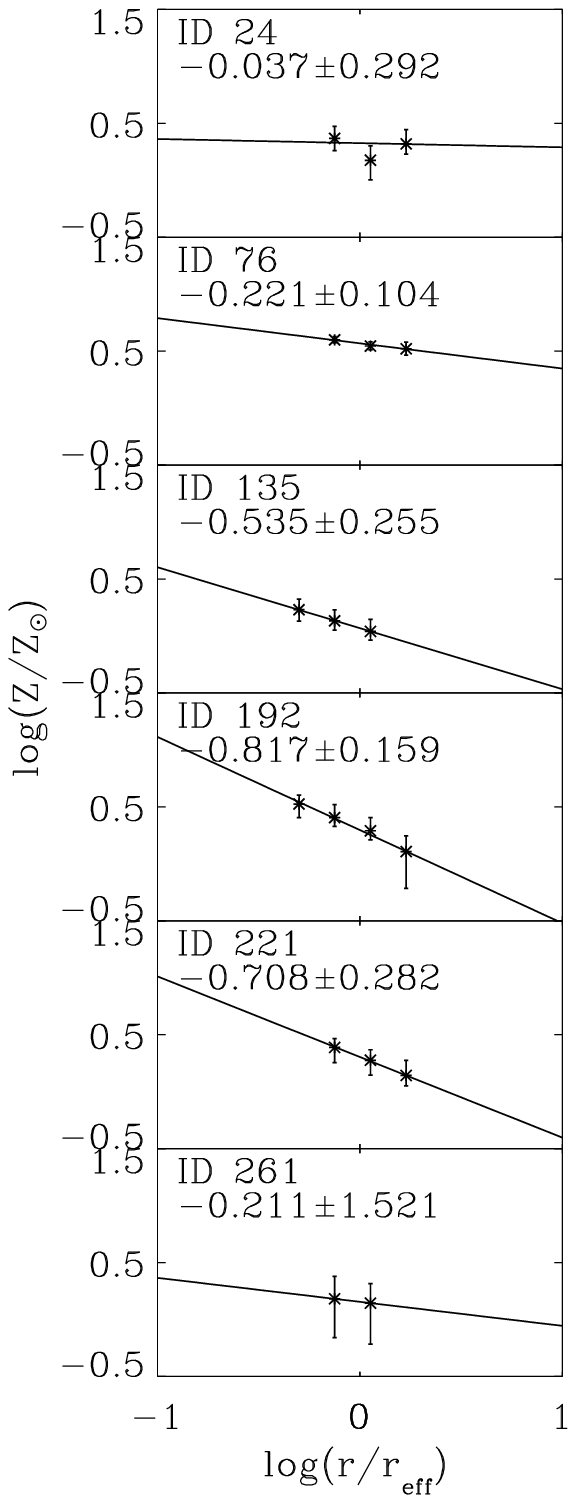}
 \caption{Age and the metallicity profiles of elliptical galaxies in the low-mass range
  ($10^{10.6} \, \mathrm{M_{\odot}} < \mathrm{M_{*}} < 10^{10.88} \, \mathrm{M_{\odot}}$). 
  The best-fit lines of the gradients are indicated with solid lines.}
\label{age_z_prof_lm}
\end{figure}

\clearpage

\begin{figure}
\epsscale{1.0}
 \includegraphics[scale=1]{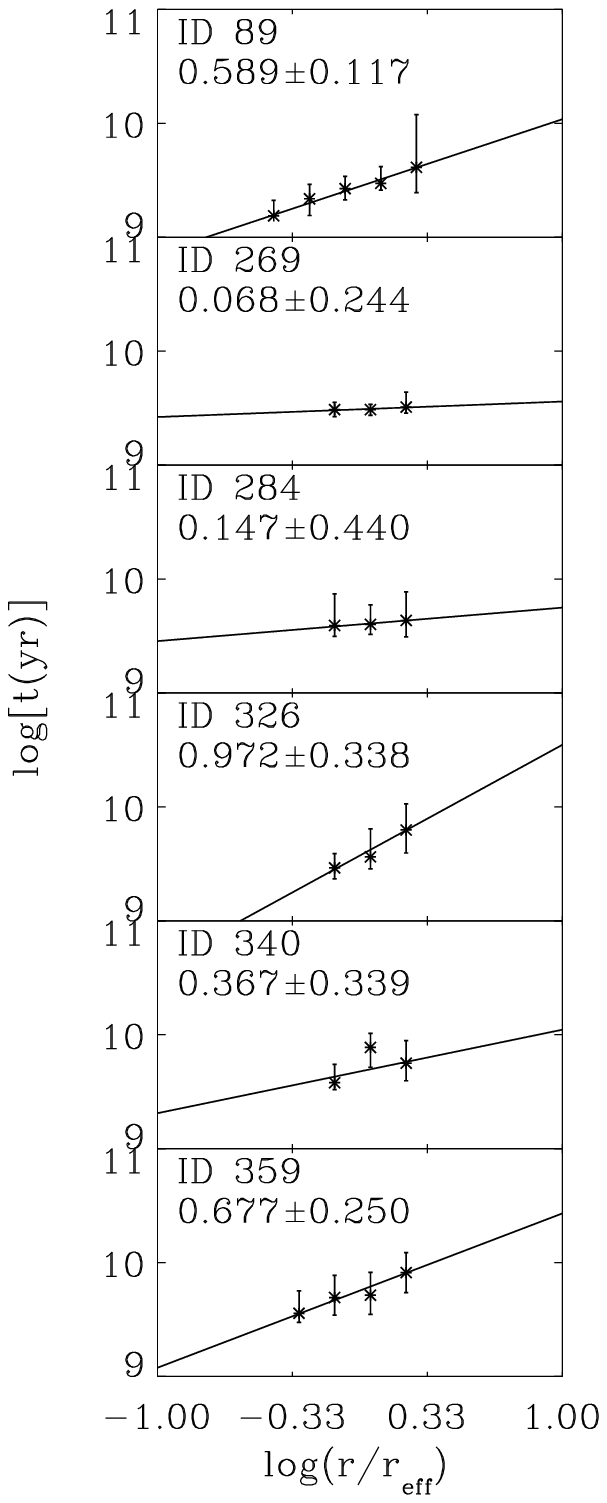}
 \includegraphics[scale=1]{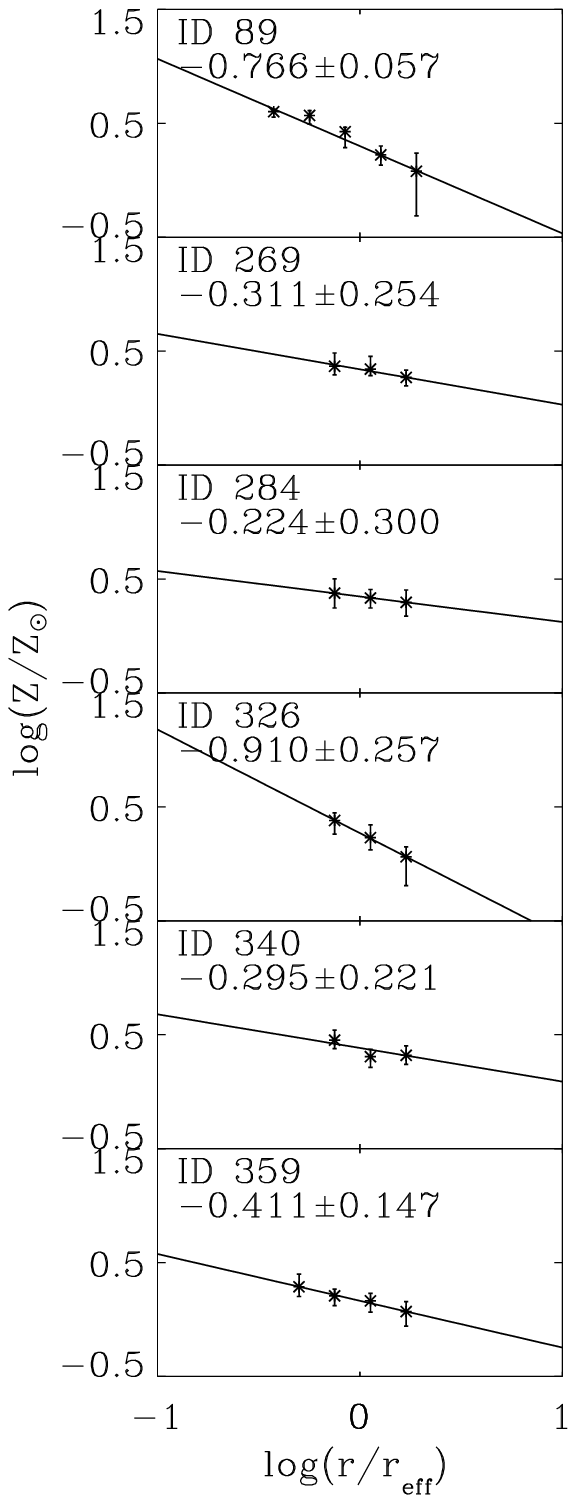}
 \caption{Same as Figure \ref{age_z_prof_lm} but for elliptical galaxies in intermediate mass range ($10^{10.88} \, \mathrm{M_{\odot}} < \mathrm{M_{*}} < 10^{11.28} \, \mathrm{M_{\odot}}$).}
\label{age_z_prof_im}
\end{figure}

\clearpage

\begin{figure}
\epsscale{1.0}
 \includegraphics[scale=1]{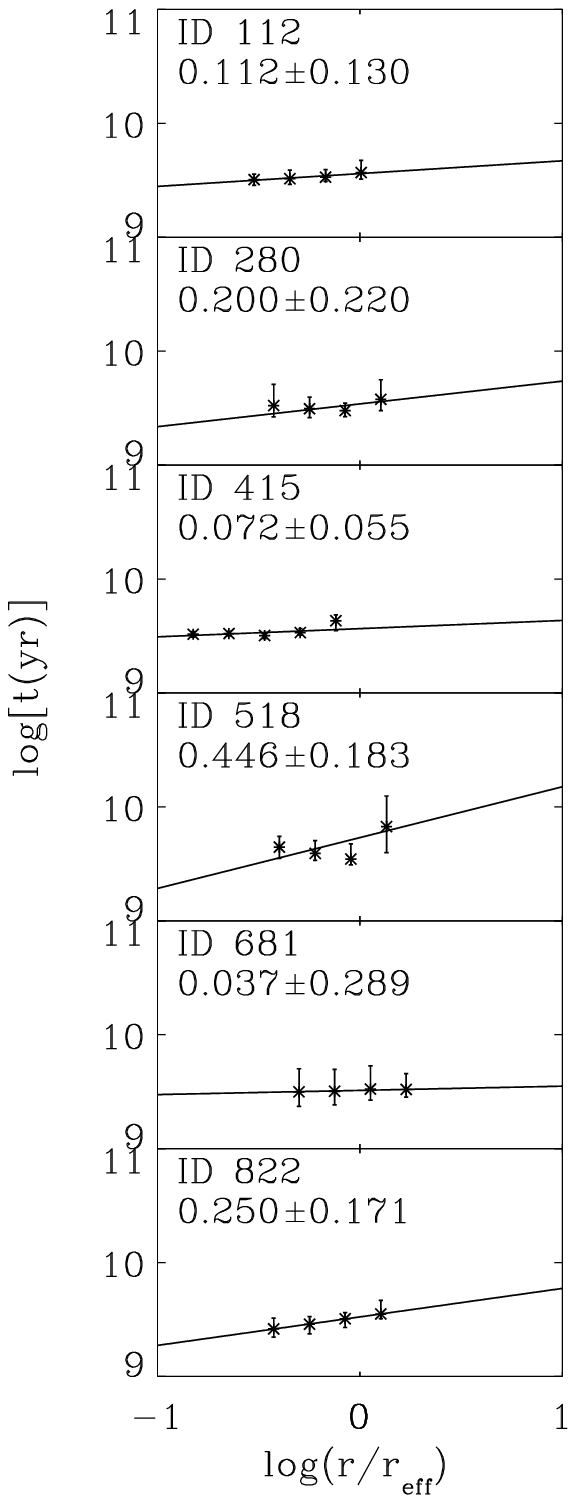}
 \includegraphics[scale=1]{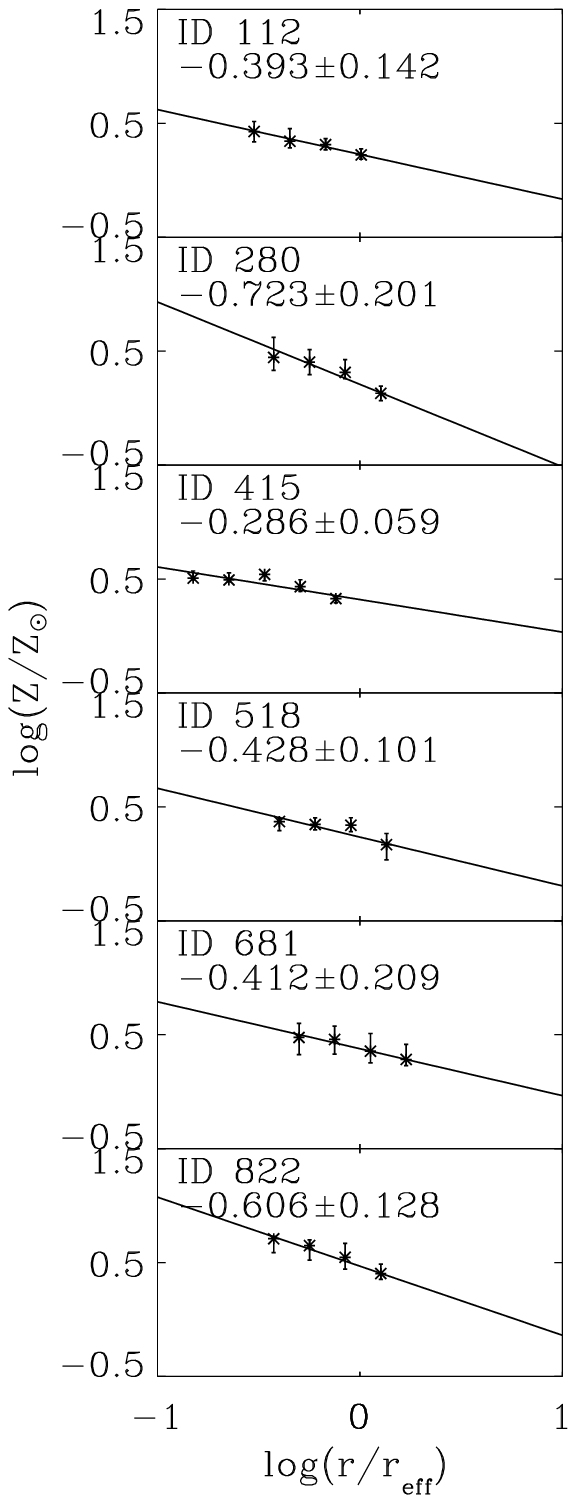}
 \caption{Same as Figure \ref{age_z_prof_lm} but for elliptical galaxies in high-mass range
 ($\mathrm{M_{*}} > 10^{11.28} \, \mathrm{M_{\odot}}$).}
\label{age_z_prof_hm}
\end{figure}

\clearpage

\begin{figure}
\epsscale{1.0}
 \includegraphics[scale=0.55]{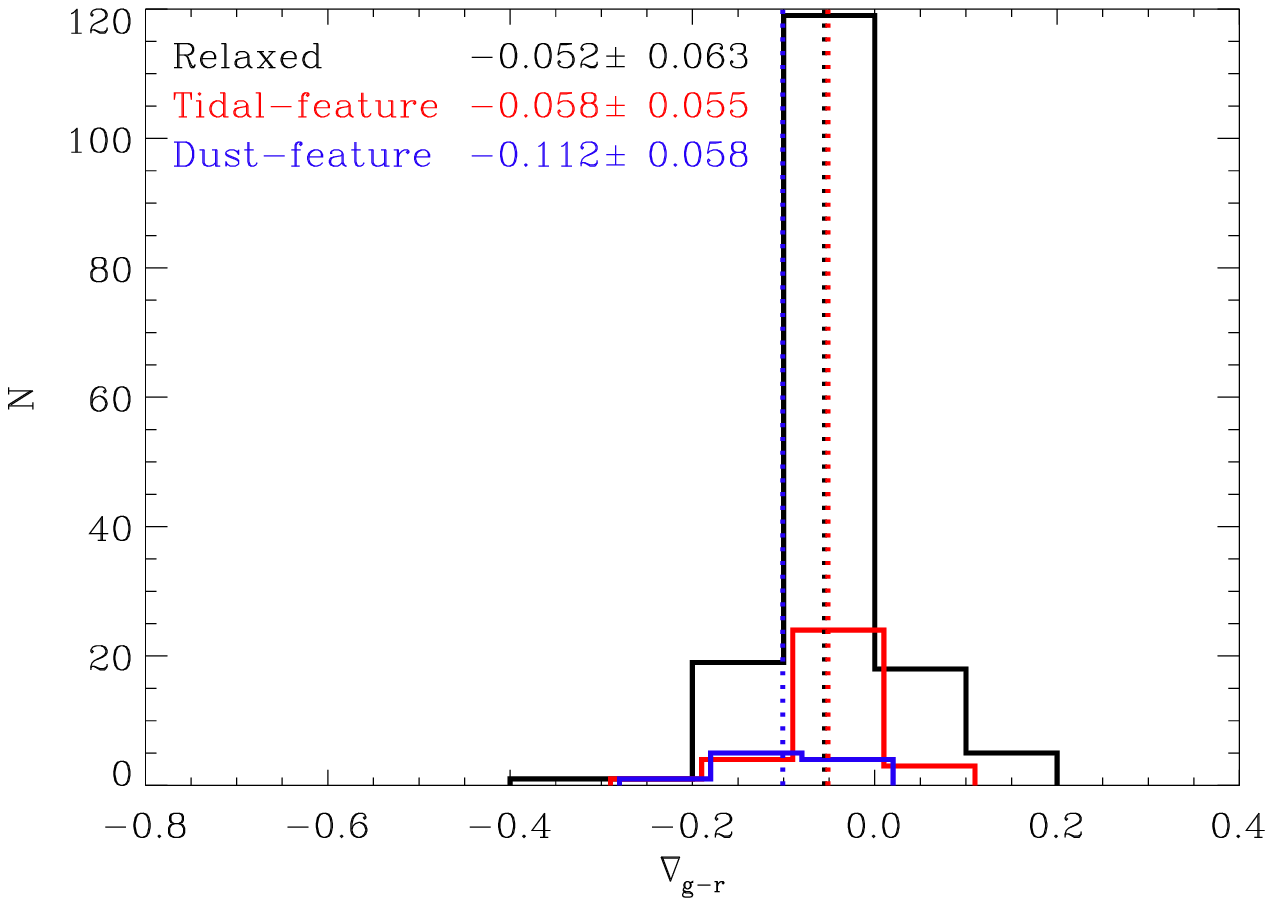}
 \includegraphics[scale=0.55]{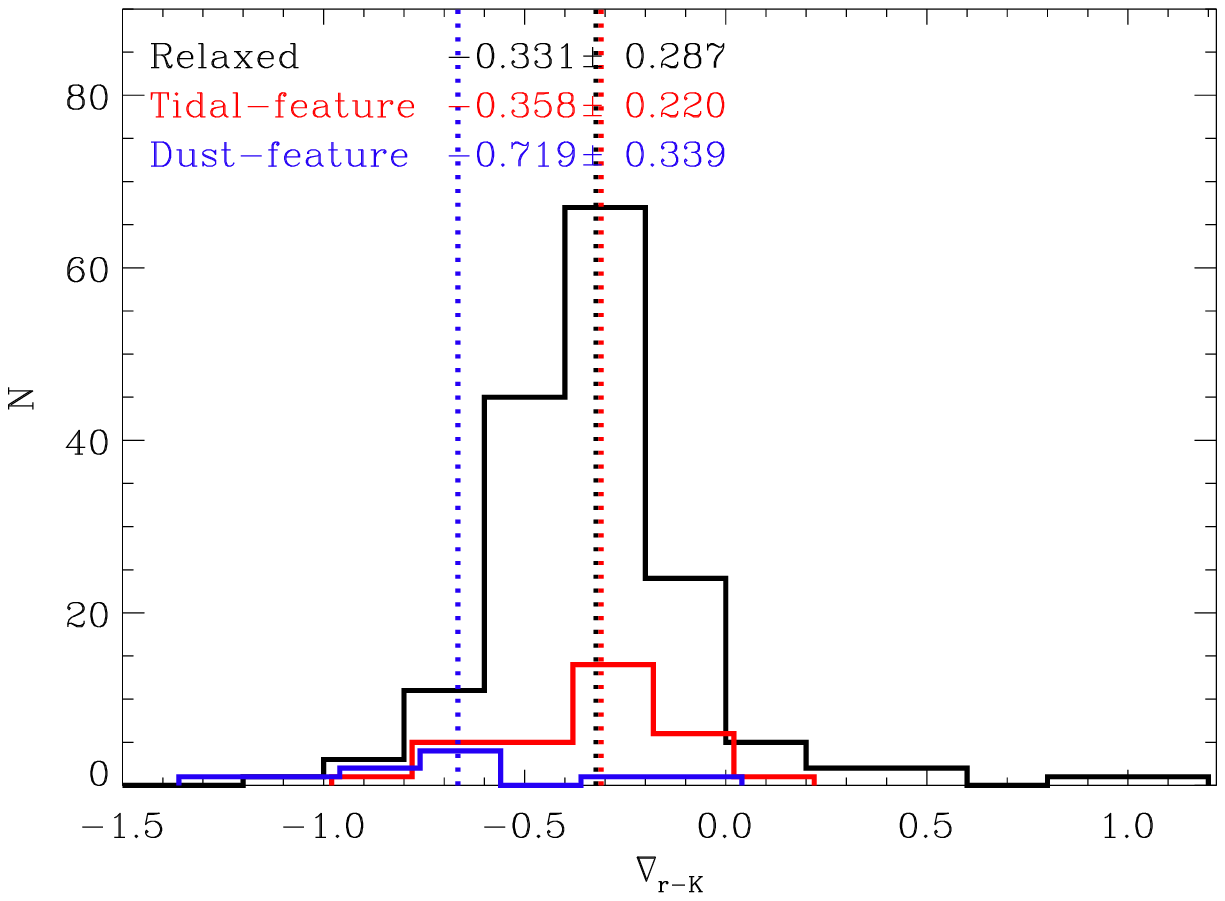}\\
 \includegraphics[scale=0.55]{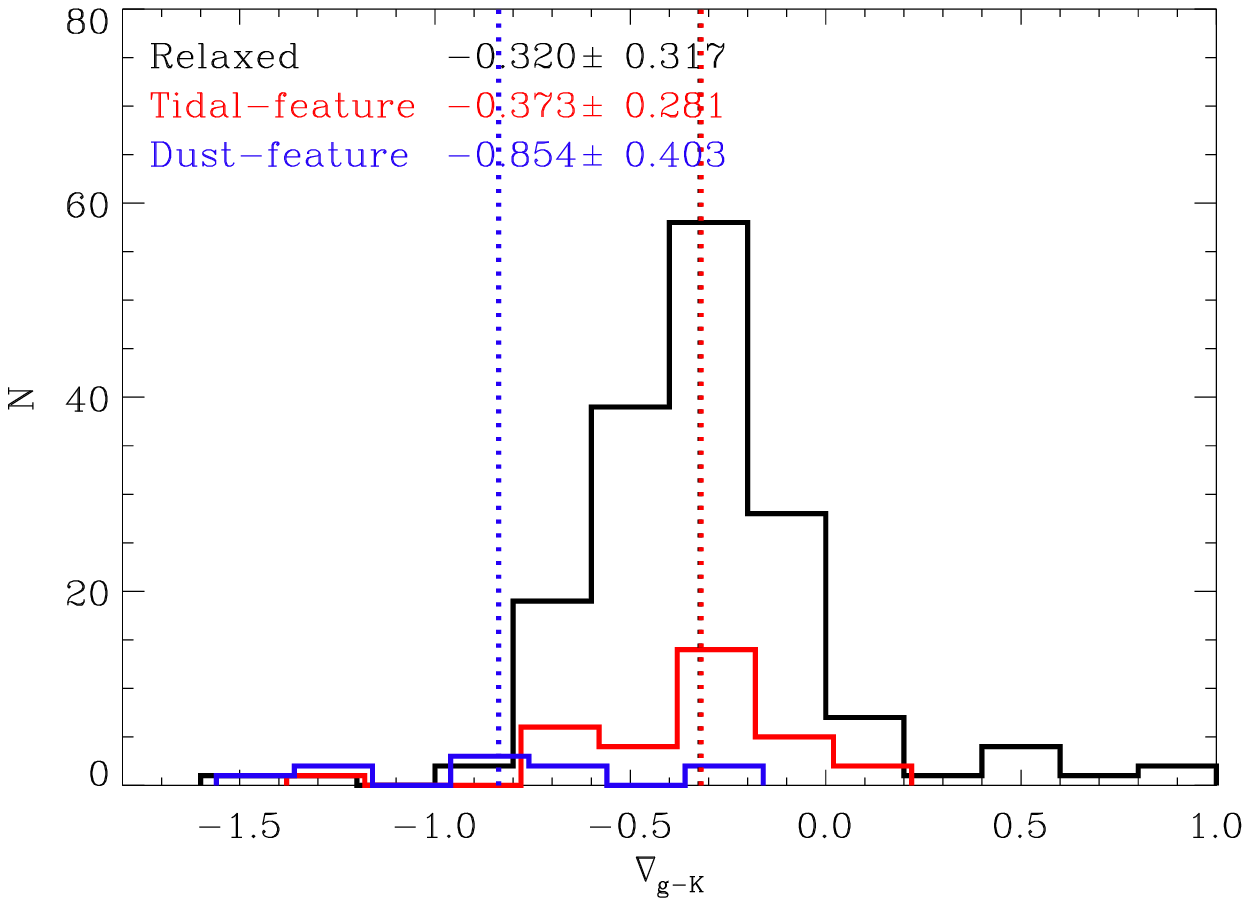}
\caption{The color gradient distributions of each type of elliptical galaxies. 
The vertical lines indicate median values for each type. We can see that 
the relaxed types and the tidal-feature types share a similar distributions,
but the dust-feature types have steeper gradients than the other two.}
\label{hist_col}
\end{figure}

\clearpage

\begin{figure}
\epsscale{0.6}
\plotone{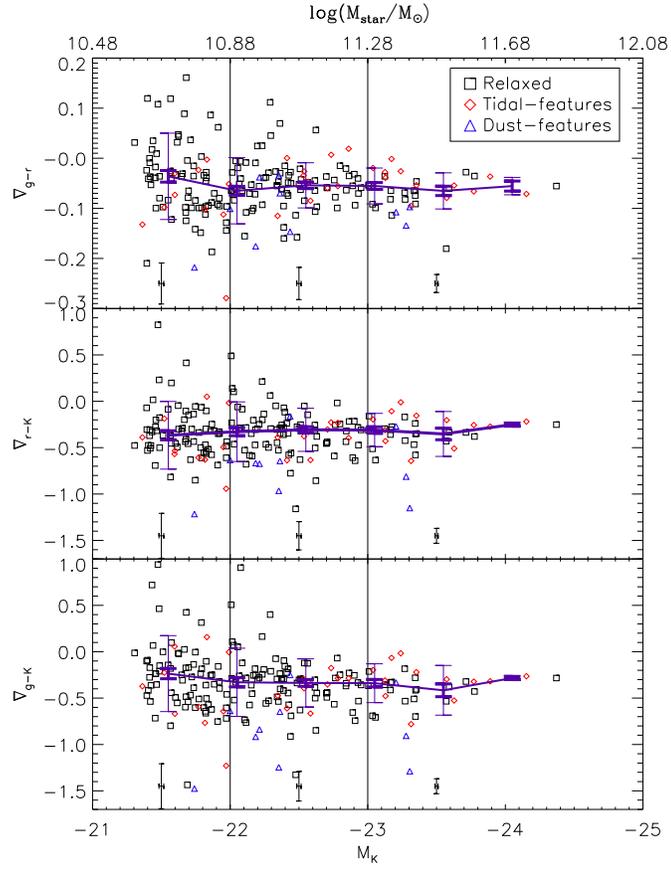}
\caption{The color gradients of the elliptical galaxies as a function of the absolute $K$-band magnitude 
($M_{K}$ or the stellar mass). The vertical lines are at $M_{K} =$ -22 and -23 which correspond to $10^{10.88}$ and $10^{11.28} \, \mathrm{M_{\odot}}$
in stellar mass. Error bars at the bottom of each panel are median measurement errors of the color gradients of objects in each bin. The thick and the thin error bars along the solid line show 
the error for the median value and the rms scatter of the color gradients in each mass bin.
The figure suggests that the color gradients are nearly constant 
but their scatter decreases at the highest mass bin over the explored mass range.}
\label{kmag_col}
\end{figure}

\clearpage

\begin{figure}
\epsscale{1.0}
\includegraphics[scale=0.6]{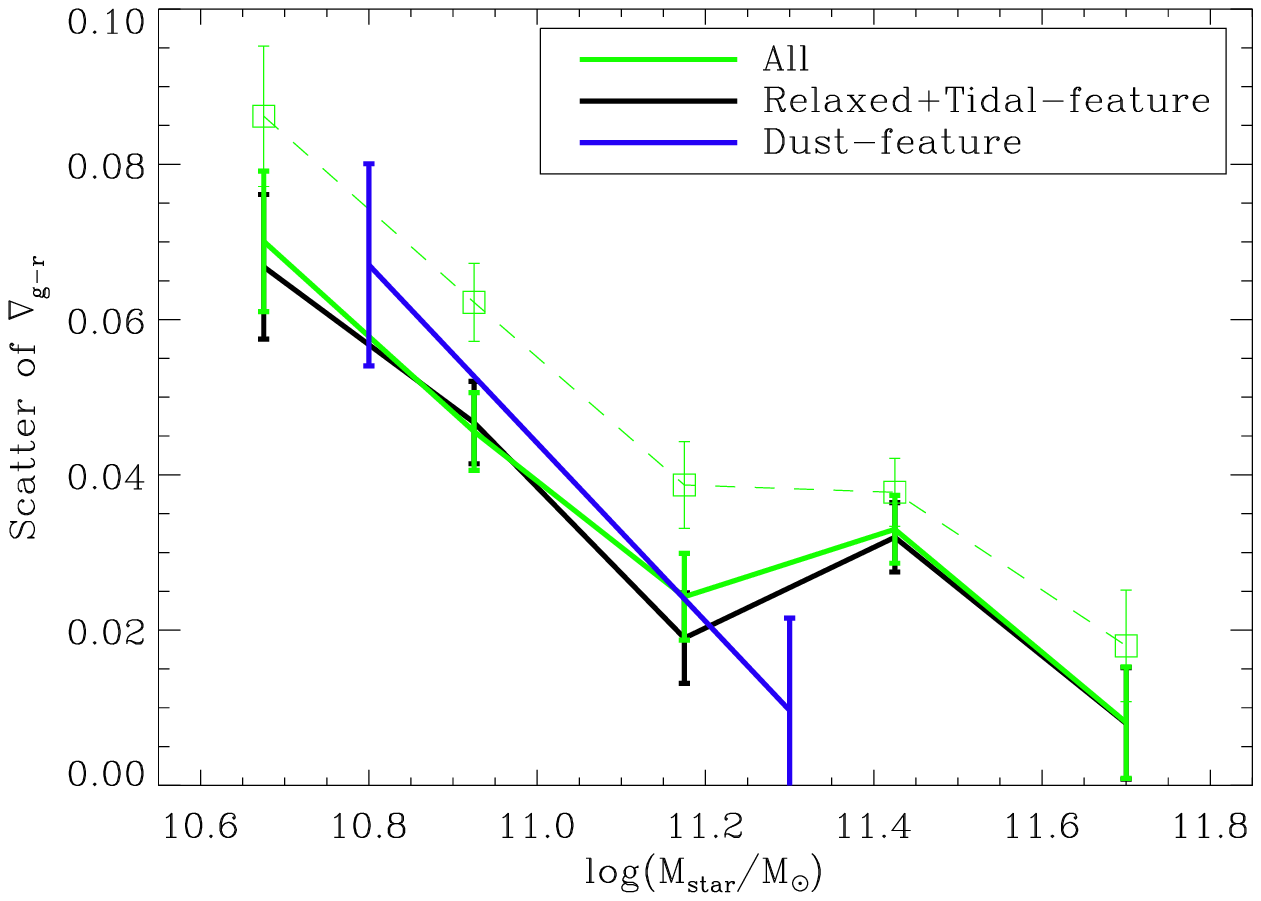}
\includegraphics[scale=0.6]{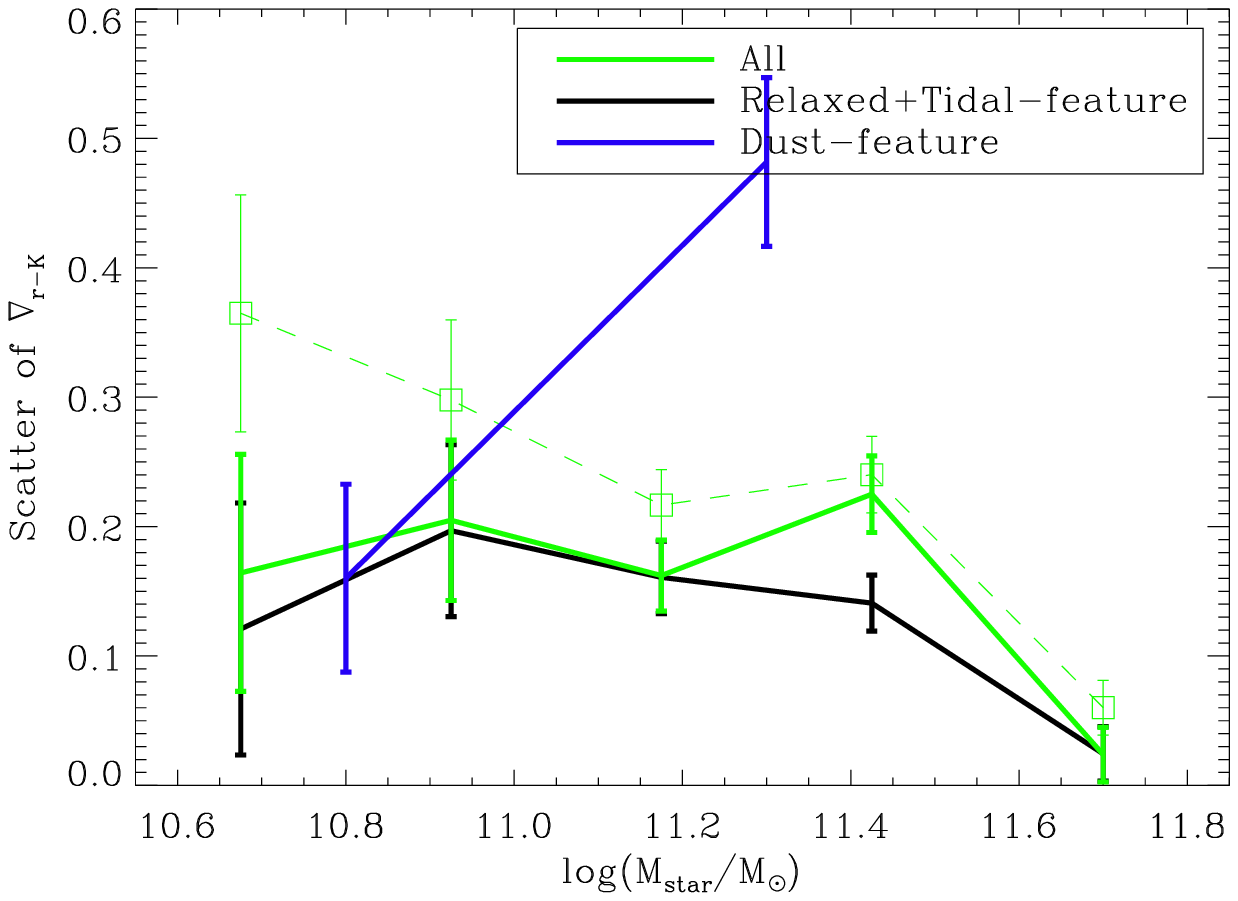}\\
\includegraphics[scale=0.6]{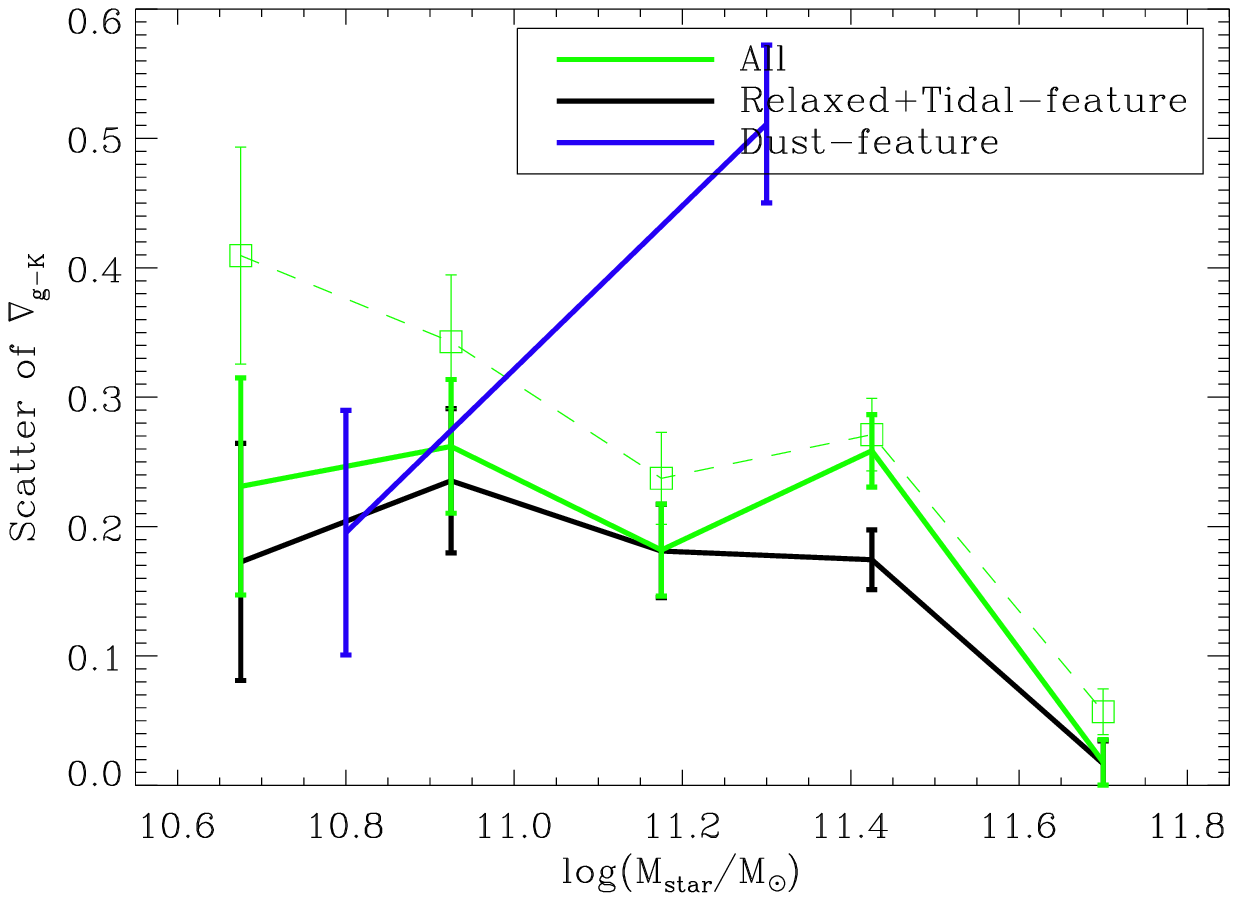}
\caption{The intrinsic scatter of color gradients of ellipticals as a function of stellar mass (solid lines). The dashed green line indicates the scatter as observed, before 
correcting for the measurement errors. The relaxed and the tidal-feature types show a reduced scatter at the highest mass bin.}
\label{scat_col}
\end{figure}

\clearpage

\begin{figure}
\epsscale{1.0}
 \includegraphics[scale=0.7]{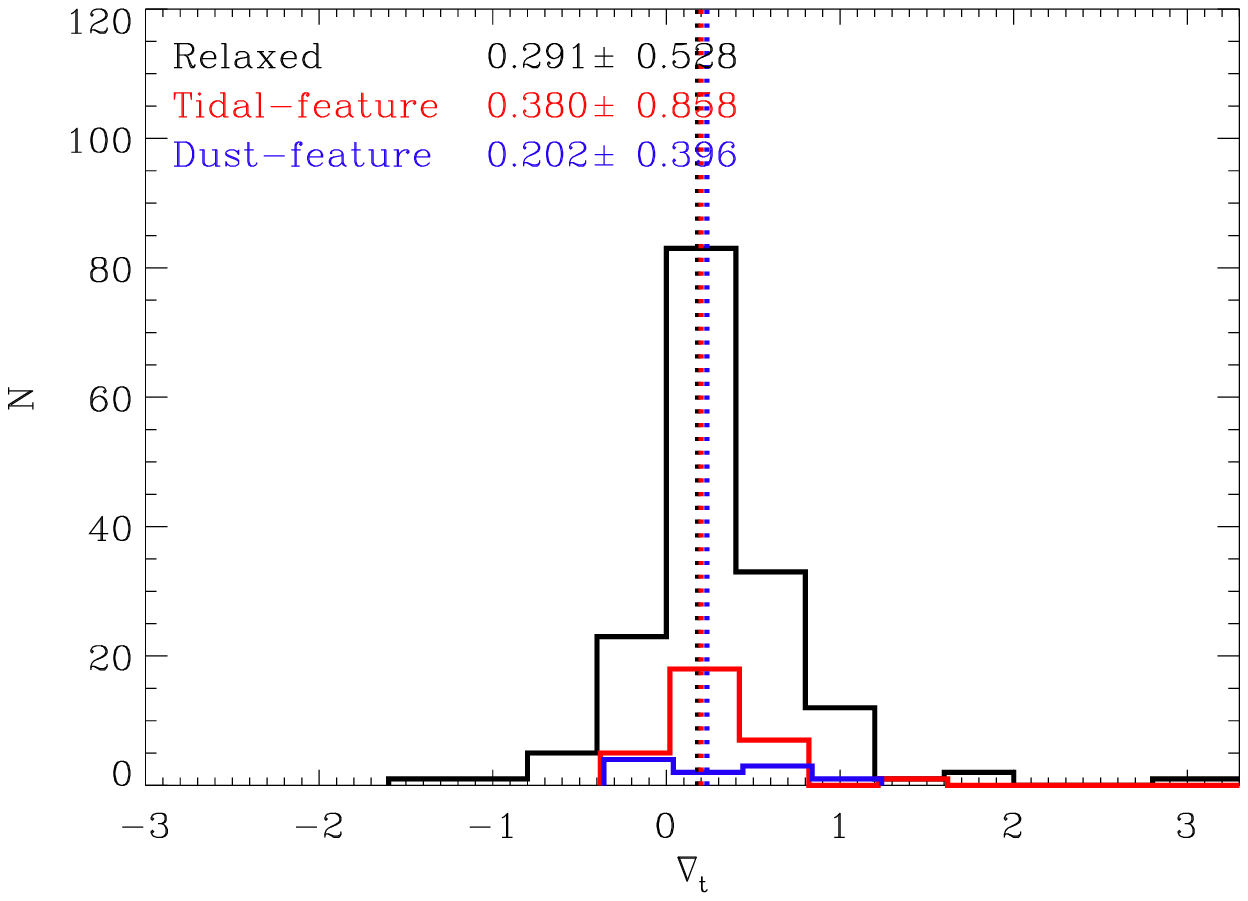}\\
 \includegraphics[scale=0.7]{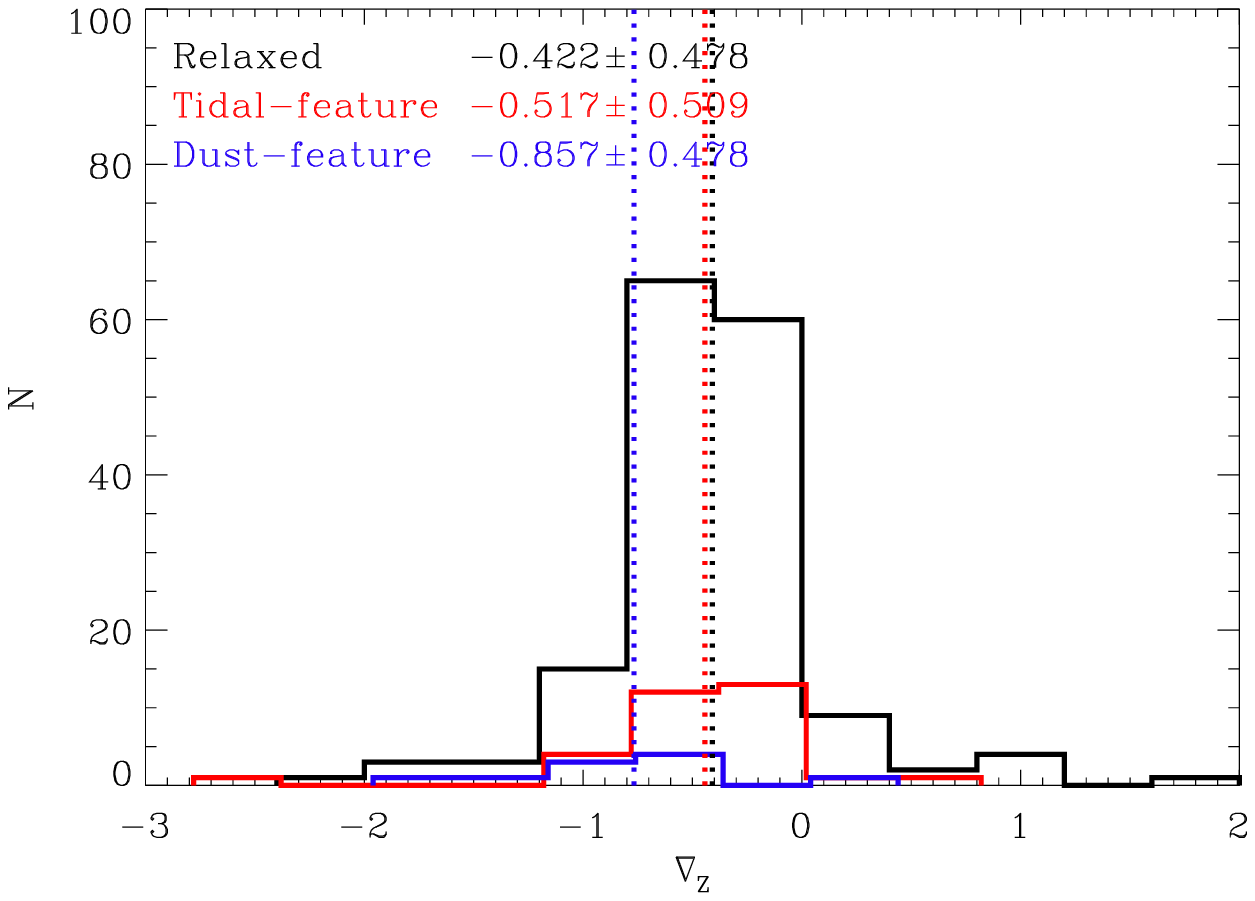}
\caption{The age (top) and the metallicity (bottom) gradient histograms of elliptical galaxies. The vertical lines indicate median values for each type. We can see that 
the relaxed types and the tidal-feature types share a similar distribution,
but the dust-feature types have steeper gradients than the other two, 
similar to the result we find for the color gradients.}
\label{hist_tz}
\end{figure}

\clearpage

\begin{figure}
\epsscale{0.6}
\plotone{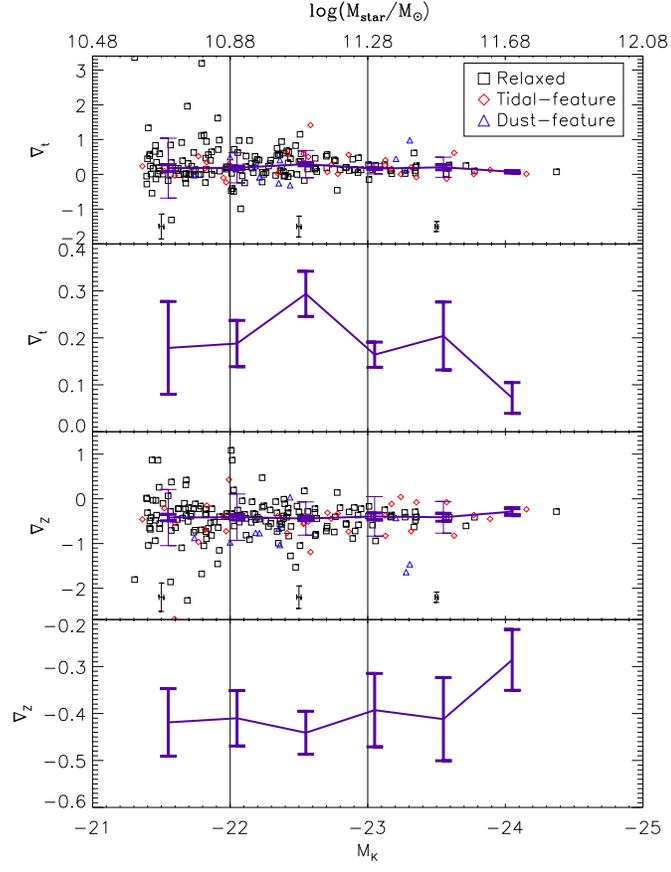}
\caption{Same as Figure \ref{kmag_col} but for the age and the metallicity gradients.
We also show expanded views of how the median values of the age and the metallicity
gradient change as a function of $M_{K}$ or stellar mass. Both the age and the metallicity gradients appear to show steepening up to $10^{11.4} \, \mathrm{M_{\odot}}$ but seem to 
flatten again at the highest mass range.}
\label{kmag_az}
\end{figure}

\clearpage

\begin{figure}
\epsscale{1.0}
\includegraphics[scale=0.7]{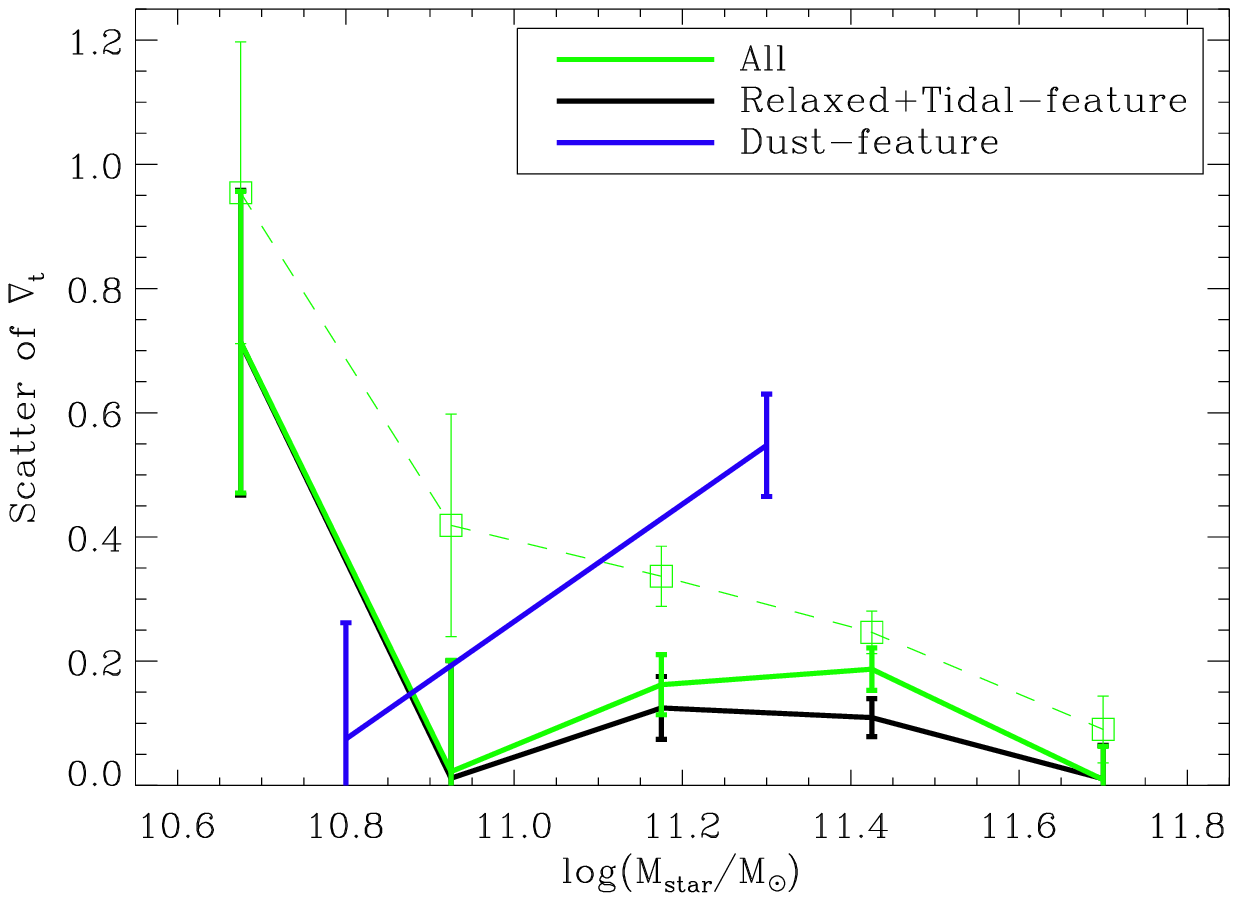}\\
\includegraphics[scale=0.7]{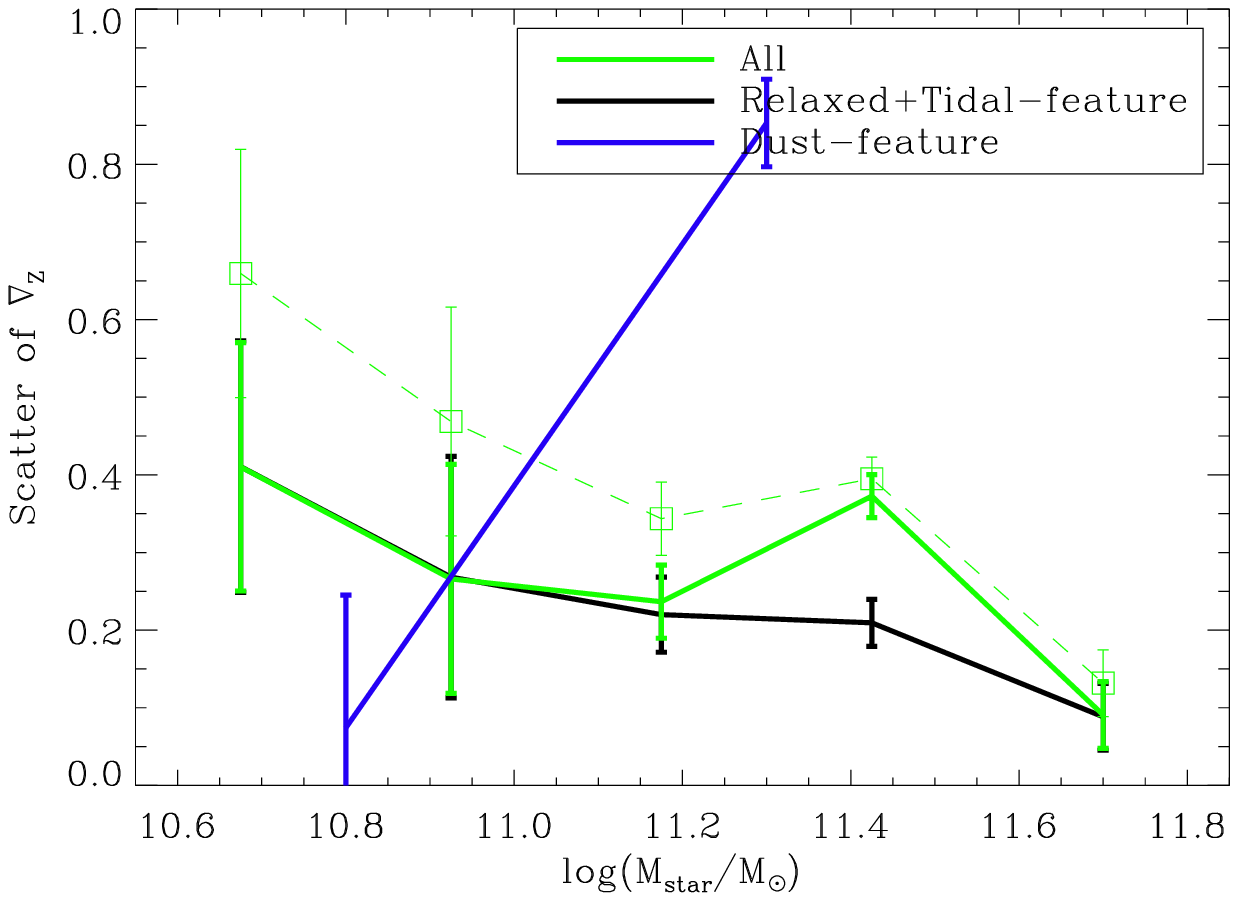}
\caption{Same as Figure \ref{scat_col} but for the age and the metallicity gradients. Like the intrinsic scatters of 
the color gradients, the relaxed and the tidal-feature types show a reduced scatter at higher 
mass which can be interpreted as a consequence due major dry merging.}
\label{scat_az}
\end{figure}

\clearpage

\begin{figure}
\epsscale{1.0}
\plotone{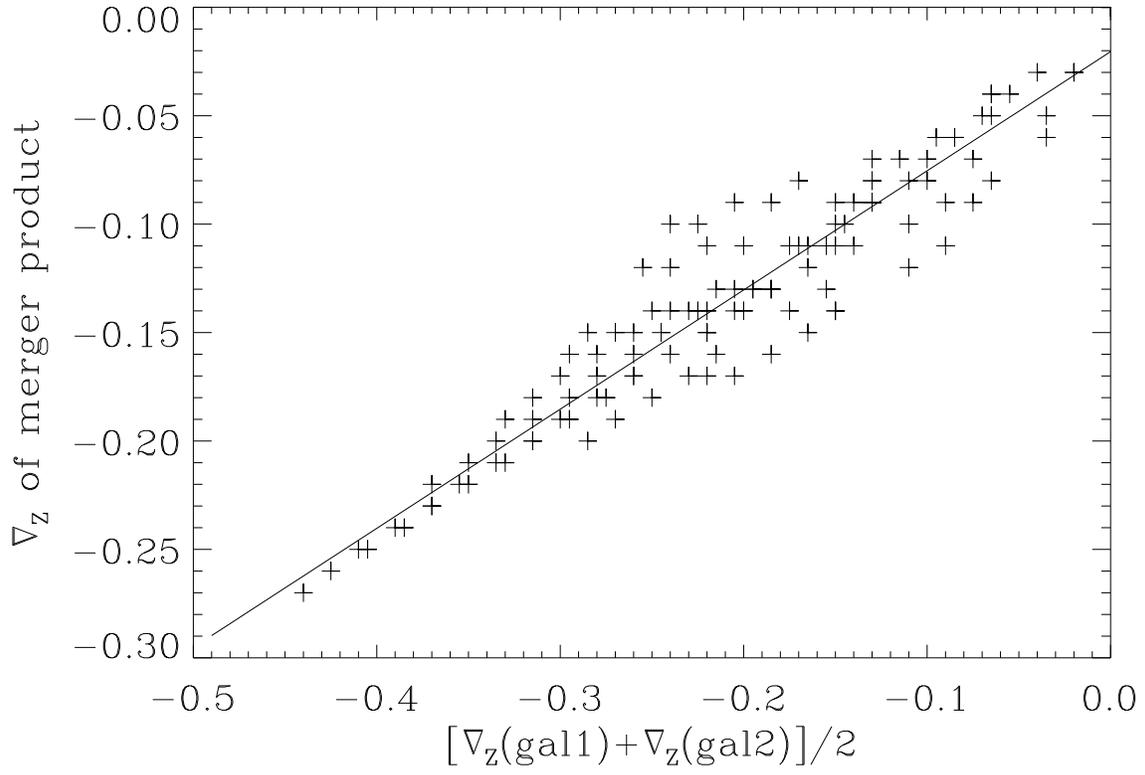}
\caption{The correlation between the average of the input $\nabla_{Z}$'s of two 
pre-mergers and the final $\nabla_{Z}$ of the merger product. The data points are
taken from a simulation by Di Matteo et al. (2010).}
\label{dmerg}
\end{figure}

\clearpage

\begin{figure}
\epsscale{1.0}
\includegraphics[scale=0.7,angle=90]{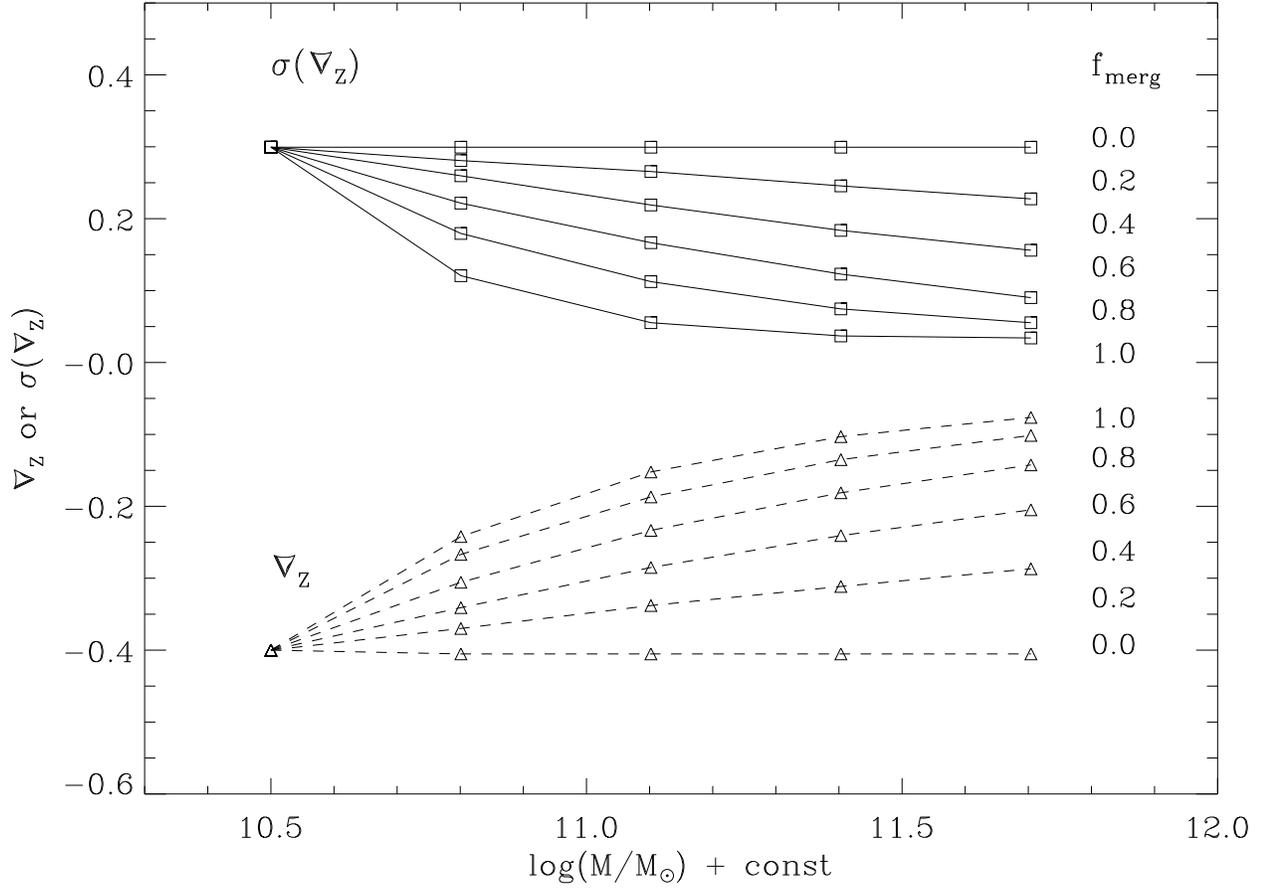}
\caption{This figure shows how the mean metallicity gradient $\langle \nabla_{Z} \rangle$ 
and its scatter $\sigma_{\nabla_{Z}}$ changes due to major dry merging,
 as a function of galaxy mass.  The fraction, $f_{merg}$ indicates the faction
 of objects that underwent major dry merging. The solid line indicates
 the change in  $\sigma_{\nabla_{Z}}$, while the dashed line show
 how $\langle \nabla_{Z} \rangle$ evolves as a result of the dry
 merging. Major dry merging reduces the scatter and the mean metallicity
 gradient value.}
\label{dm_frac}
\end{figure}

\clearpage

\begin{figure}
\epsscale{1.0}
\plotone{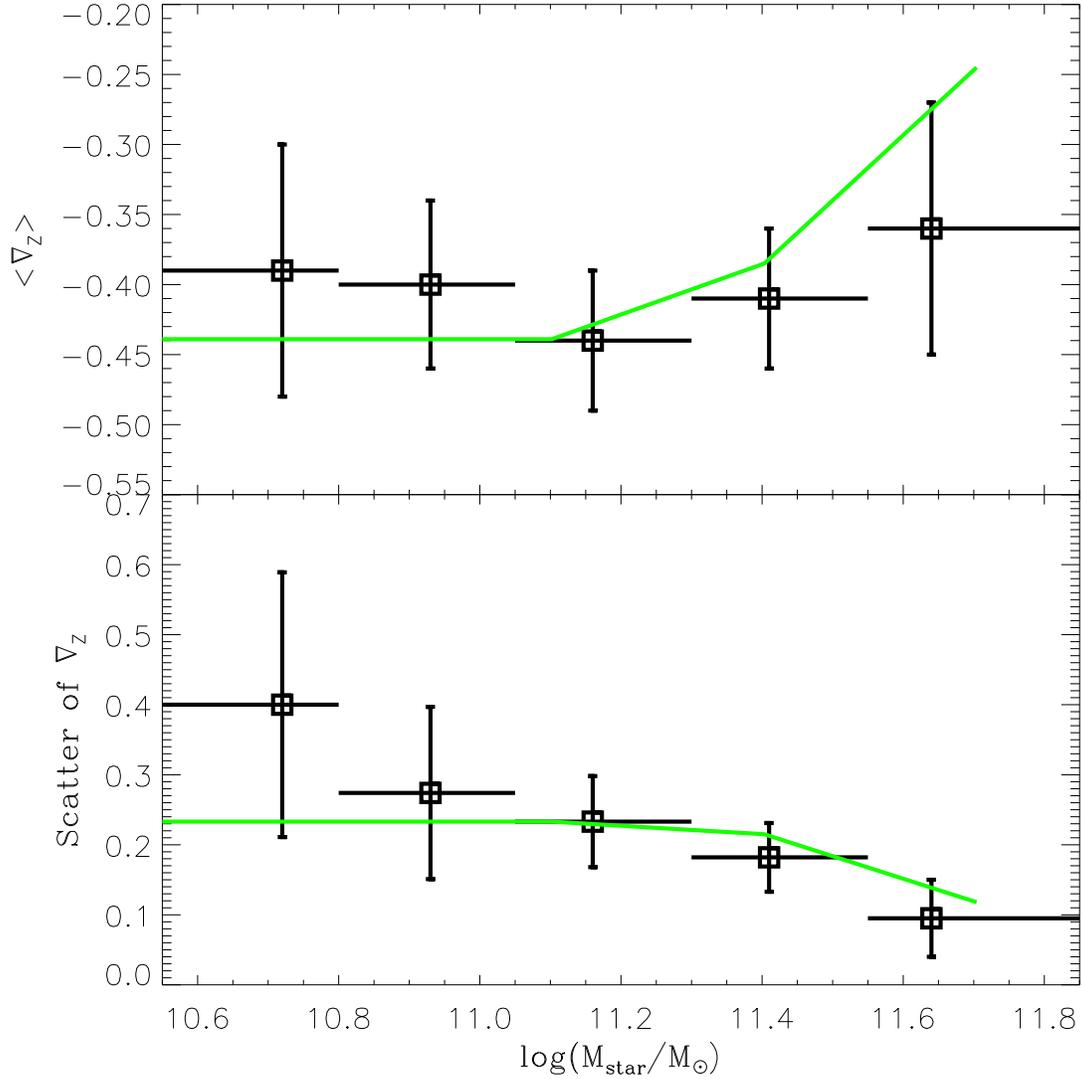}
\caption{The comparison of a simple dry merger model prediction (the green solid line) against
  the observed median $\nabla_{Z}$ (the points in the upper panel) and the intrinsic $\sigma_{\nabla_{Z}}$
  (the points in the lower panel) as a function of the stellar mass. The model lines are for the
  case where 30\% and 90\% of ellipticals were assumed to be created via dry merger at the two most
  massive bins (11.4 and 11.6, correspondingly). Note that a possible flattening of the metallicity gradient
  and the decrease in its scatter in the observed data, which can be explained by a significant amount
  of dry merging at $M_{star} > 11.4~ M_{\odot}$.
 }
\label{scat_sim}
\end{figure}

\clearpage

\begin{deluxetable}{ccccc}
\tabletypesize{\normalsize}
\tablecolumns{5} 
\tablecaption{Number of Ellipticals in Each Type} 
\tablewidth{0pt}
\tablehead{
\colhead{Types} & \colhead{Relaxed} & \colhead{Tidal-feature} & \colhead{Dust-feature} & \colhead{Total} }
\startdata
Number & 162 & 32 & 10 & 204
\enddata
\label{tbl1}
\end{deluxetable}

\begin{deluxetable}{ccccccc}
\tabletypesize{\normalsize}
\tablecolumns{7} 
\tablecaption{Median Gradients} 
\tablewidth{0pt}
\tablehead{
\colhead{Type} & \colhead{$\nabla_{g-r}$} & \colhead{$\nabla_{r-K}$} & \colhead{$\nabla_{g-K}$} & \colhead{$\nabla_{t}$} & \colhead{$\nabla_{Z}$}}
\startdata
Relaxed & 	$-0.055\pm 0.005$ & 	$-0.323\pm 0.022$ & 	$-0.322\pm 0.025$ & 	$0.180\pm 0.041$ & 	$-0.409\pm 0.037$ \\
Tidal-feature & $-0.051\pm 0.010$ & 	$-0.310\pm 0.039$ & 	$-0.319\pm 0.050$ &	$0.200\pm 0.152$ & 	$-0.443\pm 0.090$\\
Dust-feature & 	$-0.101\pm 0.018$ & 	$-0.666\pm 0.107$ & 	$-0.837\pm 0.127$ & 	$0.233\pm 0.125$ & 	$-0.767\pm 0.151$\\
\cline{1-6}\\
Total &		$-0.055\pm 0.004 $ & 	$-0.330\pm 0.020 $ & 	$-0.329\pm 0.023$ & 	$0.180\pm 0.041 $ & 	$-0.412\pm 0.034$\\
\enddata
\label{tab_col_grad}
\end{deluxetable}

\begin{deluxetable}{ccccccc}
\tabletypesize{\normalsize}
\tablecolumns{7} 
\tablecaption{K-S Test between Gradients of Different Types of Ellipticals} 
\tablewidth{0pt}
\tablehead{
\colhead{Types} & \colhead{$\nabla_{g-r}$} & \colhead{$\nabla_{r-K}$} & \colhead{$\nabla_{g-K}$} & \colhead{$\nabla_{t}$} & \colhead{$\nabla_{Z}$} }

\startdata
Relaxed vs. Tidal-feature & 	$0.951$ 	& $0.943$ 		& $0.916$ 		& $0.953$ 		& $0.809$\\
Relaxed vs. Dust-feature & 	$0.008$ 	& $2.8\times10^{-5}$ 		& $1.0\times10^{-4}$ 	& $0.488$ 		& $0.001$\\
\enddata
\label{tab_ks}
\end{deluxetable}

\begin{deluxetable}{ccccccc}
\tabletypesize{\normalsize}
\tablecolumns{7} 
\tablecaption{Intrinsic Scatters of Color/Age/Metallicity Gradients} 
\tablewidth{0pt}
\tablehead{
\colhead{Gradient} 			&
\colhead{Type} 			& \colhead{$\mathrm{M}_{K} < -23$}	& 
\colhead{$-22 > \mathrm{M}_{K} > -23$} 	& \colhead{$\mathrm{M}_{K} > -22$}}

\startdata
$\sigma_{\nabla_{g-r}}$ 	&	All & 			$0.029\pm0.004$ & $0.069\pm0.007$ & $0.031\pm0.004$\\ 
				&	Relaxed & 		$0.027\pm0.004$ & $0.066\pm0.008$ & $0.030\pm0.005$\\
				&	Tidal-feature & 	$0.020\pm0.007$ & $0.061\pm0.017$ & $0.022\pm0.012$\\
			&	Relaxed \& Tidal-feature & 	$0.027\pm0.004$ & $0.067\pm0.007$ & $0.029\pm0.005$\\
				&	Dust-feature & 	 	$0.035\pm0.013$	& $0.062\pm0.012$ & \\
$\sigma_{\nabla_{r-K}}$		&	All & 			$0.197\pm0.025$ & $0.125\pm0.073$ & $0.218\pm0.052$\\ 
				&	Relaxed & 		$0.089\pm0.024$ & $0.120\pm0.086$ & $0.223\pm0.059$\\
				&	Tidal-feature & 	$0.158\pm0.027$ & $0.025\pm0.105$ & $0.060\pm0.035$\\
			&	Relaxed \& Tidal-feature & 	$0.122\pm0.018$ & $0.098\pm0.077$ & $0.207\pm0.055$\\
				&	Dust-feature & 		$0.367\pm0.067$	& $0.150\pm0.089$ & \\
$\sigma_{\nabla_{g-K}}$		&	All & 			$0.231\pm0.023$ & $0.242\pm0.066$ & $0.246\pm0.045$\\ 
				&	Relaxed & 		$0.127\pm0.026$ & $0.201\pm0.080$ & $0.234\pm0.052$\\
				&	Tidal-feature & 	$0.189\pm0.029$ & $0.280\pm0.124$ & $0.072\pm0.041$\\
			&	Relaxed \& Tidal-feature & 	$0.157\pm0.020$ & $0.198\pm0.070$ & $0.223\pm0.048$\\
				&	Dust-feature & 		$0.417\pm0.063$	& $0.218\pm0.096$ & \\
$\sigma_{\nabla_{t}}$		&	All & 			$0.144\pm0.032$ & $0.521\pm0.200$ & $0.031\pm0.133$\\ 
				&	Relaxed & 		$0.004\pm0.048$ & $0.390\pm0.211$ & $0.023\pm0.161$\\
				&	Tidal-feature & 	$0.165\pm0.039$ & $1.289\pm0.546$ & $0.248\pm0.079$\\
			&	Relaxed \& Tidal-feature & 	$0.058\pm0.032$ & $0.551\pm0.202$ & $0.042\pm0.142$\\
				&	Dust-feature & 		$0.342\pm0.079$	& $0.090\pm0.131$ & \\
$\sigma_{\nabla_{Z}}$		&	All & 			$0.316\pm0.024$ & $0.323\pm0.129$ & $0.300\pm0.123$\\ 
				&	Relaxed & 		$0.082\pm0.032$ & $0.270\pm0.136$ & $0.304\pm0.139$\\
				&	Tidal-feature & 	$0.278\pm0.042$ & $0.620\pm0.372$ & $0.109\pm0.055$\\
			&	Relaxed \& Tidal-feature & 	$0.177\pm0.026$ & $0.321\pm0.131$ & $0.307\pm0.129$\\
				&	Dust-feature & 		$0.586\pm0.051$	& $0.166\pm0.090$ & \\

\enddata
\label{tab_scat}
\end{deluxetable}

\clearpage

\LongTables
\begin{landscape}

\begin{deluxetable}{ccccccccccccc}
\tabletypesize{\scriptsize}
\tablecolumns{13} 
\tablecaption{List of Ellipticals and Their Properties} 
\tablewidth{0pt}
\tablehead{
\colhead{KID\tablenotemark{a}}&
\colhead{redshift}&
\colhead{$r_{eff,r}$\tablenotemark{b}}&
\colhead{$r_{eff,K}$\tablenotemark{c}}&
\colhead{$\nabla_{g-r}$\tablenotemark{d}}&
\colhead{$\nabla_{r-K}$\tablenotemark{e}}&
\colhead{$\nabla_{g-K}$\tablenotemark{f}}&
\colhead{$\nabla_{t}$\tablenotemark{g}}&
\colhead{$\nabla_{Z}$\tablenotemark{h}}&
\colhead{$\mathrm{M}_{K}$\tablenotemark{i}}&
\colhead{$\mathrm{M}_{\ast}$\tablenotemark{j}}&
\colhead{B/T\tablenotemark{k}}&
\colhead{Type\tablenotemark{l}}
}
\startdata
24 & 0.0350 & 5.28 & 3.21 & $-0.10\pm0.04$ & $-0.20\pm0.15$ & $-0.24\pm0.16$ & $-0.13\pm0.40$ & $-0.04\pm0.29$ & $-21.46\pm0.02$ & $10.66\pm0.01$ & $1.00$ & R\\
76 & 0.0431 & 2.35 & 3.21 & $-0.10\pm0.04$ & $-0.34\pm0.22$ & $-0.51\pm0.22$ & $-0.01\pm0.00$ & $-0.22\pm0.10$ & $-21.77\pm0.02$ & $10.79\pm0.01$ & $0.93$ & R\\
89 & 0.0275 & 7.92 & 6.42 & $0.00\pm0.04$ & $-0.64\pm0.06$ & $-0.61\pm0.07$ & $0.59\pm0.12$ & $-0.77\pm0.06$ & $-22.41\pm0.01$ & $11.05\pm0.00$ & $0.65$ & T\\
112 & 0.0313 & 11.88 & 8.03 & $-0.08\pm0.01$ & $-0.36\pm0.08$ & $-0.41\pm0.08$ & $0.11\pm0.13$ & $-0.39\pm0.14$ & $-23.17\pm0.01$ & $11.35\pm0.00$ & $0.73$ & R\\
135 & 0.0302 & 7.92 & 4.82 & $-0.07\pm0.03$ & $-0.53\pm0.14$ & $-0.67\pm0.14$ & $-0.04\pm0.38$ & $-0.54\pm0.25$ & $-21.60\pm0.02$ & $10.72\pm0.01$ & $0.52$ & T\\
192 & 0.0267 & 5.28 & 4.82 & $-0.09\pm0.02$ & $-0.58\pm0.12$ & $-0.62\pm0.12$ & $0.65\pm0.14$ & $-0.82\pm0.16$ & $-21.86\pm0.01$ & $10.83\pm0.01$ & $0.52$ & R\\
221 & 0.0306 & 5.28 & 3.21 & $-0.04\pm0.04$ & $-0.45\pm0.19$ & $-0.44\pm0.19$ & $0.34\pm0.36$ & $-0.71\pm0.28$ & $-21.57\pm0.02$ & $10.71\pm0.01$ & $0.79$ & R\\
261 & 0.0422 & 5.28 & 3.21 & $-0.10\pm0.03$ & $-0.19\pm0.50$ & $-0.23\pm0.50$ & $0.09\pm2.20$ & $-0.21\pm1.52$ & $-21.52\pm0.03$ & $10.69\pm0.01$ & $0.68$ & T\\
269 & 0.0444 & 5.28 & 3.21 & $-0.06\pm0.02$ & $-0.30\pm0.13$ & $-0.35\pm0.14$ & $0.07\pm0.24$ & $-0.31\pm0.25$ & $-22.71\pm0.01$ & $11.16\pm0.01$ & $0.77$ & T\\
272 & 0.0443 & 3.52 & 3.21 & $0.06\pm0.04$ & $-0.35\pm0.32$ & $-0.16\pm0.33$ & $0.71\pm0.48$ & $-0.64\pm0.30$ & $-21.91\pm0.02$ & $10.85\pm0.01$ & $0.67$ & R\\
280 & 0.0448 & 11.88 & 6.42 & $-0.09\pm0.02$ & $-0.64\pm0.10$ & $-0.78\pm0.10$ & $0.20\pm0.22$ & $-0.72\pm0.20$ & $-23.31\pm0.01$ & $11.41\pm0.00$ & $0.52$ & T\\
284 & 0.0500 & 3.52 & 3.21 & $-0.06\pm0.03$ & $-0.31\pm0.19$ & $-0.38\pm0.19$ & $0.15\pm0.44$ & $-0.22\pm0.30$ & $-22.62\pm0.02$ & $11.13\pm0.01$ & $1.00$ & R\\
326 & 0.0395 & 5.28 & 3.21 & $-0.07\pm0.04$ & $-0.58\pm0.14$ & $-0.57\pm0.15$ & $0.97\pm0.34$ & $-0.91\pm0.26$ & $-22.14\pm0.02$ & $10.94\pm0.01$ & $0.58$ & R\\
340 & 0.0483 & 5.28 & 3.21 & $-0.05\pm0.03$ & $-0.29\pm0.16$ & $-0.30\pm0.16$ & $0.37\pm0.34$ & $-0.30\pm0.22$ & $-22.50\pm0.02$ & $11.08\pm0.01$ & $0.57$ & R\\
346 & 0.0497 & 2.81 & 3.21 & $-0.08\pm0.14$ & $-0.09\pm0.53$ & $-0.14\pm0.52$ & $-0.18\pm1.28$ & $0.07\pm0.85$ & $-21.55\pm0.02$ & $10.70\pm0.01$ & $0.80$ & R\\
354 & 0.0380 & 1.68 & 1.61 & $-0.03\pm0.09$ & $-0.57\pm0.75$ & $0.06\pm0.78$ & $4.76\pm1.92$ & $-2.69\pm1.41$ & $-21.60\pm0.02$ & $10.72\pm0.01$ & $0.58$ & T\\
359 & 0.0334 & 7.92 & 4.82 & $-0.07\pm0.03$ & $-0.24\pm0.07$ & $-0.22\pm0.08$ & $0.68\pm0.25$ & $-0.41\pm0.15$ & $-22.41\pm0.01$ & $11.04\pm0.01$ & $0.52$ & R\\
415 & 0.0453 & 17.82 & 16.05 & $-0.06\pm0.01$ & $-0.25\pm0.03$ & $-0.28\pm0.03$ & $0.07\pm0.05$ & $-0.29\pm0.06$ & $-24.37\pm0.02$ & $11.83\pm0.01$ & $0.82$ & R\\
448 & 0.0274 & 7.92 & 4.82 & $-0.01\pm0.01$ & $-0.31\pm0.08$ & $-0.33\pm0.08$ & $0.15\pm0.12$ & $-0.43\pm0.09$ & $-22.04\pm0.01$ & $10.90\pm0.00$ & $0.69$ & R\\
477 & 0.0475 & 4.22 & 3.21 & $-0.14\pm0.07$ & $-0.85\pm0.64$ & $-0.60\pm0.65$ & $0.00\pm0.00$ & $-0.70\pm0.35$ & $-21.75\pm0.02$ & $10.78\pm0.01$ & $0.53$ & R\\
481 & 0.0177 & 17.82 & 14.45 & $-0.11\pm0.01$ & $-0.49\pm0.06$ & $-0.64\pm0.06$ & $-0.10\pm0.22$ & $-0.39\pm0.13$ & $-21.95\pm0.02$ & $10.86\pm0.01$ & $0.68$ & T\\
483 & 0.0353 & 3.52 & 3.21 & $-0.07\pm0.06$ & $-0.15\pm0.19$ & $-0.17\pm0.19$ & $0.00\pm0.00$ & $-0.20\pm0.34$ & $-21.70\pm0.01$ & $10.76\pm0.00$ & $0.56$ & R\\
506 & 0.0476 & 7.92 & 4.82 & $-0.10\pm0.03$ & $-0.02\pm0.14$ & $-0.07\pm0.13$ & $-0.46\pm0.63$ & $0.14\pm0.29$ & $-22.78\pm0.02$ & $11.19\pm0.01$ & $0.56$ & R\\
518 & 0.0407 & 7.92 & 6.02 & $-0.11\pm0.01$ & $-0.27\pm0.07$ & $-0.33\pm0.07$ & $0.45\pm0.18$ & $-0.43\pm0.10$ & $-23.21\pm0.01$ & $11.36\pm0.00$ & $0.71$ & D\\
528 & 0.0412 & 11.88 & 4.41 & $-0.04\pm0.03$ & $-0.67\pm0.16$ & $-0.84\pm0.14$ & $-0.09\pm0.29$ & $-0.77\pm0.23$ & $-22.21\pm0.02$ & $10.97\pm0.01$ & $0.50$ & D\\
536 & 0.0387 & 5.28 & 4.82 & $-0.10\pm0.07$ & $-0.63\pm0.37$ & $-0.64\pm0.38$ & $0.49\pm0.62$ & $-0.97\pm0.59$ & $-22.00\pm0.02$ & $10.88\pm0.01$ & $0.58$ & D\\
537 & 0.0369 & 7.92 & 4.82 & $-0.09\pm0.02$ & $-0.20\pm0.11$ & $-0.24\pm0.11$ & $0.06\pm0.43$ & $-0.19\pm0.18$ & $-22.23\pm0.02$ & $10.97\pm0.01$ & $0.56$ & R\\
556 & 0.0326 & 5.28 & 3.21 & $0.02\pm0.03$ & $-0.30\pm0.13$ & $-0.30\pm0.13$ & $0.43\pm0.26$ & $-0.41\pm0.19$ & $-22.20\pm0.02$ & $10.96\pm0.01$ & $0.87$ & R\\
560 & 0.0235 & 5.28 & 4.82 & $-0.21\pm0.04$ & $-0.30\pm0.12$ & $-0.47\pm0.12$ & $-0.28\pm0.22$ & $0.00\pm0.16$ & $-21.39\pm0.01$ & $10.64\pm0.00$ & $0.62$ & R\\
561 & 0.0494 & 3.52 & 3.21 & $0.01\pm0.06$ & $-0.23\pm0.19$ & $-0.18\pm0.21$ & $0.27\pm0.54$ & $-0.44\pm0.40$ & $-22.73\pm0.02$ & $11.17\pm0.01$ & $0.75$ & T\\
597 & 0.0460 & 3.52 & 3.21 & $-0.01\pm0.05$ & $-0.29\pm0.24$ & $-0.30\pm0.25$ & $0.13\pm0.32$ & $-0.36\pm0.16$ & $-22.07\pm0.02$ & $10.91\pm0.01$ & $0.59$ & R\\
601 & 0.0474 & 5.28 & 3.21 & $0.05\pm0.05$ & $-0.29\pm0.20$ & $-0.21\pm0.21$ & $0.41\pm0.41$ & $-0.55\pm0.29$ & $-22.30\pm0.03$ & $11.00\pm0.01$ & $0.74$ & R\\
606 & 0.0468 & 5.28 & 4.82 & $-0.06\pm0.03$ & $-0.61\pm0.12$ & $-0.66\pm0.12$ & $0.43\pm0.18$ & $-0.89\pm0.17$ & $-22.89\pm0.01$ & $11.24\pm0.00$ & $0.73$ & R\\
624 & 0.0426 & 3.52 & 3.21 & $-0.08\pm0.07$ & $-0.33\pm0.29$ & $-0.32\pm0.28$ & $0.98\pm1.20$ & $-1.15\pm0.72$ & $-21.92\pm0.02$ & $10.85\pm0.01$ & $0.64$ & R\\
639 & 0.0337 & 5.28 & 3.21 & $-0.04\pm0.05$ & $-0.47\pm0.25$ & $-0.42\pm0.26$ & $1.34\pm0.47$ & $-1.04\pm0.35$ & $-21.40\pm0.02$ & $10.64\pm0.01$ & $0.70$ & R\\
640 & 0.0418 & 5.50 & 8.03 & $-0.07\pm0.03$ & $-0.47\pm0.12$ & $-0.18\pm0.39$ & $0.52\pm1.73$ & $-0.34\pm1.16$ & $-22.18\pm0.05$ & $10.95\pm0.02$ & $0.64$ & R\\
675 & 0.0366 & 5.28 & 4.82 & $-0.06\pm0.03$ & $-0.45\pm0.10$ & $-0.49\pm0.10$ & $0.75\pm0.19$ & $-0.73\pm0.14$ & $-22.34\pm0.02$ & $11.02\pm0.01$ & $0.74$ & R\\
681 & 0.0427 & 7.92 & 4.82 & $-0.09\pm0.03$ & $-0.35\pm0.10$ & $-0.42\pm0.10$ & $0.04\pm0.29$ & $-0.41\pm0.21$ & $-23.35\pm0.01$ & $11.42\pm0.00$ & $0.94$ & R\\
723 & 0.0473 & 3.52 & 3.21 & $-0.05\pm0.06$ & $-0.33\pm0.14$ & $-0.33\pm0.14$ & $0.50\pm0.43$ & $-0.51\pm0.31$ & $-22.57\pm0.02$ & $11.11\pm0.01$ & $0.87$ & T\\
731 & 0.0456 & 2.99 & 3.21 & $0.12\pm0.08$ & $-0.53\pm0.41$ & $-0.09\pm0.40$ & $0.45\pm0.81$ & $-0.31\pm0.54$ & $-21.40\pm0.03$ & $10.64\pm0.01$ & $0.71$ & R\\
739 & 0.0297 & 7.92 & 4.82 & $-0.04\pm0.10$ & $1.03\pm0.88$ & $0.91\pm0.84$ & $-0.99\pm3.21$ & $1.64\pm2.27$ & $-22.08\pm0.02$ & $10.91\pm0.01$ & $0.64$ & R\\
745 & 0.0475 & 5.28 & 4.82 & $-0.11\pm0.03$ & $-0.50\pm0.15$ & $-0.50\pm0.16$ & $0.46\pm0.39$ & $-0.82\pm0.29$ & $-22.62\pm0.02$ & $11.13\pm0.01$ & $0.77$ & R\\
776 & 0.0467 & 4.24 & 3.21 & $-0.10\pm0.04$ & $-0.59\pm0.34$ & $-0.45\pm0.34$ & $0.18\pm0.53$ & $-0.07\pm0.42$ & $-22.16\pm0.03$ & $10.94\pm0.01$ & $1.00$ & R\\
789 & 0.0405 & 5.28 & 4.82 & $-0.04\pm0.02$ & $-0.28\pm0.11$ & $-0.33\pm0.11$ & $0.05\pm0.22$ & $-0.36\pm0.20$ & $-22.93\pm0.01$ & $11.25\pm0.01$ & $0.78$ & R\\
822 & 0.0479 & 11.88 & 6.42 & $-0.03\pm0.03$ & $-0.34\pm0.08$ & $-0.32\pm0.08$ & $0.25\pm0.17$ & $-0.61\pm0.13$ & $-23.71\pm0.01$ & $11.57\pm0.00$ & $0.71$ & R\\
825 & 0.0236 & 17.82 & 6.02 & $-0.15\pm0.03$ & $-0.16\pm0.08$ & $-0.25\pm0.08$ & $-0.31\pm0.18$ & $0.03\pm0.13$ & $-22.44\pm0.01$ & $11.05\pm0.00$ & $0.55$ & D\\
829 & 0.0485 & 11.88 & 6.42 & $-0.05\pm0.02$ & $-0.51\pm0.10$ & $-0.53\pm0.10$ & $0.62\pm0.17$ & $-0.82\pm0.18$ & $-23.63\pm0.01$ & $11.53\pm0.00$ & $0.57$ & T\\
831 & 0.0490 & 6.50 & 4.82 & $-0.06\pm0.01$ & $-0.31\pm0.06$ & $-0.36\pm0.06$ & $0.13\pm0.15$ & $-0.22\pm0.08$ & $-22.95\pm0.01$ & $11.26\pm0.00$ & $0.72$ & R\\
844 & 0.0346 & 5.28 & 4.82 & $-0.08\pm0.02$ & $-0.05\pm0.11$ & $-0.12\pm0.11$ & $-0.18\pm0.23$ & $0.18\pm0.22$ & $-22.54\pm0.02$ & $11.10\pm0.01$ & $0.90$ & R\\
900 & 0.0440 & 5.28 & 3.21 & $0.02\pm0.04$ & $-0.40\pm0.22$ & $-0.29\pm0.22$ & $0.56\pm0.30$ & $-0.73\pm0.25$ & $-22.86\pm0.02$ & $11.22\pm0.01$ & $0.66$ & T\\
947 & 0.0443 & 5.28 & 3.21 & $-0.10\pm0.03$ & $-0.38\pm0.10$ & $-0.45\pm0.10$ & $0.08\pm0.26$ & $-0.44\pm0.21$ & $-22.63\pm0.01$ & $11.13\pm0.00$ & $0.69$ & R\\
983 & 0.0127 & 11.88 & 14.45 & $-0.06\pm0.02$ & $-0.25\pm0.03$ & $-0.28\pm0.03$ & $0.41\pm0.09$ & $-0.45\pm0.05$ & $-22.21\pm0.01$ & $10.96\pm0.00$ & $0.96$ & R\\
1006 & 0.0500 & 2.88 & 3.21 & $-0.09\pm0.06$ & $-0.69\pm0.67$ & $-0.75\pm0.68$ & $1.62\pm2.17$ & $-1.46\pm1.25$ & $-21.91\pm0.02$ & $10.84\pm0.01$ & $0.95$ & R\\
1025 & 0.0480 & 2.35 & 1.61 & $0.11\pm0.15$ & $0.82\pm1.06$ & $0.94\pm1.07$ & $0.00\pm0.00$ & $0.86\pm0.98$ & $-21.47\pm0.02$ & $10.67\pm0.01$ & $0.51$ & R\\
1029 & 0.0498 & 3.58 & 3.21 & $-0.02\pm0.10$ & $-0.44\pm0.23$ & $-0.37\pm0.41$ & $0.19\pm0.98$ & $-0.55\pm0.60$ & $-22.84\pm0.01$ & $11.22\pm0.00$ & $0.84$ & R\\
1098 & 0.0472 & 3.01 & 3.21 & $0.04\pm0.06$ & $-0.30\pm0.53$ & $-0.02\pm0.53$ & $-0.01\pm0.00$ & $-0.21\pm0.95$ & $-21.63\pm0.04$ & $10.73\pm0.02$ & $0.81$ & R\\
1103 & 0.0392 & 7.92 & 3.21 & $-0.10\pm0.14$ & $-0.20\pm0.65$ & $-0.25\pm0.69$ & $0.05\pm2.87$ & $-0.33\pm1.73$ & $-22.17\pm0.01$ & $10.95\pm0.01$ & $0.96$ & R\\
1108 & 0.0468 & 4.18 & 3.21 & $0.09\pm0.03$ & $-0.20\pm0.60$ & $0.03\pm0.61$ & $0.01\pm0.00$ & $-0.10\pm0.84$ & $-21.67\pm0.02$ & $10.75\pm0.01$ & $0.63$ & R\\
1114 & 0.0440 & 5.28 & 4.82 & $-0.03\pm0.04$ & $-0.58\pm0.21$ & $-0.57\pm0.21$ & $0.65\pm0.29$ & $-0.83\pm0.26$ & $-22.43\pm0.01$ & $11.05\pm0.01$ & $0.93$ & R\\
1115 & 0.0446 & 2.92 & 3.21 & $-0.03\pm0.08$ & $0.14\pm0.37$ & $0.11\pm0.37$ & $-0.41\pm0.82$ & $0.86\pm0.60$ & $-22.02\pm0.02$ & $10.89\pm0.01$ & $0.55$ & R\\
1116 & 0.0443 & 2.35 & 3.21 & $-0.06\pm0.08$ & $0.49\pm0.23$ & $0.51\pm0.23$ & $-0.47\pm0.72$ & $1.08\pm0.42$ & $-22.01\pm0.02$ & $10.88\pm0.01$ & $0.78$ & R\\
1177 & 0.0441 & 3.52 & 4.82 & $-0.01\pm0.03$ & $-0.31\pm0.20$ & $-0.26\pm0.20$ & $0.32\pm0.25$ & $-0.47\pm0.25$ & $-22.18\pm0.02$ & $10.95\pm0.01$ & $0.53$ & R\\
1186 & 0.0421 & 5.28 & 4.82 & $-0.11\pm0.03$ & $-0.68\pm0.20$ & $-0.76\pm0.20$ & $0.52\pm0.19$ & $-0.81\pm0.24$ & $-22.12\pm0.03$ & $10.93\pm0.01$ & $0.95$ & R\\
1191 & 0.0431 & 3.52 & 3.21 & $0.00\pm0.00$ & $-0.37\pm0.27$ & $-0.28\pm0.27$ & $0.53\pm0.36$ & $-0.65\pm0.25$ & $-21.41\pm0.02$ & $10.65\pm0.01$ & $1.00$ & R\\
1203 & 0.0435 & 4.18 & 3.21 & $-0.10\pm0.04$ & $-0.37\pm0.55$ & $-0.00\pm0.00$ & $0.01\pm0.00$ & $-0.10\pm0.29$ & $-21.51\pm0.03$ & $10.68\pm0.01$ & $0.73$ & R\\
1217 & 0.0232 & 11.88 & 6.02 & $-0.22\pm0.04$ & $-1.22\pm0.08$ & $-1.48\pm0.08$ & $-0.03\pm0.13$ & $-0.87\pm0.04$ & $-21.74\pm0.01$ & $10.78\pm0.00$ & $0.61$ & D\\
1221 & 0.0432 & 3.52 & 4.82 & $-0.10\pm0.05$ & $-0.36\pm0.20$ & $-0.42\pm0.21$ & $0.07\pm0.30$ & $-0.44\pm0.25$ & $-22.39\pm0.01$ & $11.04\pm0.01$ & $0.57$ & R\\
1228 & 0.0214 & 11.88 & 8.03 & $-0.06\pm0.01$ & $-0.25\pm0.04$ & $-0.29\pm0.04$ & $0.37\pm0.15$ & $-0.40\pm0.10$ & $-22.51\pm0.01$ & $11.08\pm0.00$ & $0.52$ & R\\
1238 & 0.0233 & 7.92 & 4.82 & $0.01\pm0.03$ & $-0.32\pm0.09$ & $-0.35\pm0.09$ & $0.35\pm0.30$ & $-0.37\pm0.18$ & $-21.47\pm0.02$ & $10.67\pm0.01$ & $0.57$ & R\\
1254 & 0.0495 & 5.28 & 3.21 & $-0.04\pm0.04$ & $-0.46\pm0.32$ & $-0.56\pm0.31$ & $0.05\pm0.44$ & $-0.50\pm0.43$ & $-22.26\pm0.03$ & $10.98\pm0.01$ & $0.69$ & R\\
1280 & 0.0275 & 5.28 & 3.21 & $-0.09\pm0.03$ & $-0.52\pm0.17$ & $-0.57\pm0.17$ & $0.57\pm0.39$ & $-0.66\pm0.29$ & $-21.84\pm0.01$ & $10.82\pm0.00$ & $0.85$ & R\\
1287 & 0.0446 & 5.28 & 3.21 & $-0.02\pm0.04$ & $-0.06\pm0.15$ & $0.05\pm0.16$ & $0.63\pm0.59$ & $-0.25\pm0.34$ & $-22.06\pm0.02$ & $10.90\pm0.01$ & $0.63$ & R\\
1322 & 0.0437 & 11.88 & 9.63 & $-0.06\pm0.01$ & $-0.47\pm0.08$ & $-0.51\pm0.08$ & $0.61\pm0.13$ & $-0.63\pm0.12$ & $-23.34\pm0.02$ & $11.41\pm0.01$ & $0.59$ & R\\
1330 & 0.0419 & 3.52 & 2.81 & $-0.03\pm0.03$ & $-0.45\pm0.48$ & $-0.49\pm0.48$ & $0.17\pm0.43$ & $-0.64\pm0.54$ & $-21.60\pm0.02$ & $10.72\pm0.01$ & $0.89$ & R\\
1335 & 0.0454 & 7.92 & 4.82 & $-0.07\pm0.02$ & $-0.47\pm0.12$ & $-0.53\pm0.12$ & $0.47\pm0.30$ & $-0.58\pm0.22$ & $-22.93\pm0.01$ & $11.25\pm0.01$ & $0.53$ & R\\
1339 & 0.0443 & 5.28 & 4.82 & $-0.05\pm0.03$ & $-0.30\pm0.13$ & $-0.29\pm0.12$ & $0.54\pm0.38$ & $-0.31\pm0.28$ & $-22.34\pm0.02$ & $11.02\pm0.01$ & $0.99$ & R\\
1352 & 0.0361 & 5.28 & 4.82 & $-0.02\pm0.03$ & $-0.43\pm0.20$ & $-0.39\pm0.19$ & $0.63\pm0.28$ & $-0.75\pm0.29$ & $-21.74\pm0.05$ & $10.77\pm0.02$ & $0.73$ & R\\
1367 & 0.0359 & 5.28 & 3.21 & $-0.16\pm0.04$ & $-0.53\pm0.16$ & $-0.73\pm0.16$ & $-0.25\pm0.36$ & $-0.35\pm0.28$ & $-22.11\pm0.01$ & $10.92\pm0.00$ & $0.65$ & R\\
1377 & 0.0421 & 4.24 & 3.21 & $-0.19\pm0.03$ & $-0.25\pm0.27$ & $-0.25\pm0.28$ & $-0.00\pm0.00$ & $-0.22\pm0.36$ & $-21.87\pm0.02$ & $10.83\pm0.01$ & $0.58$ & R\\
1406 & 0.0493 & 5.28 & 6.42 & $-0.00\pm0.03$ & $-0.11\pm0.18$ & $-0.07\pm0.18$ & $0.16\pm0.27$ & $-0.11\pm0.26$ & $-23.17\pm0.03$ & $11.35\pm0.01$ & $0.74$ & T\\
1421 & 0.0248 & 5.28 & 4.82 & $-0.01\pm0.04$ & $-0.31\pm0.18$ & $-0.17\pm0.18$ & $0.58\pm0.21$ & $-0.49\pm0.15$ & $-21.42\pm0.01$ & $10.65\pm0.01$ & $0.64$ & R\\
1436 & 0.0364 & 7.92 & 6.42 & $-0.03\pm0.03$ & $-0.31\pm0.09$ & $-0.31\pm0.09$ & $0.26\pm0.16$ & $-0.37\pm0.09$ & $-23.13\pm0.01$ & $11.33\pm0.01$ & $0.57$ & T\\
1440 & 0.0489 & 5.28 & 3.21 & $-0.09\pm0.04$ & $-0.51\pm0.15$ & $-0.69\pm0.16$ & $0.13\pm0.78$ & $-0.59\pm0.32$ & $-23.03\pm0.01$ & $11.29\pm0.01$ & $0.82$ & R\\
1451 & 0.0466 & 3.52 & 3.21 & $-0.08\pm0.07$ & $-0.06\pm0.22$ & $-0.00\pm0.20$ & $0.13\pm0.36$ & $-0.11\pm0.32$ & $-21.75\pm0.01$ & $10.78\pm0.01$ & $0.53$ & R\\
1461 & 0.0419 & 5.28 & 4.82 & $-0.01\pm0.04$ & $-0.28\pm0.22$ & $-0.28\pm0.23$ & $0.22\pm0.33$ & $-0.44\pm0.28$ & $-22.11\pm0.02$ & $10.93\pm0.01$ & $0.59$ & R\\
1462 & 0.0451 & 7.92 & 4.82 & $-0.10\pm0.02$ & $-0.63\pm0.08$ & $-0.71\pm0.08$ & $0.26\pm0.15$ & $-0.94\pm0.12$ & $-23.06\pm0.01$ & $11.30\pm0.00$ & $0.60$ & R\\
1470 & 0.0472 & 3.94 & 3.21 & $0.07\pm0.02$ & $-0.29\pm0.12$ & $-0.24\pm0.12$ & $0.56\pm0.13$ & $-0.56\pm0.09$ & $-22.36\pm0.01$ & $11.02\pm0.00$ & $0.57$ & R\\
1477 & 0.0443 & 5.28 & 4.82 & $-0.09\pm0.02$ & $-0.64\pm0.20$ & $-0.74\pm0.20$ & $0.13\pm0.22$ & $-0.72\pm0.31$ & $-22.91\pm0.01$ & $11.24\pm0.00$ & $0.65$ & R\\
1526 & 0.0370 & 5.18 & 4.82 & $0.03\pm0.03$ & $-0.48\pm0.35$ & $-0.01\pm0.37$ & $3.36\pm1.22$ & $-1.81\pm0.73$ & $-21.31\pm0.04$ & $10.60\pm0.02$ & $0.60$ & R\\
1551 & 0.0439 & 3.17 & 3.21 & $0.03\pm0.12$ & $-0.49\pm0.51$ & $-0.55\pm0.50$ & $0.10\pm0.51$ & $-0.46\pm0.48$ & $-21.62\pm0.03$ & $10.73\pm0.01$ & $0.65$ & R\\
1670 & 0.0431 & 5.28 & 4.82 & $-0.05\pm0.06$ & $-0.15\pm0.11$ & $-0.13\pm0.12$ & $0.15\pm0.34$ & $-0.28\pm0.21$ & $-22.56\pm0.01$ & $11.11\pm0.00$ & $0.60$ & R\\
1694 & 0.0370 & 3.38 & 3.21 & $-0.04\pm0.03$ & $-0.50\pm0.20$ & $-0.15\pm0.22$ & $0.59\pm0.45$ & $-0.55\pm0.25$ & $-21.49\pm0.02$ & $10.67\pm0.01$ & $0.76$ & R\\
1698 & 0.0393 & 5.28 & 4.82 & $-0.04\pm0.03$ & $-0.22\pm0.07$ & $-0.23\pm0.07$ & $0.11\pm0.22$ & $-0.29\pm0.14$ & $-22.86\pm0.03$ & $11.23\pm0.01$ & $0.61$ & R\\
1737 & 0.0462 & 5.28 & 4.82 & $-0.02\pm0.03$ & $-0.12\pm0.16$ & $-0.16\pm0.16$ & $-0.03\pm0.32$ & $-0.11\pm0.24$ & $-22.40\pm0.02$ & $11.04\pm0.01$ & $0.64$ & R\\
1775 & 0.0411 & 3.52 & 3.21 & $-0.08\pm0.03$ & $-0.18\pm0.20$ & $-0.21\pm0.20$ & $0.10\pm0.17$ & $-0.66\pm0.20$ & $-21.49\pm0.01$ & $10.67\pm0.00$ & $1.00$ & R\\
1777 & 0.0407 & 11.88 & 6.42 & $-0.05\pm0.01$ & $-0.35\pm0.04$ & $-0.41\pm0.04$ & $0.25\pm0.09$ & $-0.41\pm0.06$ & $-23.35\pm0.01$ & $11.42\pm0.01$ & $0.63$ & R\\
1779 & 0.0407 & 5.28 & 4.82 & $-0.04\pm0.04$ & $-0.26\pm0.08$ & $-0.30\pm0.09$ & $0.18\pm0.21$ & $-0.31\pm0.13$ & $-22.62\pm0.01$ & $11.13\pm0.00$ & $0.69$ & R\\
1790 & 0.0424 & 4.23 & 3.21 & $0.16\pm0.03$ & $0.41\pm0.65$ & $0.42\pm0.65$ & $0.01\pm0.00$ & $0.42\pm0.29$ & $-21.68\pm0.02$ & $10.75\pm0.01$ & $0.62$ & R\\
1791 & 0.0414 & 3.52 & 3.21 & $-0.00\pm0.07$ & $0.05\pm0.23$ & $0.16\pm0.25$ & $0.35\pm0.66$ & $-0.15\pm0.72$ & $-21.83\pm0.02$ & $10.81\pm0.01$ & $0.60$ & T\\
1792 & 0.0419 & 5.28 & 4.82 & $-0.08\pm0.04$ & $-0.46\pm0.16$ & $-0.52\pm0.16$ & $0.72\pm0.55$ & $-0.61\pm0.34$ & $-22.03\pm0.03$ & $10.89\pm0.01$ & $0.73$ & R\\
1808 & 0.0418 & 7.92 & 6.42 & $-0.06\pm0.02$ & $-0.32\pm0.11$ & $-0.34\pm0.11$ & $0.20\pm0.29$ & $-0.57\pm0.24$ & $-22.84\pm0.02$ & $11.22\pm0.01$ & $0.56$ & R\\
1818 & 0.0450 & 5.28 & 2.81 & $-0.07\pm0.04$ & $-0.64\pm0.30$ & $-0.65\pm0.30$ & $0.41\pm0.32$ & $-1.03\pm0.39$ & $-22.36\pm0.01$ & $11.02\pm0.00$ & $0.84$ & D\\
1823 & 0.0387 & 11.88 & 6.42 & $-0.04\pm0.02$ & $-0.27\pm0.03$ & $-0.32\pm0.03$ & $0.13\pm0.11$ & $-0.45\pm0.09$ & $-23.89\pm0.01$ & $11.64\pm0.00$ & $0.69$ & T\\
1862 & 0.0224 & 5.28 & 3.21 & $-0.02\pm0.02$ & $-0.07\pm0.13$ & $-0.10\pm0.13$ & $-0.05\pm0.27$ & $0.02\pm0.21$ & $-21.39\pm0.01$ & $10.64\pm0.00$ & $0.51$ & R\\
1864 & 0.0473 & 2.74 & 3.21 & $-0.03\pm0.08$ & $-0.55\pm0.32$ & $-0.60\pm0.32$ & $1.13\pm0.87$ & $-1.05\pm0.56$ & $-21.81\pm0.02$ & $10.80\pm0.01$ & $0.94$ & R\\
1865 & 0.0499 & 5.28 & 3.21 & $-0.02\pm0.03$ & $-0.25\pm0.20$ & $-0.25\pm0.20$ & $0.13\pm0.26$ & $-0.20\pm0.26$ & $-22.80\pm0.01$ & $11.20\pm0.00$ & $0.61$ & R\\
1874 & 0.0419 & 4.20 & 3.21 & $-0.12\pm0.03$ & $0.20\pm0.59$ & $0.10\pm0.59$ & $-1.31\pm2.23$ & $0.37\pm1.68$ & $-21.57\pm0.02$ & $10.71\pm0.01$ & $0.78$ & R\\
1880 & 0.0465 & 7.92 & 3.21 & $-0.08\pm0.02$ & $-0.63\pm0.10$ & $-0.67\pm0.10$ & $1.41\pm0.18$ & $-1.19\pm0.14$ & $-22.58\pm0.01$ & $11.11\pm0.00$ & $1.00$ & T\\
1890 & 0.0254 & 5.28 & 4.82 & $-0.10\pm0.02$ & $-0.56\pm0.11$ & $-0.66\pm0.11$ & $0.47\pm0.15$ & $-0.73\pm0.12$ & $-21.82\pm0.02$ & $10.81\pm0.01$ & $0.59$ & R\\
1968 & 0.0376 & 7.92 & 6.42 & $-0.18\pm0.02$ & $-0.43\pm0.04$ & $-0.53\pm0.04$ & $0.02\pm0.08$ & $-0.42\pm0.08$ & $-23.57\pm0.01$ & $11.51\pm0.00$ & $0.92$ & R\\
1977 & 0.0381 & 3.52 & 3.21 & $-0.02\pm0.06$ & $-0.61\pm0.25$ & $-0.59\pm0.24$ & $0.52\pm0.44$ & $-0.97\pm0.26$ & $-21.77\pm0.01$ & $10.79\pm0.01$ & $0.87$ & T\\
1988 & 0.0359 & 3.52 & 3.21 & $-0.10\pm0.05$ & $-0.30\pm0.18$ & $-0.27\pm0.18$ & $0.23\pm0.35$ & $-0.34\pm0.16$ & $-21.72\pm0.01$ & $10.77\pm0.00$ & $0.89$ & R\\
2003 & 0.0220 & 17.82 & 17.66 & $-0.11\pm0.02$ & $-0.27\pm0.03$ & $-0.33\pm0.03$ & $0.02\pm0.09$ & $-0.36\pm0.09$ & $-23.17\pm0.01$ & $11.35\pm0.01$ & $0.59$ & R\\
2027 & 0.0430 & 11.88 & 8.03 & $-0.04\pm0.01$ & $-0.38\pm0.05$ & $-0.43\pm0.05$ & $0.10\pm0.08$ & $-0.41\pm0.10$ & $-23.78\pm0.01$ & $11.59\pm0.00$ & $0.77$ & R\\
2040 & 0.0418 & 5.28 & 3.21 & $-0.05\pm0.03$ & $-0.56\pm0.18$ & $-0.69\pm0.18$ & $0.03\pm0.23$ & $-0.62\pm0.26$ & $-22.44\pm0.01$ & $11.06\pm0.00$ & $0.50$ & R\\
2064 & 0.0407 & 2.95 & 3.21 & $-0.13\pm0.07$ & $-0.61\pm0.41$ & $-0.72\pm0.42$ & $0.15\pm1.41$ & $-0.38\pm0.88$ & $-21.97\pm0.01$ & $10.87\pm0.01$ & $0.86$ & R\\
2068 & 0.0418 & 11.88 & 8.03 & $-0.05\pm0.02$ & $-0.16\pm0.09$ & $-0.22\pm0.09$ & $-0.08\pm0.16$ & $-0.08\pm0.17$ & $-23.35\pm0.01$ & $11.42\pm0.00$ & $0.57$ & T\\
2069 & 0.0432 & 5.28 & 4.82 & $-0.03\pm0.05$ & $-0.10\pm0.28$ & $-0.10\pm0.27$ & $0.14\pm0.43$ & $-0.33\pm0.36$ & $-21.83\pm0.02$ & $10.81\pm0.01$ & $0.52$ & R\\
2070 & 0.0419 & 3.52 & 3.21 & $-0.10\pm0.04$ & $-0.44\pm0.16$ & $-0.46\pm0.16$ & $0.19\pm0.33$ & $-0.33\pm0.25$ & $-21.98\pm0.02$ & $10.87\pm0.01$ & $0.70$ & R\\
2099 & 0.0249 & 13.86 & 6.42 & $-0.14\pm0.03$ & $-0.48\pm0.07$ & $-0.61\pm0.07$ & $0.22\pm0.32$ & $-0.40\pm0.11$ & $-21.97\pm0.01$ & $10.87\pm0.00$ & $0.62$ & R\\
2102 & 0.0254 & 5.28 & 4.82 & $-0.05\pm0.07$ & $-0.30\pm0.16$ & $-0.36\pm0.16$ & $0.13\pm0.40$ & $-0.23\pm0.19$ & $-21.82\pm0.01$ & $10.81\pm0.00$ & $0.58$ & R\\
2116 & 0.0416 & 11.88 & 6.02 & $-0.10\pm0.03$ & $-1.15\pm0.20$ & $-1.29\pm0.19$ & $0.98\pm0.21$ & $-1.46\pm0.20$ & $-23.31\pm0.01$ & $11.40\pm0.00$ & $0.51$ & D\\
2136 & 0.0371 & 7.92 & 6.42 & $-0.11\pm0.03$ & $-0.47\pm0.15$ & $-0.64\pm0.15$ & $-0.24\pm0.65$ & $-0.36\pm0.34$ & $-21.82\pm0.02$ & $10.81\pm0.01$ & $0.71$ & R\\
2146 & 0.0378 & 26.73 & 9.63 & $-0.06\pm0.01$ & $-0.33\pm0.04$ & $-0.38\pm0.04$ & $0.27\pm0.14$ & $-0.41\pm0.06$ & $-23.59\pm0.01$ & $11.52\pm0.00$ & $0.68$ & R\\
2165 & 0.0317 & 5.28 & 4.82 & $0.11\pm0.03$ & $0.06\pm0.18$ & $0.40\pm0.18$ & $0.31\pm0.29$ & $-0.25\pm0.10$ & $-22.29\pm0.02$ & $11.00\pm0.01$ & $0.90$ & R\\
2176 & 0.0470 & 4.50 & 3.21 & $-0.10\pm0.04$ & $-0.61\pm0.54$ & $-0.18\pm0.51$ & $3.19\pm2.47$ & $-1.68\pm1.46$ & $-21.79\pm0.02$ & $10.80\pm0.01$ & $1.00$ & R\\
2196 & 0.0391 & 11.88 & 8.03 & $-0.07\pm0.03$ & $-0.58\pm0.14$ & $-0.70\pm0.14$ & $0.20\pm0.23$ & $-0.48\pm0.25$ & $-23.35\pm0.02$ & $11.42\pm0.01$ & $0.56$ & R\\
2201 & 0.0496 & 5.28 & 4.82 & $-0.05\pm0.04$ & $-0.17\pm0.22$ & $-0.19\pm0.22$ & $0.33\pm0.56$ & $-0.21\pm0.42$ & $-22.42\pm0.03$ & $11.05\pm0.01$ & $0.71$ & R\\
2298 & 0.0336 & 5.28 & 3.21 & $0.09\pm0.03$ & $-0.32\pm0.19$ & $-0.26\pm0.20$ & $0.40\pm0.26$ & $-0.54\pm0.21$ & $-21.86\pm0.02$ & $10.82\pm0.01$ & $0.74$ & R\\
2331 & 0.0418 & 7.92 & 4.82 & $-0.05\pm0.04$ & $-0.31\pm0.14$ & $-0.37\pm0.15$ & $0.28\pm0.46$ & $-0.39\pm0.28$ & $-23.00\pm0.01$ & $11.28\pm0.01$ & $0.63$ & R\\
2345 & 0.0458 & 4.14 & 3.21 & $-0.07\pm0.03$ & $0.10\pm0.58$ & $0.07\pm0.57$ & $-0.49\pm1.69$ & $0.20\pm1.03$ & $-22.03\pm0.02$ & $10.89\pm0.01$ & $1.00$ & R\\
2346 & 0.0422 & 3.52 & 3.21 & $-0.05\pm0.04$ & $-0.35\pm0.15$ & $-0.36\pm0.16$ & $0.31\pm0.38$ & $-0.53\pm0.27$ & $-22.15\pm0.02$ & $10.94\pm0.01$ & $0.92$ & R\\
2361 & 0.0423 & 2.85 & 3.21 & $0.12\pm0.08$ & $-0.82\pm0.34$ & $-0.80\pm0.36$ & $0.84\pm0.47$ & $-1.86\pm1.04$ & $-21.57\pm0.02$ & $10.71\pm0.01$ & $0.84$ & R\\
2362 & 0.0283 & 5.28 & 3.21 & $-0.08\pm0.02$ & $-0.46\pm0.11$ & $-0.54\pm0.11$ & $0.31\pm0.17$ & $-0.50\pm0.13$ & $-21.69\pm0.01$ & $10.76\pm0.01$ & $1.00$ & R\\
2366 & 0.0380 & 3.52 & 3.21 & $-0.04\pm0.05$ & $-0.54\pm0.24$ & $-0.50\pm0.23$ & $0.94\pm0.55$ & $-0.89\pm0.45$ & $-21.83\pm0.01$ & $10.81\pm0.00$ & $0.51$ & R\\
2368 & 0.0294 & 5.28 & 4.82 & $-0.11\pm0.04$ & $-0.10\pm0.10$ & $-0.17\pm0.10$ & $-0.25\pm0.21$ & $0.31\pm0.14$ & $-21.68\pm0.01$ & $10.75\pm0.00$ & $0.52$ & R\\
2371 & 0.0135 & 17.82 & 12.44 & $-0.18\pm0.01$ & $-0.67\pm0.03$ & $-0.92\pm0.03$ & $0.23\pm0.09$ & $-0.76\pm0.05$ & $-22.19\pm0.01$ & $10.95\pm0.00$ & $0.75$ & D\\
2377 & 0.0391 & 4.06 & 3.21 & $-0.15\pm0.04$ & $-0.50\pm0.28$ & $-0.24\pm0.29$ & $-0.00\pm0.00$ & $-0.16\pm0.33$ & $-21.79\pm0.02$ & $10.79\pm0.01$ & $0.90$ & R\\
2381 & 0.0374 & 5.94 & 3.21 & $-0.28\pm0.05$ & $-0.94\pm0.30$ & $-1.23\pm0.30$ & $-0.23\pm0.53$ & $-0.72\pm0.39$ & $-21.97\pm0.02$ & $10.87\pm0.01$ & $0.93$ & T\\
2389 & 0.0407 & 3.52 & 3.21 & $0.05\pm0.05$ & $-0.05\pm0.20$ & $-0.02\pm0.22$ & $0.18\pm0.34$ & $-0.22\pm0.30$ & $-21.62\pm0.02$ & $10.73\pm0.01$ & $0.73$ & R\\
2422 & 0.0439 & 4.17 & 3.21 & $-0.03\pm0.04$ & $-0.37\pm0.66$ & $-0.18\pm0.67$ & $0.00\pm0.00$ & $-0.28\pm0.85$ & $-21.57\pm0.03$ & $10.71\pm0.01$ & $0.61$ & R\\
2424 & 0.0488 & 5.28 & 4.82 & $-0.06\pm0.04$ & $-0.44\pm0.17$ & $-0.44\pm0.17$ & $0.57\pm0.34$ & $-0.61\pm0.29$ & $-22.70\pm0.02$ & $11.16\pm0.01$ & $0.51$ & R\\
2431 & 0.0419 & 3.52 & 3.21 & $-0.06\pm0.02$ & $-0.45\pm0.15$ & $-0.49\pm0.15$ & $0.18\pm0.21$ & $-0.65\pm0.21$ & $-22.70\pm0.01$ & $11.16\pm0.00$ & $0.95$ & R\\
2438 & 0.0440 & 7.92 & 4.82 & $-0.04\pm0.02$ & $-0.46\pm0.21$ & $-0.45\pm0.21$ & $0.52\pm0.26$ & $-0.72\pm0.31$ & $-22.70\pm0.02$ & $11.16\pm0.01$ & $0.68$ & R\\
2440 & 0.0437 & 4.24 & 3.21 & $0.02\pm0.05$ & $0.21\pm0.38$ & $0.16\pm0.38$ & $-0.22\pm1.20$ & $0.47\pm0.71$ & $-22.23\pm0.03$ & $10.97\pm0.01$ & $0.82$ & R\\
2454 & 0.0442 & 5.28 & 4.41 & $-0.03\pm0.04$ & $-0.96\pm0.36$ & $-1.25\pm0.35$ & $-0.26\pm0.57$ & $-0.67\pm0.54$ & $-22.35\pm0.02$ & $11.02\pm0.01$ & $0.54$ & D\\
2467 & 0.0452 & 4.06 & 3.21 & $0.04\pm0.05$ & $-0.63\pm0.27$ & $-0.47\pm0.26$ & $-0.00\pm0.00$ & $-0.39\pm0.21$ & $-21.71\pm0.01$ & $10.76\pm0.00$ & $0.57$ & R\\
2494 & 0.0402 & 5.28 & 3.21 & $-0.06\pm0.03$ & $-0.16\pm0.12$ & $-0.21\pm0.12$ & $-0.15\pm0.33$ & $-0.17\pm0.15$ & $-23.01\pm0.01$ & $11.28\pm0.00$ & $0.71$ & R\\
2495 & 0.0242 & 7.92 & 6.42 & $-0.02\pm0.02$ & $-0.19\pm0.04$ & $-0.20\pm0.05$ & $0.15\pm0.13$ & $-0.39\pm0.10$ & $-23.04\pm0.01$ & $11.30\pm0.00$ & $0.84$ & T\\
2528 & 0.0443 & 5.28 & 4.82 & $0.06\pm0.03$ & $-0.86\pm0.13$ & $-0.85\pm0.13$ & $0.19\pm0.10$ & $-0.79\pm0.09$ & $-22.62\pm0.02$ & $11.13\pm0.01$ & $0.81$ & R\\
2548 & 0.0281 & 11.88 & 6.42 & $-0.04\pm0.01$ & $-0.43\pm0.04$ & $-0.48\pm0.04$ & $0.41\pm0.06$ & $-0.83\pm0.07$ & $-23.13\pm0.01$ & $11.33\pm0.00$ & $0.85$ & T\\
2585 & 0.0437 & 2.15 & 1.61 & $-0.04\pm0.07$ & $-0.32\pm0.50$ & $-0.25\pm0.49$ & $0.91\pm0.58$ & $-1.10\pm0.54$ & $-21.65\pm0.02$ & $10.74\pm0.01$ & $0.61$ & R\\
2587 & 0.0437 & 6.60 & 4.82 & $-0.03\pm0.03$ & $-0.78\pm0.27$ & $-0.91\pm0.52$ & $0.45\pm0.53$ & $-1.28\pm0.63$ & $-22.44\pm0.02$ & $11.06\pm0.01$ & $0.75$ & R\\
2648 & 0.0466 & 5.28 & 3.21 & $-0.04\pm0.02$ & $-0.59\pm0.14$ & $-0.57\pm0.14$ & $0.41\pm0.18$ & $-0.74\pm0.14$ & $-22.76\pm0.02$ & $11.18\pm0.01$ & $0.74$ & R\\
2664 & 0.0415 & 11.88 & 6.42 & $-0.06\pm0.02$ & $-0.23\pm0.15$ & $-0.29\pm0.15$ & $0.01\pm0.43$ & $-0.33\pm0.23$ & $-22.78\pm0.02$ & $11.19\pm0.01$ & $0.67$ & T\\
2676 & 0.0313 & 5.28 & 4.82 & $-0.10\pm0.03$ & $-0.35\pm0.10$ & $-0.40\pm0.10$ & $0.09\pm0.19$ & $-0.40\pm0.18$ & $-21.92\pm0.01$ & $10.85\pm0.00$ & $0.54$ & R\\
2684 & 0.0308 & 5.28 & 3.21 & $-0.12\pm0.05$ & $-0.41\pm0.12$ & $-0.49\pm0.12$ & $0.02\pm0.28$ & $-0.38\pm0.22$ & $-21.68\pm0.01$ & $10.75\pm0.00$ & $0.93$ & R\\
2696 & 0.0400 & 5.28 & 3.21 & $-0.03\pm0.03$ & $-0.50\pm0.28$ & $-0.53\pm0.28$ & $0.24\pm0.27$ & $-0.62\pm0.36$ & $-22.28\pm0.01$ & $10.99\pm0.01$ & $0.71$ & R\\
2712 & 0.0421 & 11.88 & 9.63 & $-0.08\pm0.02$ & $-0.20\pm0.05$ & $-0.30\pm0.05$ & $-0.13\pm0.12$ & $-0.07\pm0.10$ & $-23.57\pm0.01$ & $11.51\pm0.00$ & $0.50$ & T\\
2756 & 0.0398 & 3.52 & 3.21 & $-0.16\pm0.08$ & $-0.07\pm0.17$ & $-0.10\pm0.16$ & $-0.05\pm0.56$ & $-0.01\pm0.32$ & $-22.39\pm0.01$ & $11.04\pm0.00$ & $0.59$ & R\\
2760 & 0.0371 & 5.28 & 3.21 & $-0.02\pm0.04$ & $-0.63\pm0.12$ & $-0.63\pm0.13$ & $1.15\pm0.28$ & $-1.04\pm0.26$ & $-22.51\pm0.01$ & $11.08\pm0.00$ & $0.70$ & R\\
2783 & 0.0430 & 7.92 & 4.82 & $-0.07\pm0.03$ & $-0.17\pm0.14$ & $-0.18\pm0.14$ & $0.11\pm0.22$ & $-0.30\pm0.20$ & $-23.27\pm0.01$ & $11.39\pm0.00$ & $0.72$ & R\\
2820 & 0.0270 & 7.92 & 4.82 & $-0.12\pm0.04$ & $-0.39\pm0.07$ & $-0.47\pm0.08$ & $0.02\pm0.35$ & $-0.38\pm0.16$ & $-22.34\pm0.01$ & $11.02\pm0.00$ & $0.83$ & T\\
2829 & 0.0231 & 7.92 & 3.21 & $-0.10\pm0.06$ & $-0.43\pm0.14$ & $-0.60\pm0.14$ & $-0.13\pm0.58$ & $-0.23\pm0.32$ & $-21.67\pm0.01$ & $10.75\pm0.00$ & $0.88$ & R\\
2836 & 0.0341 & 3.52 & 3.21 & $-0.04\pm0.03$ & $-0.14\pm0.20$ & $-0.14\pm0.20$ & $0.09\pm0.24$ & $-0.22\pm0.25$ & $-21.73\pm0.01$ & $10.77\pm0.00$ & $0.70$ & R\\
2838 & 0.0458 & 5.28 & 3.21 & $-0.03\pm0.03$ & $-0.29\pm0.16$ & $-0.29\pm0.17$ & $0.59\pm0.35$ & $-0.56\pm0.26$ & $-22.53\pm0.02$ & $11.09\pm0.01$ & $0.63$ & T\\
2877 & 0.0386 & 3.52 & 3.21 & $-0.02\pm0.08$ & $-0.28\pm0.21$ & $-0.24\pm0.21$ & $0.83\pm0.89$ & $-0.48\pm0.54$ & $-21.46\pm0.02$ & $10.66\pm0.01$ & $0.56$ & R\\
2931 & 0.0180 & 17.82 & 9.63 & $-0.03\pm0.01$ & $-0.29\pm0.07$ & $-0.31\pm0.07$ & $0.39\pm0.19$ & $-0.43\pm0.14$ & $-21.65\pm0.02$ & $10.74\pm0.01$ & $0.56$ & R\\
2971 & 0.0360 & 5.28 & 3.21 & $-0.05\pm0.05$ & $-0.29\pm0.16$ & $-0.26\pm0.17$ & $0.97\pm0.61$ & $-0.52\pm0.32$ & $-22.07\pm0.01$ & $10.91\pm0.00$ & $1.00$ & R\\
2982 & 0.0425 & 7.92 & 4.82 & $-0.08\pm0.02$ & $-0.42\pm0.11$ & $-0.53\pm0.11$ & $0.02\pm0.14$ & $-0.41\pm0.14$ & $-23.27\pm0.01$ & $11.39\pm0.00$ & $0.71$ & R\\
2983 & 0.0446 & 5.28 & 3.21 & $-0.16\pm0.05$ & $-0.27\pm0.17$ & $-0.34\pm0.17$ & $-0.00\pm0.00$ & $-0.16\pm0.31$ & $-22.49\pm0.02$ & $11.07\pm0.01$ & $0.90$ & R\\
3049 & 0.0449 & 4.17 & 3.21 & $-0.05\pm0.05$ & $-0.48\pm0.52$ & $-0.36\pm0.53$ & $0.01\pm0.00$ & $-0.42\pm0.50$ & $-21.42\pm0.03$ & $10.65\pm0.01$ & $0.62$ & R\\
3073 & 0.0229 & 11.88 & 6.42 & $-0.03\pm0.02$ & $-0.35\pm0.03$ & $-0.40\pm0.04$ & $0.10\pm0.11$ & $-0.48\pm0.09$ & $-22.36\pm0.01$ & $11.03\pm0.00$ & $0.54$ & R\\
3097 & 0.0438 & 3.52 & 3.21 & $0.03\pm0.07$ & $0.02\pm0.36$ & $0.07\pm0.40$ & $0.17\pm0.73$ & $-0.03\pm0.68$ & $-21.44\pm0.02$ & $10.66\pm0.01$ & $0.50$ & R\\
3122 & 0.0487 & 3.22 & 3.21 & $-0.32\pm0.07$ & $0.23\pm0.29$ & $0.46\pm0.29$ & $0.19\pm0.97$ & $0.26\pm0.58$ & $-21.48\pm0.02$ & $10.67\pm0.01$ & $0.52$ & R\\
3124 & 0.0415 & 7.92 & 8.03 & $-0.03\pm0.01$ & $-0.01\pm0.05$ & $-0.02\pm0.05$ & $0.00\pm0.00$ & $0.04\pm0.09$ & $-23.24\pm0.02$ & $11.38\pm0.01$ & $0.56$ & T\\
3131 & 0.0437 & 7.92 & 4.82 & $-0.04\pm0.02$ & $-0.32\pm0.17$ & $-0.31\pm0.17$ & $0.59\pm0.38$ & $-0.44\pm0.28$ & $-22.52\pm0.02$ & $11.09\pm0.01$ & $0.50$ & R\\
3148 & 0.0291 & 11.88 & 8.03 & $-0.06\pm0.01$ & $-0.08\pm0.07$ & $-0.12\pm0.07$ & $-0.05\pm0.15$ & $-0.05\pm0.11$ & $-22.09\pm0.02$ & $10.91\pm0.01$ & $0.59$ & R\\
3191 & 0.0442 & 5.28 & 4.82 & $-0.05\pm0.05$ & $-0.02\pm0.46$ & $-0.00\pm0.53$ & $0.30\pm0.58$ & $0.43\pm0.54$ & $-21.99\pm0.03$ & $10.88\pm0.01$ & $0.64$ & T\\
3218 & 0.0431 & 17.82 & 11.24 & $-0.06\pm0.01$ & $-0.31\pm0.11$ & $-0.33\pm0.11$ & $0.12\pm0.16$ & $-0.41\pm0.17$ & $-23.17\pm0.03$ & $11.35\pm0.01$ & $0.69$ & R\\
3247 & 0.0450 & 5.28 & 3.21 & $-0.05\pm0.04$ & $-0.53\pm0.15$ & $-0.54\pm0.17$ & $0.57\pm0.37$ & $-0.99\pm0.22$ & $-22.35\pm0.01$ & $11.02\pm0.01$ & $1.00$ & R\\
3308 & 0.0345 & 3.52 & 3.21 & $-0.13\pm0.07$ & $-0.39\pm0.64$ & $-0.37\pm0.63$ & $0.24\pm0.72$ & $-0.46\pm0.46$ & $-21.36\pm0.01$ & $10.62\pm0.01$ & $0.87$ & T\\
3343 & 0.0456 & 5.28 & 3.21 & $-0.06\pm0.04$ & $-0.30\pm0.18$ & $-0.30\pm0.19$ & $0.94\pm0.74$ & $-0.64\pm0.36$ & $-22.37\pm0.03$ & $11.03\pm0.01$ & $0.62$ & R\\
3374 & 0.0268 & 11.88 & 4.82 & $-0.10\pm0.02$ & $-0.63\pm0.08$ & $-0.77\pm0.09$ & $0.17\pm0.15$ & $-0.69\pm0.13$ & $-21.82\pm0.01$ & $10.81\pm0.00$ & $0.57$ & T\\
3392 & 0.0388 & 11.88 & 6.42 & $-0.03\pm0.01$ & $-0.22\pm0.07$ & $-0.25\pm0.07$ & $0.14\pm0.17$ & $-0.20\pm0.12$ & $-22.90\pm0.02$ & $11.24\pm0.01$ & $0.61$ & R\\
3402 & 0.0437 & 4.00 & 3.21 & $-0.06\pm0.03$ & $-0.26\pm0.15$ & $-0.29\pm0.15$ & $0.18\pm0.22$ & $-0.57\pm0.16$ & $-22.38\pm0.01$ & $11.03\pm0.01$ & $0.91$ & R\\
3418 & 0.0448 & 17.82 & 12.84 & $-0.07\pm0.01$ & $-0.22\pm0.06$ & $-0.26\pm0.05$ & $0.01\pm0.09$ & $-0.23\pm0.11$ & $-24.15\pm0.01$ & $11.74\pm0.00$ & $0.99$ & T\\
3438 & 0.0250 & 3.52 & 3.21 & $-0.04\pm0.03$ & $-0.69\pm0.15$ & $-0.72\pm0.15$ & $0.66\pm0.21$ & $-0.94\pm0.13$ & $-21.46\pm0.01$ & $10.66\pm0.00$ & $0.91$ & R\\
3440 & 0.0293 & 11.88 & 8.03 & $-0.07\pm0.01$ & $-0.24\pm0.04$ & $-0.28\pm0.04$ & $0.11\pm0.11$ & $-0.29\pm0.07$ & $-23.35\pm0.01$ & $11.42\pm0.00$ & $0.66$ & R\\
3442 & 0.0182 & 7.92 & 6.42 & $-0.07\pm0.02$ & $-0.32\pm0.06$ & $-0.35\pm0.06$ & $0.11\pm0.12$ & $-0.52\pm0.10$ & $-21.88\pm0.01$ & $10.83\pm0.00$ & $0.56$ & R\\
3457 & 0.0493 & 3.10 & 3.21 & $0.02\pm0.12$ & $-0.33\pm0.23$ & $0.72\pm0.70$ & $-0.54\pm2.72$ & $0.87\pm2.03$ & $-21.43\pm0.02$ & $10.65\pm0.01$ & $0.68$ & R\\
3468 & 0.0188 & 7.92 & 6.42 & $-0.04\pm0.04$ & $-0.14\pm0.06$ & $-0.19\pm0.06$ & $-0.00\pm0.00$ & $-0.16\pm0.08$ & $-22.24\pm0.01$ & $10.98\pm0.00$ & $0.80$ & R\\
3471 & 0.0176 & 11.88 & 4.82 & $-0.03\pm0.01$ & $-0.38\pm0.04$ & $-0.39\pm0.04$ & $0.33\pm0.08$ & $-0.54\pm0.06$ & $-22.54\pm0.00$ & $11.09\pm0.00$ & $0.75$ & T\\
3499 & 0.0283 & 11.88 & 6.42 & $-0.07\pm0.02$ & $-0.35\pm0.10$ & $-0.45\pm0.10$ & $-0.07\pm0.16$ & $-0.31\pm0.12$ & $-22.42\pm0.01$ & $11.05\pm0.01$ & $0.75$ & R\\
3523 & 0.0498 & 2.35 & 3.21 & $-0.14\pm0.07$ & $-0.28\pm0.16$ & $-0.29\pm0.17$ & $0.61\pm0.63$ & $-0.38\pm0.29$ & $-22.39\pm0.01$ & $11.04\pm0.00$ & $0.64$ & R\\
3551 & 0.0478 & 5.35 & 3.21 & $-0.00\pm0.03$ & $-1.74\pm0.77$ & $-1.44\pm0.77$ & $1.96\pm1.98$ & $-2.27\pm1.29$ & $-21.69\pm0.03$ & $10.76\pm0.01$ & $0.59$ & R\\
3582 & 0.0182 & 7.92 & 4.82 & $-0.04\pm0.01$ & $-0.51\pm0.05$ & $-0.54\pm0.05$ & $0.24\pm0.11$ & $-0.68\pm0.06$ & $-21.42\pm0.01$ & $10.65\pm0.00$ & $0.87$ & R\\
3620 & 0.0431 & 7.92 & 4.82 & $-0.08\pm0.02$ & $-0.25\pm0.17$ & $-0.28\pm0.17$ & $0.39\pm0.27$ & $-0.40\pm0.20$ & $-22.76\pm0.02$ & $11.19\pm0.01$ & $0.70$ & R\\
3644 & 0.0445 & 5.28 & 3.21 & $0.00\pm0.04$ & $-0.06\pm0.45$ & $0.31\pm0.43$ & $1.11\pm0.70$ & $-0.69\pm0.46$ & $-21.79\pm0.02$ & $10.80\pm0.01$ & $1.00$ & R\\
3657 & 0.0280 & 17.82 & 7.62 & $-0.13\pm0.02$ & $-0.81\pm0.08$ & $-0.91\pm0.07$ & $0.14\pm0.08$ & $-1.64\pm0.06$ & $-23.28\pm0.01$ & $11.39\pm0.00$ & $0.85$ & D\\
3676 & 0.0365 & 3.52 & 3.21 & $-0.06\pm0.04$ & $-0.23\pm0.15$ & $-0.29\pm0.16$ & $0.03\pm0.35$ & $-0.34\pm0.31$ & $-22.01\pm0.01$ & $10.88\pm0.00$ & $0.94$ & R\\
3689 & 0.0419 & 7.92 & 6.42 & $-0.07\pm0.02$ & $-0.26\pm0.09$ & $-0.32\pm0.09$ & $0.00\pm0.08$ & $-0.36\pm0.10$ & $-23.77\pm0.01$ & $11.59\pm0.00$ & $0.85$ & T\\
3710 & 0.0179 & 7.92 & 6.42 & $-0.03\pm0.01$ & $-0.05\pm0.03$ & $-0.08\pm0.03$ & $-0.03\pm0.10$ & $-0.06\pm0.07$ & $-22.33\pm0.00$ & $11.01\pm0.00$ & $0.96$ & R\\
3725 & 0.0499 & 11.88 & 8.03 & $-0.04\pm0.02$ & $-0.30\pm0.08$ & $-0.36\pm0.08$ & $0.27\pm0.35$ & $-0.25\pm0.17$ & $-23.12\pm0.02$ & $11.33\pm0.01$ & $0.59$ & R\\
3726 & 0.0494 & 5.10 & 3.21 & $-0.11\pm0.04$ & $-1.16\pm0.25$ & $-1.33\pm0.25$ & $0.83\pm0.30$ & $-1.53\pm0.34$ & $-22.48\pm0.01$ & $11.07\pm0.00$ & $0.62$ & R\\
\enddata
\tablenotetext{a}{Object ID from the catalog of \citet{kav10}}
\tablenotetext{b}{Effective radius in $r$-band (arcsec)}
\tablenotetext{c}{Effective radius in $K$-band (arcsec)}
\tablenotetext{d}{Color gradient in $g-r$}
\tablenotetext{e}{Color gradient in $r-K$}
\tablenotetext{f}{Color gradient in $g-K$}
\tablenotetext{g}{Age gradient}
\tablenotetext{h}{Metallicity gradient}
\tablenotetext{i}{Absolute $K$-band magnitude (AB)}
\tablenotetext{j}{Stellar mass from absolute $K$-band magnitude}
\tablenotetext{k}{Bulge to total light ratio}
\tablenotetext{l}{Morphological type (R:relaxed, T:tidal-feature, D:dust-feature)}
\label{tab_sp}
\end{deluxetable}
\clearpage
\end{landscape}

\end{document}